\documentclass{aa}
\pdfoutput=1
\usepackage{amsmath,bm} %amsthm
\usepackage[varg]{txfonts}
\usepackage{natbib}
\usepackage{chemformula}

\usepackage{afterpage}
\usepackage[markup=bfit]{changes}
\usepackage{xcolor}

\usepackage{pdflscape}
\usepackage{afterpage}
\usepackage{mathtools}

\usepackage{booktabs}

\usepackage{color}
\usepackage{natbib,twoopt}
\usepackage[hyphenbreaks]{breakurl}
\usepackage[breaklinks]{hyperref}  

\begin{document} 
   \title{The galaxy cluster AC114\\ III. The role of galaxy clusters in the mass-metallicity relation}
   %\titlerunning{The galaxy cluster AC114 III: The role of galaxy clusters in the Mass-Metallicity relation}
   %\subtitle{ssss}
   
   \author{A. Andrade \inst{1,2} 
            \and I. Saviane\inst{2} 
            \and L. Monaco \inst{3}
            \and I. Yegorova\inst{2}
            \and D. Proust\inst{4}
          }

   \institute{Universidad Andres Bello, Facultad de Ciencias Exactas, Departamento de Ciencias Físicas - Instituto de Astrofísica, Fernández            Concha 700, Las Condes, Santiago, Chile. \\   
   \email{a.andradevalenzuela$@$uandresbello.edu}
    \and
        European Southern Observatory, Alonso de Cordova 3107, Vitacura, Casilla 19001, Santiago de Chile 19, Chile.
    \and 
        Universidad Andres Bello, Facultad de Ciencias Exactas, Departamento de Ciencias Físicas - Instituto de Astrofísica, Autopista
        Concepción-Talcahuano, 7100, Talcahuano, Chile.
    \and
        Observatoire de Paris - PSL, GEPI, F-92195 MEUDON, France.
        }

   \date{Received XX XX, 2024; accepted XX XX, 2024}

% \abstract{}{}{}{}{} 
% 5 {} token are mandatory
 
  \abstract
   {The mass-metallicity relation (MZR) is a powerful tool to constrain internal physical processes that drive the chemical evolution of galaxies. However, the construction of this relation is carried out with field star-forming galaxies in big data surveys where environmental effects are either negligible or not studied in detail.}
   {We study the role of galaxy clusters in the MZR and its evolution at z=0.317 with star-forming members of AC114 (ABELL S1077). The purpose of this work is to understand how both the environmental effects and dynamical events modify the chemical evolution in this galaxy cluster.}
   {Spectroscopic VIMOS/VLT data was used to select cluster members and classify the galaxy sample in star-forming and passive galaxies. Gas-phase metallicities were estimated by using the strong-line method O3N2 calibrated on Te-based oxygen abundances. Available optical and near-infrared (NIR) photometry from DECaLS DR10 and the VIKING DR4 ESO survey was used to derive the stellar mass of the galaxy sample.}
   {AC114 is dominated by passive galaxies located in the central region of the cluster, whereas the star-forming members tend to be located outside this region. The constructed MZR from the latter indicates that star-forming galaxies have a lower metal content than foreground galaxies (spanning redshifts up to z=0.28), and the same or even lower metallicities with respect to background galaxies (spanning redshifts 0.34 to 0.70). Additionally, it shows a higher scatter of $\sigma = 0.17$ dex, consistent with MZRs of galaxy clusters reported in the literature. The MZR at z=0.317 is downshifted by 0.19 dex on average with respect to local galaxies. Comparing the AC114-MZR with the field MZR at the same redshift, two galaxies are found to be more metal-rich than the field ones by $\sim 0.10$ dex. Likely as a result of ram-pressure stripping, star-forming galaxies deviate more from the MZR than field galaxies at the same redshift. Star-forming galaxies in the cluster are in general metal-poorer than field galaxies at the same redshfit up to $\sim 0.22$ dex, and show a MZR that is slightly shallower in slope compared with that of field galaxies. With a redshift analysis, three substructures were identified: star-forming galaxies in the main component show a higher scatter of 0.20 dex in metallicity than both the front and back ones, with a scatter of 0.07 and 0.11 dex, respectively. Star-forming galaxies located outside the central region of AC114 are driving the shallower slope of the cluster MZR.}
    {The slightly shallower slope and high scatter of AC114 with respect to foreground and background galaxies in the mass-metallicity plane indicates that galaxies are suffering from environmental and dynamical effects. Ram-pressure stripping and strangulation are likely the main drivers in increasing the metallicities of at least two star-forming members with respect to the field MZR at the same redshift. However, the lower metallicities of the star-forming members, which drive the flatter slope of the AC114-MZR, can be explained by strong metal-poor inflows triggered by galaxy-galaxy interactions. In fact, the downshift reported for these galaxies is consistent with other observations and simulations, as a result of mergers and/or flybys, which dilute the gas-phase metallicities from metal-poor inflows.
    The mass of a galaxy cluster appears to be a key variable in determining the importance of environmental effects in the evolution of cluster members, where massive galaxy clusters ($M_{vir}>10^{15}M_{\odot}$) show changes in the slope of the MZR.}


   \maketitle
%
% -------------------------------------------------------------------
\section{Introduction}
The evolution of galaxies is associated with global features, such as morphology and kinematics, as well as properties that drive the evolution of the stellar populations that they have (e.g., chemical composition, star formation rates (SFRs), initial mass, and luminosity functions). Physical phenomena such as inflow and outflow processes are also known to be key in driving the chemical evolution of galaxies.

Seminal works in the 70s showed that supernovae (SNe) explosions and stellar winds play a significant role in the chemical evolution. Supernovae explosions bring enough energy to ionize and accelerate particles in the interstellar medium, heating the environment and expelling the gas component through galactic winds. This process is more efficient in low-mass galaxies, where reaching the escape velocity is easier because of their shallow potential wells (\citeauthor{larson1} \citeyear{larson1}, \citeauthor{lequeux} \citeyear{lequeux}). In addition, inflows of pristine gas can dilute metals, increasing star formation and leading to the synthesis of new heavy elements \citep{finlator}. The balance between these processes naturally produces the mass-metallicity relation (MZR), according to which high-mass galaxies present high gas-phase metallicities.

In the last two decades, the MZR has shown itself to be a powerful tool to constrain the chemical evolution of galaxies. \citet{tremonti}, using a large sample of star-forming (SF) galaxies, constructed the MZR of the local Universe, showing statistically robust evidence of the existence of this relation and confirming the suggestion introduced by \citet{larson1} that low abundances and a tight ($\pm0.1$dex) MZR are driven by efficient gas loss. On the other hand, the MZR tends to flatten for massive galaxies with $M\geq 3\times10^{10} M_{\odot}$ (\citealt{kauffmann}) because of galactic downsizing, where the closed-box-like behavior begins (\citealt{cowie}, \citealt{neistein}, \citealt{sanchez}, \citealt{spitoni}). The MZR also evolves in redshift: both the gas-phase metal content and stellar masses of galaxies increase across cosmic time, where the higher-redshift MZR ($z<3.5$) presents a higher downshift with respect to the local Universe (\citealt{savaglio}, \citealt{erb}, \citealt{maiolino}, \citealt{moustakas}, \citealt{zahid}, \citealt{gao}). Evidence of smooth metallicity gradients and the weak dependence on the SFRs in SF galaxies at $z\sim3$ show that processes involving material flows are stronger at higher redshifts, naturally resulting in a shallower slope across cosmic time (\citealt{troncoso}). On the other hand, the formation processes of galaxies in the hierarchical growth scenario and the key role of SNe feedback, which acts as the main regulator of the star formation and morphology of the galaxies, proves that mergers depending on the gas fractions of their galaxies drive changes in the evolution of the MZR (\citealt{tissera}, \citealt{derossi}, \citealt{kobayashi}, \citealt{scannapieco}). 

The internal phenomena of SF galaxies that drive the MZR have been extensively studied (see \citealt{maiolinomanucci} for a detailed review). However, a similar investigation of the physical processes intervening in the chemical evolution of galaxies in dense environments such as groups and galaxy clusters is missing. Therefore, constructing the MZRs in a galaxy cluster allows the exploration of the chemical evolution of its members in a broad range of masses at a given redshift, under a variety of dynamical and environmental effects. 

Clusters of galaxies present deep potential wells originating in the interaction of smaller-scale structures or the evolution of overdensities in the early Universe (\citealt{press}, \citealt{springel}). In the former scenario, the “preprocessing” can affect galaxies' chemical evolution as well as their morphology and kinematics (\citealt{Fujita}). In a recent work, \citet{kim} showed that in galaxy clusters with ongoing dynamical activity, early infallers will stop the star formation activity due to strong strangulation processes, while the recent infallers will suffer from ram-pressure stripping (RPS) when galaxies interact with the intracluster medium (ICM). Moreover, the scenario will be even more chaotic when infall makes physical mechanisms such as accretion, tidal interactions, and mergers more likely (\citealt{barton}, \citealt{alonso},  \citealt{ellison}, \citeyear{ellison2}). 

Up to date, the general picture of the role of the physical phenomena driving the evolution of galaxies in clusters is still unclear. However, the virial mass of galaxy clusters plays a crucial role, underlying differences in the chemical content of cluster members with respect to field galaxies \citep{petropoulou}. This allows the exploration of environmental mechanisms that affect the chemical evolution of galaxies and to investigate the origin of the quenching of galaxies in more detail. For example, in high-mass-density environments, galaxies found at the cluster center or group center will experience enhancements in the gas-phase abundances up to $\sim 0.05$ dex \citep{ellison3}. This could be produced (i) by a metal-poor inflow suppression toward the galactic center, (ii) preprocessed gas from the ICM, (iii) tidal interactions producing starbursts, (iv) strangulation, and (v) RPS (\citealt{gunn}) acting on galaxies with nonzero metallicity gradients (\citealt{ellison3},  \citealt{petropoulou2}, \citeyear{petropoulou}, \citealt{gupta}). On the other hand, although galaxy clusters with  $M\sim 10^{14}M_{\odot}$ also contain SF galaxies suffering environmental effects, these do not seem to be strong enough to produce significant changes in the chemical content of the cluster's members. However, a weak metallicity dependence on clustercentric radii is seen (\citealt{petropoulou}, \citealt{gupta},\citealt{vaughan}, \citealt{lara-lopez}), showing the relevance of galaxy cluster masses in the exploration of the phenomena that shape the chemical evolution of galaxies.

In this work, we study the role of galaxy clusters in the MZR, by using SF galaxies at redshift $z\sim 0.3$ belonging to AC114 (ABELL S1077). AC114 ($\alpha = 22^{h} 58^{m} 52^{s}.3$, $\delta = -34^{o}46'55''$; J2000) has been widely studied near the cluster core, $r_{core} = 0.27 h^{-1}\mathrm{Mpc}$ (\citealt{couch1}). The cluster core shows strong lensing arcs produced by the central dominant (cD) galaxy of AC114 (\citealt{smail}, \citealt{natarajan}, \citealt{campusano}). AC114 hosts a higher fraction of blue, late-type galaxies than lower-redshift galaxy clusters (\citealt{couch2}, \citealt{sereno}), which is a typical symptom of the Butcher-Oemler effect \citep{butcher}. Roughly $60\%$ of galaxies are found outside the core region, and its red sequence is separated from the blue galaxies (\citealt{proust}). A soft extended X-ray tail originates close to the cluster core, extending $\sim1.5'$ toward the southeast ($\sim 400$ kpc), suggesting that the substructure could be part of an ongoing merger event \citep{defilippis}.
\\

This is the third paper in a series. In paper I, \citet{proust} revisited the dynamical state of AC114. They find that the large velocity dispersion of $ 1893_{-82}^{+73} \ \mathrm{km \ s^{-1}}$  and the shape of its infalling pattern is indicative of AC114 being in a radial relaxation phase, showing a very elongated radial filament structure spanning $12000$ km s$^{-1}$. From the dynamical activity, they find that AC114 is a massive galaxy cluster with a virial mass of $M_{vir} =(4.3\pm 0.7)\times 10^{15} h^{-1} M_{\odot}$. In paper II, \citet{saviane23}, by studying stellar populations of passive galaxies and emission line galaxies (ELGs), find that the stellar MZR is steeper than the local Universe analogue. Moreover, the stellar evolution of galaxies in the center is faster than that found in those on the outskirts of AC114, where the latter are still forming stars, a signal that strangulation processes are stronger in the central region of AC114.

In this work, we study the AC114-MZR through a chemical analysis of the gas-phase component of its SF galaxies. In Section 2, we present the spectroscopic and photometric observations used, together with the data reduction. In section 3, we describe the data analysis, including the redshift, mass, and metallicity measurements and estimates. In section 4, we present the MZRs constructed for the foreground, cluster members, and background galaxies, together with their properties. In Sections 5 and 6 we discuss the results that we obtain and presents our conclusions. 

In this paper, we adopt the following cosmological parameters: $H_{0} = 69.6 \ \mathrm{km \ s^{-1} \ Mpc ^{-1}}$, $\Omega_{M} = 0.286$, and $\Omega_{\Lambda} = 0.714$ in a flat Universe according to \citet{cosmocalc}.

%--------------------------------------------------------------------
\section{Observations and data reduction}

\subsection{VIMOS spectroscopic data}
\label{subsection:vimos}
In this work, we use the same spectroscopic data as \citet{proust} and \citet{saviane23}, hereafter P15 and S23. The data was obtained with the VIsible MultiObject Spectrograph (VIMOS, \citealt{lefevre}) mounted on the ESO Very Large Telescope (VLT), under the program 083.A-0566 (PI: I. Saviane), between August 16 and September 25, 2009. VIMOS was a visible (360 to 1000 nm) wide field imager and multi-object spectrograph mounted on the Nasmyth focus B of UT3 Melipal. The instrument is made of four identical arms with a field of view (FOV) of $7' \times 8'$ with a $0.205''$ pixel size and a gap between each quadrant of $\sim 2'$. Seven exposures each for the HR-red and MR grisms (630 - 870 nm, $R=2500$ and 500 - 1000 nm, $R= 580$, respectively, for a $1''$ slit) were obtained, for a total observing time of $\sim14$h.

The galaxy selection was made from the pre-imaging frames of the cluster. Nonstellar objects in the region were selected by eye in order to punch the maximum number of slits into each of the four quadrants. Such a visual inspection allowed us to discriminate between extended objects and stars.\\

The VIMOS Pipeline in the EsoReflex environment \footnote{https://www.eso.org/sci/software/pipelines/vimos/} (\citealt{esoreflex}) was used to perform bias subtraction, flat normalization and cosmic ray removal in each scientific frame. Then, sky subtraction was done, taking the median of the sky spectrum located between the galaxy spectra in the two-dimensional frames. The spectral flux was calibrated by using two standard stars from \citeauthor{hamuy}  (\citeyear{hamuy}, \citeyear{hamuy2}): the F-type LTT1788 ($V = 13.16$, $B-V = 0.47$) and DA-type LTT7987 ($V = 12.23$, $B-V = 0.05$).

Once the galaxy spectra were reduced, we proceeded with emission line corrections to get proper flux measurements. We subtracted stellar continuum spectra by fitting Chebyshev polynomials with the Specutils package \citep{specutils}. Emission line fluxes from collisional de-excitation ($O^{+}, O^{++},N^{+},S^{++}$) were then measured with a Gaussian fitting. To correct for dust and reddening effects, H$\alpha$ and H$\beta$ fluxes were measured by Gaussian fitting as well, and the observed hydrogen line ratio ($F_{H\alpha}/F_{H\beta}$) was compared with the theoretical expectations. The H line ratio, $I_{H\alpha}/I_{H\beta}$= 2.85, was adopted from case B ($T_{e}\simeq 10000$K) following \citet{hummer}. The reddening constant was calculated as $C_{H\beta} = [\log{(I_{H\alpha}/I_{H\beta})} - \log{(F_{H\alpha}/F_{H\beta})}]/[f(\lambda)- f(H\beta)]$, where $f(\lambda) = \langle A(\lambda)/A(V) \rangle$ is the extinction law at a given wavelength of the modified version of \citet{cardelli} in the optical domain, obtained from \citet{odonell}. We also assumed that $R_{V}=3.1$. Finally, emission line measurements were corrected by applying $I_{\lambda}/I_{H\beta} = (F_{\lambda}/F_{H\beta})\times 10^{C_{H{\beta}}[f(\lambda)-f_{H\beta}]}$. 

\begin{figure}
   \centering
    \includegraphics[width=\hsize]{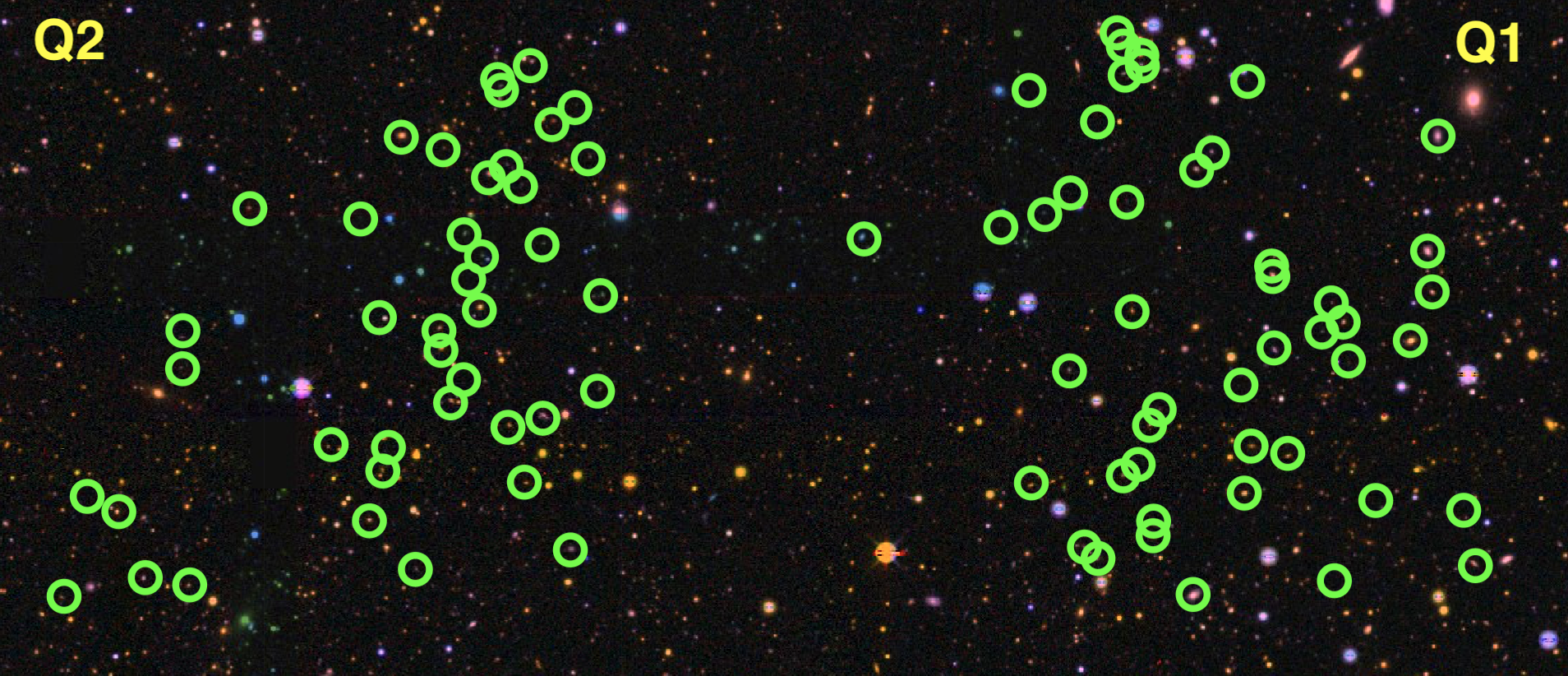}
    \includegraphics[width=\hsize]{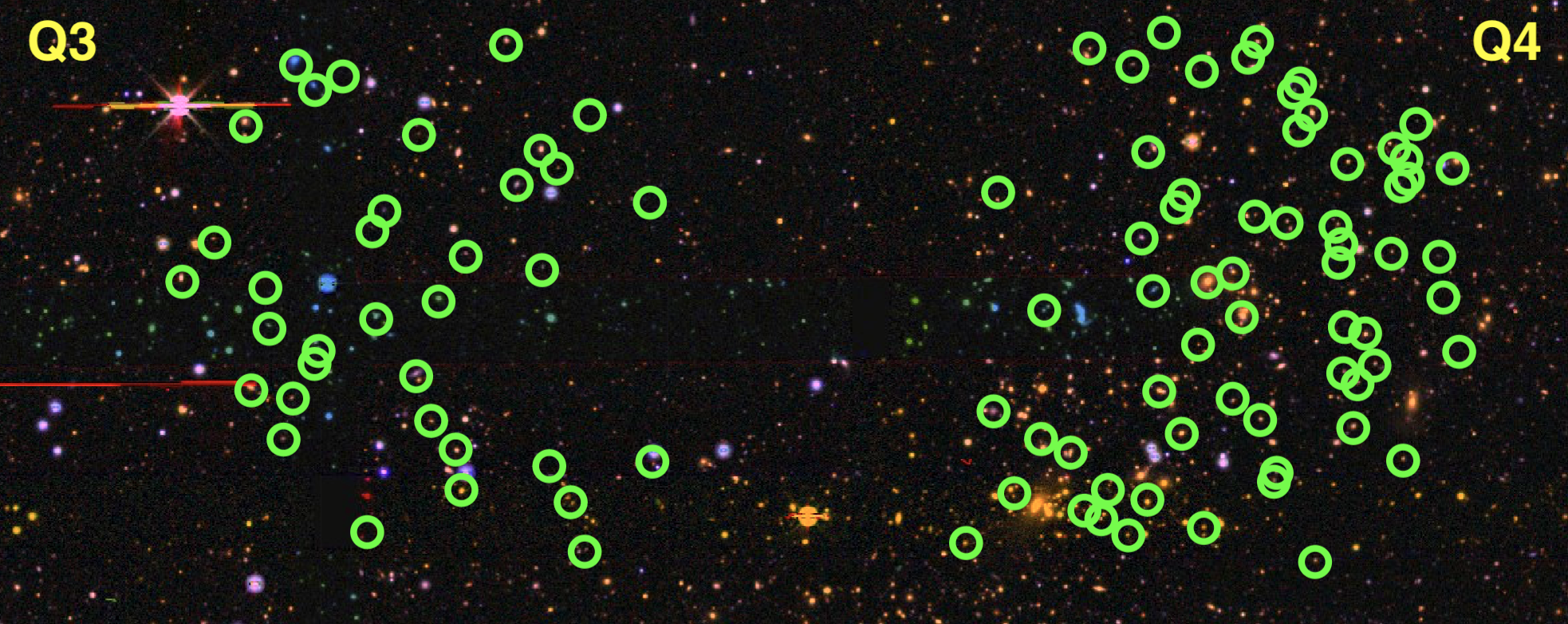}
      \caption{Observed galaxies (colored circles) on top of an image taken from the DECaLS DR10 viewer for the AC114 cluster. Quadrants Q1, Q2, Q3, and Q4 are counterclockwise, with Q4 in the bottom right clump.}
         \label{fig0}
   \end{figure}

Recombination H lines can be affected by underlying stellar absorption, producing changes in the metallicity estimations due to the use of Balmer lines in the most common calibrators. However, the spectra of our sample are contaminated by fringing patterns in  $\lambda>7500\AA$, so it is difficult to make an appropriate correction for the stellar absorption at $z\sim 0.3$, the H$\alpha$ emission line falls at $\sim 8500 \AA$).
\\ 
To evaluate a proper correction of this effect, we employed the \citet{hopkins} correction, which was tested in galaxies of the SDSS survey. The authors show that an appropriate stellar absorption correction is about $\mathrm{EW=1.3\AA}$, which in this work is translated in a flux correction of $F_{\lambda} \sim 0.001$ and $\sim 0.002 \times 10^{-16}$ erg s$^{-1}$ cm$^{-2}$ on average for H$\alpha$ and H$\beta$, respectively. Therefore, these corrections were applied to the emission line measurements of the Balmer lines.\\
\\
Our sample consists of 184 spectra of galaxies, where 25 ($13\%$) are not reported in earlier works with the same data, because in this paper the observations were reduced with the latest version (4.1.8) of the VIMOS pipeline. The sample is listed in Table \ref{table:1}, with galaxies labeled according to the quadrant and slit of extraction. Additional galaxies are listed in Table \ref{table:1} with the syntax $\#$quadrant-A$\#$. Our sample is shown on top of an image of AC114 in Figure \ref{fig0}.

\begin{figure}
   \centering
    \includegraphics[width=\hsize]{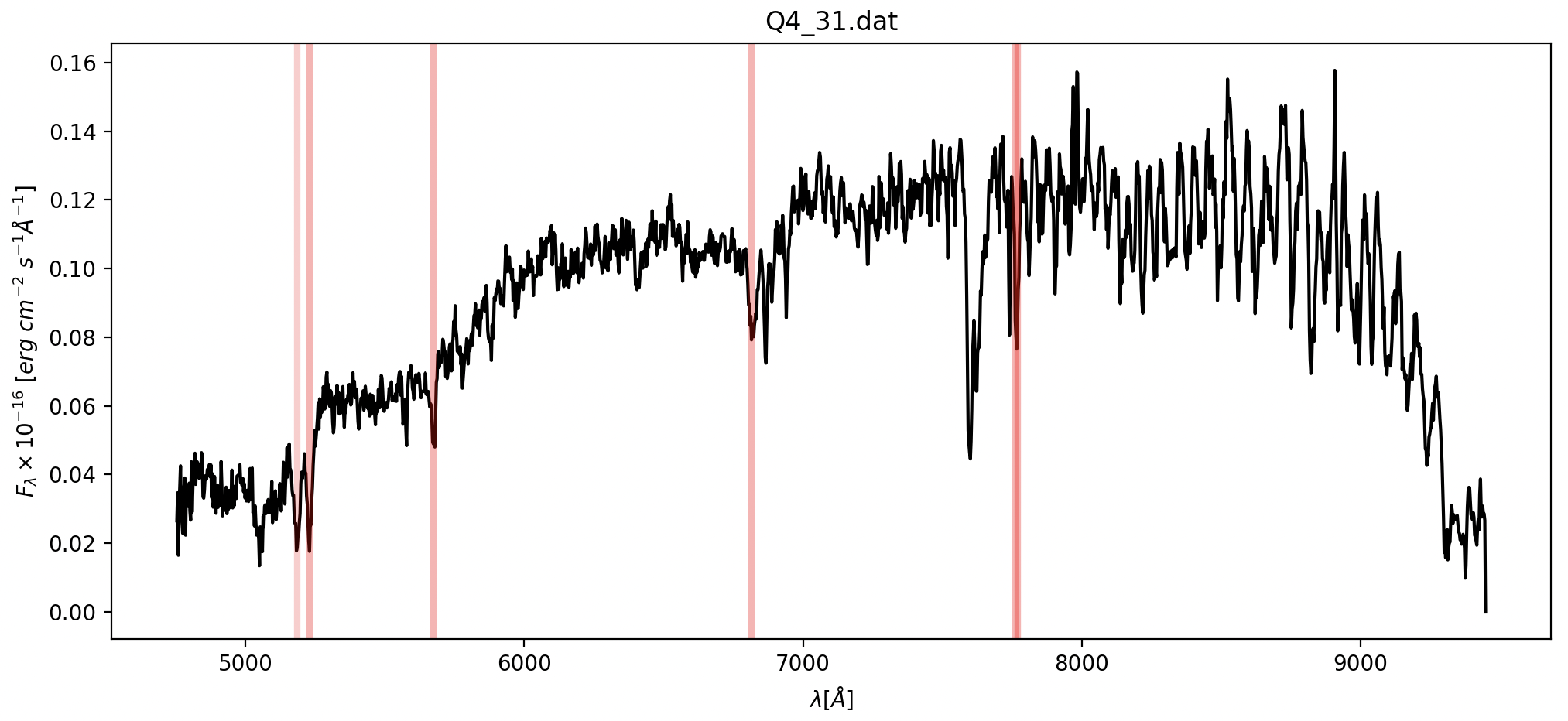}
    \includegraphics[width=\hsize]{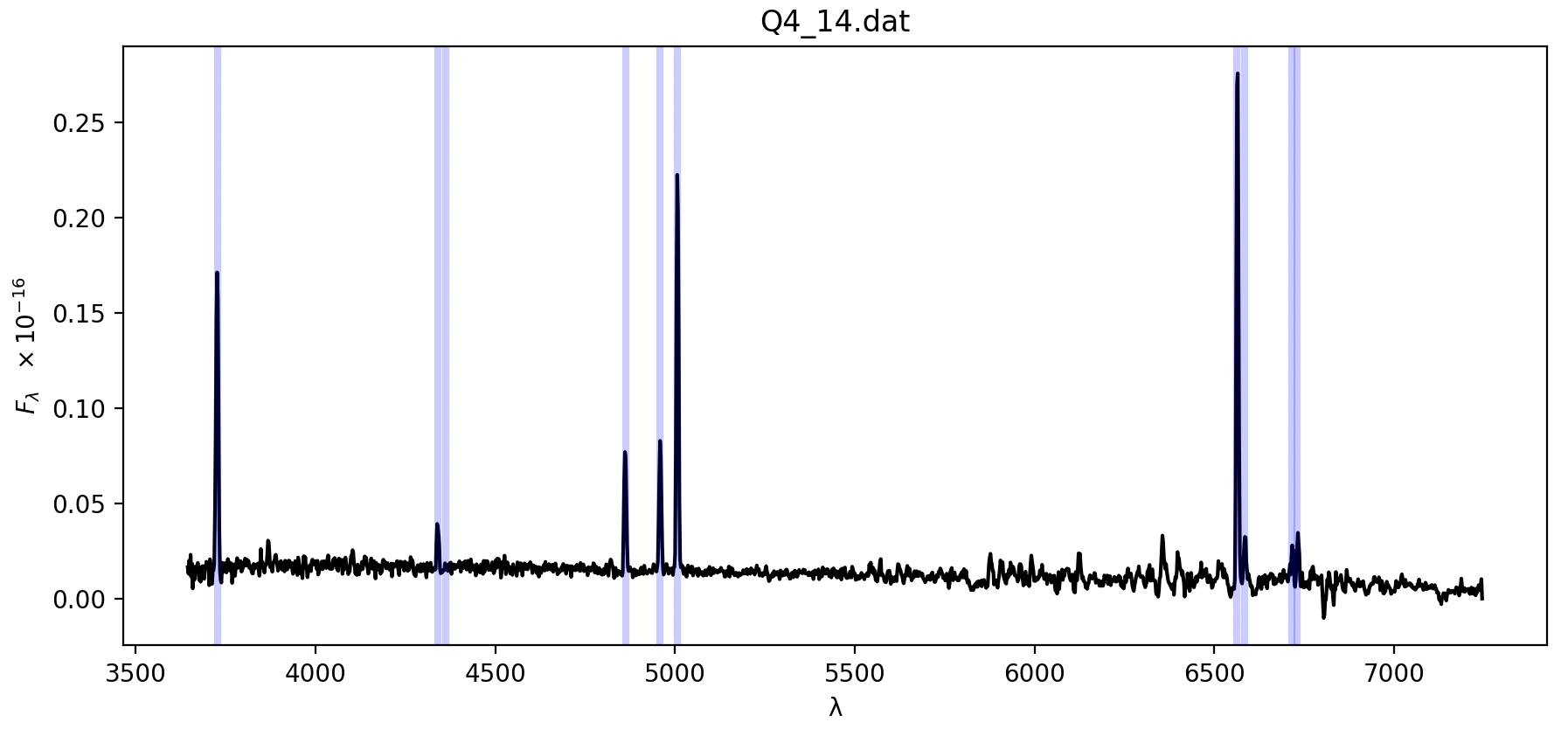}
      \caption{Reduced spectra of a passive galaxy (top panel) and an ELG (bottom panel). Red lines indicate absorption features, whereas blue lines indicate emission features.}
         \label{fig1}
   \end{figure}

Applying the same criteria as in S23, a manual inspection of the reduced spectra was conducted to classify the sample into passive galaxies and ELGs, based on the presence or absence of the [OII]$\lambda\lambda 3727$, [OIII]$\lambda\lambda 4959,5007$, and/or H$\gamma$, H$\beta$ emission lines. Figure \ref{fig1} shows the difference between a passive galaxy and an ELG in the sample. We identified 103 ($56\%$) passive and 81 ($44\%$) ELGs. The top panel shows the spectrum of the passive galaxy Q4-31 ($\#quadrant \ \#slit$). Only absorption features can be identified in the spectrum, such as CaII H $\&$ K, the CH G band, the Mg triplet, and the Na D doublet, shown with red lines from left to right, respectively. The bottom panel of Figure \ref{fig1} shows the spectrum of the ELG Q4-14. In this case, only emission lines are present, such as [OII]$\lambda\lambda 3727$, H$\gamma$,$\mathrm{[OIII]\lambda4363}$, H$\beta$, [OIII]$\lambda\lambda 4959,5007$, H$\alpha$, [NII]$\lambda 6584$, and [SII]$\lambda\lambda 6716,6731$ shown with blue lines from left to right, respectively. 

Most of the galaxies are in Q4 (62 galaxies, bottom right in Figure \ref{fig0}), which covers the central region of AC114. The other 35, 41, and 46 galaxies are grouped in the Q1, Q2, and Q3 quadrants, respectively (counterclockwise). The clumps are produced by the VIMOS CDD-chips gap of $\sim 2'$. 

% plot RA DEC del sample 

\subsection{Optical and near-infrared photometric data}
\label{subsection:photometry}
To construct the MZR, photometric data is needed to compute the stellar masses of the galaxies. Optical photometry was found for the VIMOS FoV of AC114 in the DECaLS DR10 Survey (see \citealt{decals} for an overview of the DESI Legacy Imaging Surveys). The DECaLS program made use of the Dark Energy Camera (DECam) on the Blanco 4m telescope, located at the Cerro Tololo Inter-American Observatory. DR10 covers the South Galactic Cap region at Dec$\leq 34^{\mathrm{o}}$ in the $g,r,z,i$ bands. 121 ($66\%$) and 139 ($76\%$) galaxies in our sample have available photometry in the $g$ and $i$ bands, respectively. However, no photometry is available in the $r$ and $z$ bands. 

 \begin{figure}
   \centering
   \includegraphics[width=\hsize]{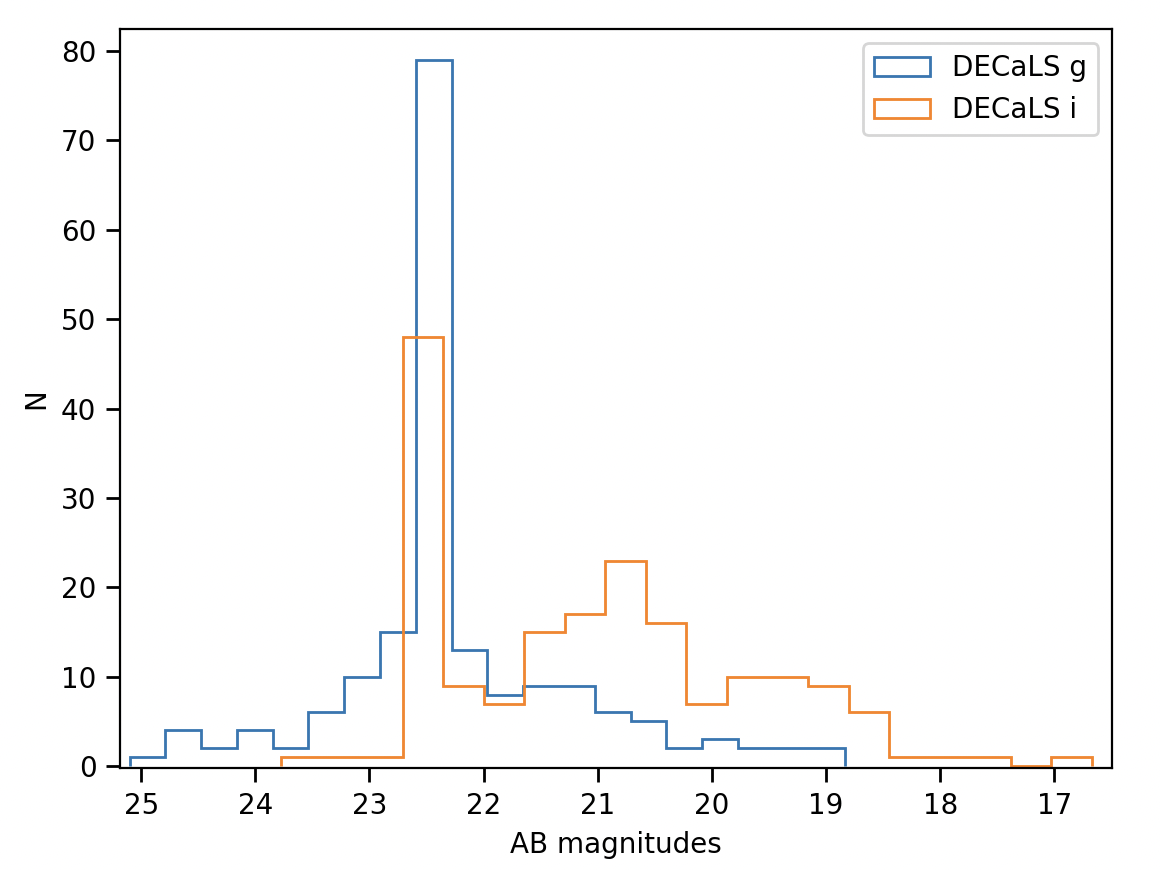}
   \includegraphics[width=\hsize]{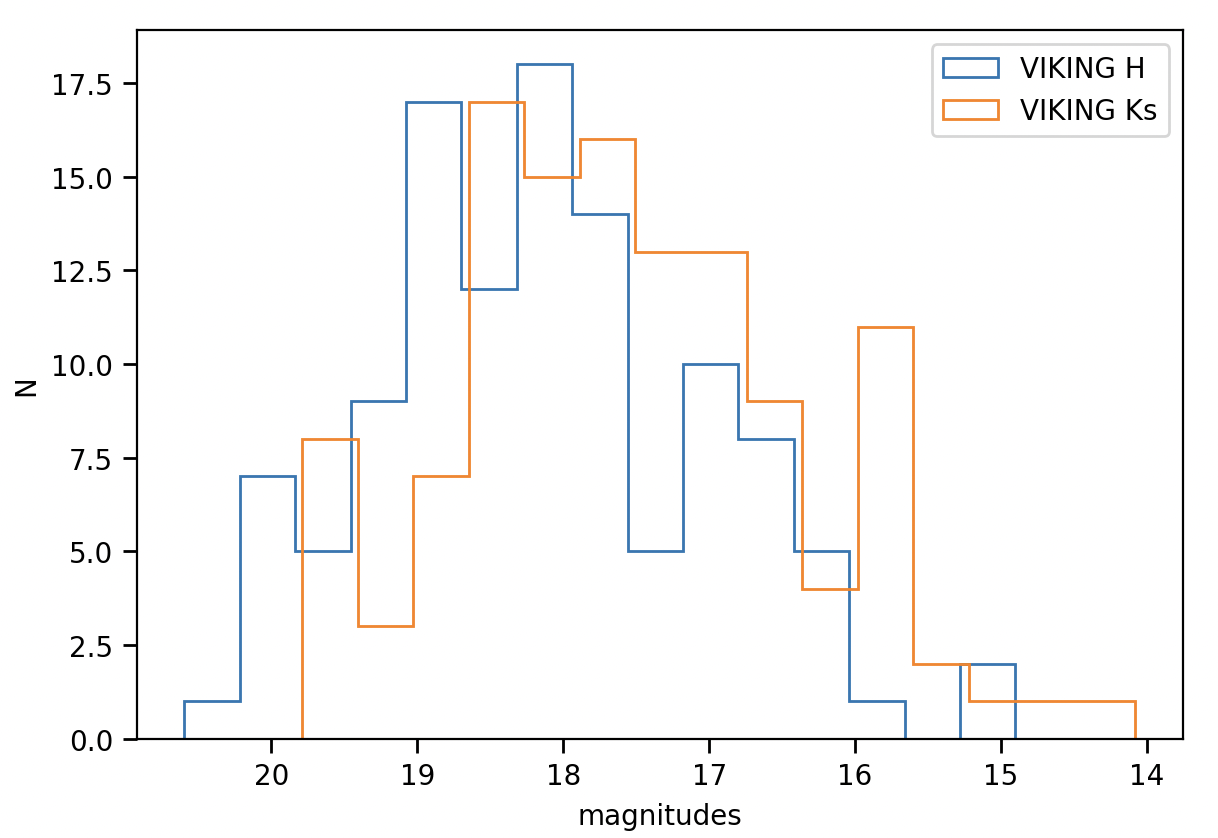}
      \caption{Top panel: Distribution of photometric data collected from DECaLS DR10 survey. The blue line indicates the distribution of $g-$band magnitudes and the orange line indicates the distribution of $i-$band magnitudes. Bottom panel: Distribution of photometric data collected by the VIKING DR4 survey. The blue line indicates the distribution of $H-$band magnitudes and the orange line indicates the distribution of $K_{s}-$band magnitudes.
              }
         \label{fig2}
   \end{figure}

The top panel of Figure \ref{fig2} shows the distribution of apparent magnitudes for both $g$ and $i$ band passes with blue and orange lines, respectively. Galaxies observed in the $g$ band cover a magnitude range from 18.83 to 25.10, with a clear peak at 22.46, while the respective $i$ band magnitudes cover a range of 16.67 to 23.77, with three prominent clumps centered on 22.24, 20.76, and 19.32.

\citet{Bell}, studying stellar mass-to-light ratios in optical and near-infrared (NIR) band passes, found that NIR luminosities are less sensitive to the past star formation history, smoothing out variations with respect to the color. Therefore, they concluded that NIR luminosities lead to more accurate stellar masses. With this motivation, we also searched for available NIR photometric data, which we found in the VIKING DR4 ESO survey. The VISTA Kilo-degree Infrared Galaxy Survey (VIKING)\footnote{https://www.eso.org/rm/api/v1/public/releaseDescriptions/135} is a wide-area (covering a final area of 1350 square degrees), intermediate-depth ($5\sigma$ detection limit $J\sim21$ on Vega system) NIR imaging survey in the five broadband filters, $Z$, $Y$, $J$, $H$, and $K_{s}$. The sky coverage is at high galactic latitudes and includes two main stripes of $~\sim 70 \times 70^{\mathrm{o}}$ each: one in the South Galactic cap near Dec$\simeq -30^{\mathrm{o}}$ and one near Dec$\sim 0^{\mathrm{o}}$. 
A cone search of $25'$ was conducted, centered on the cluster, and 121 ($66\%$) of the galaxies in our data set were found. Near-infrared photometry in VIKING DR4 was only available in the $H$ and $K_{s}$ bands.

The bottom panel of Figure \ref{fig2} shows the distribution of apparent magnitudes for our galaxy sample for both $H$ and $K_{s}$ bands, shown with blue and orange lines, respectively. Galaxies observed in the $H$ and $K_{s}$ bands cover a range of magnitudes from 14.90 to 20.60 and from 14.10 to 19.79, respectively.\\
\\
As is discussed in P15 and S23, our galaxy sample reaches luminosities as faint as $R\sim 21$, but it starts to suffer incompleteness at $R\sim20$. We searched in the galaxy sample for objects with values $R\sim 20$ and found that the galaxy sample is incomplete at $g \sim 21.4$ and $i \sim 19.8$ mags in the optical bands, whereas the same is seen at $H\sim 17.7$ and $K_{s} \sim 17.3$ mags in the NIR band passes. The photometry is used to compute the stellar masses via mass-to-light ratios in the galaxy sample. The incompleteness starts to be significant in galaxies with $\sim10^{10} M_{\odot}$.

The cluster members in the galaxy sample are well distributed in terms of mass, spanning values from $\sim10^{9}M_{\odot}$ to $\sim10^{11} M_{\odot}$, as is expected for a construction of the MZR. Therefore, the incompleteness does not affect our results.

%--------------------------------------------------------------------

\section{Data analysis}
\subsection{Redshift estimations}
\label{subsection:redshifts}
\begin{figure}
   \centering
   \includegraphics[width=\hsize]{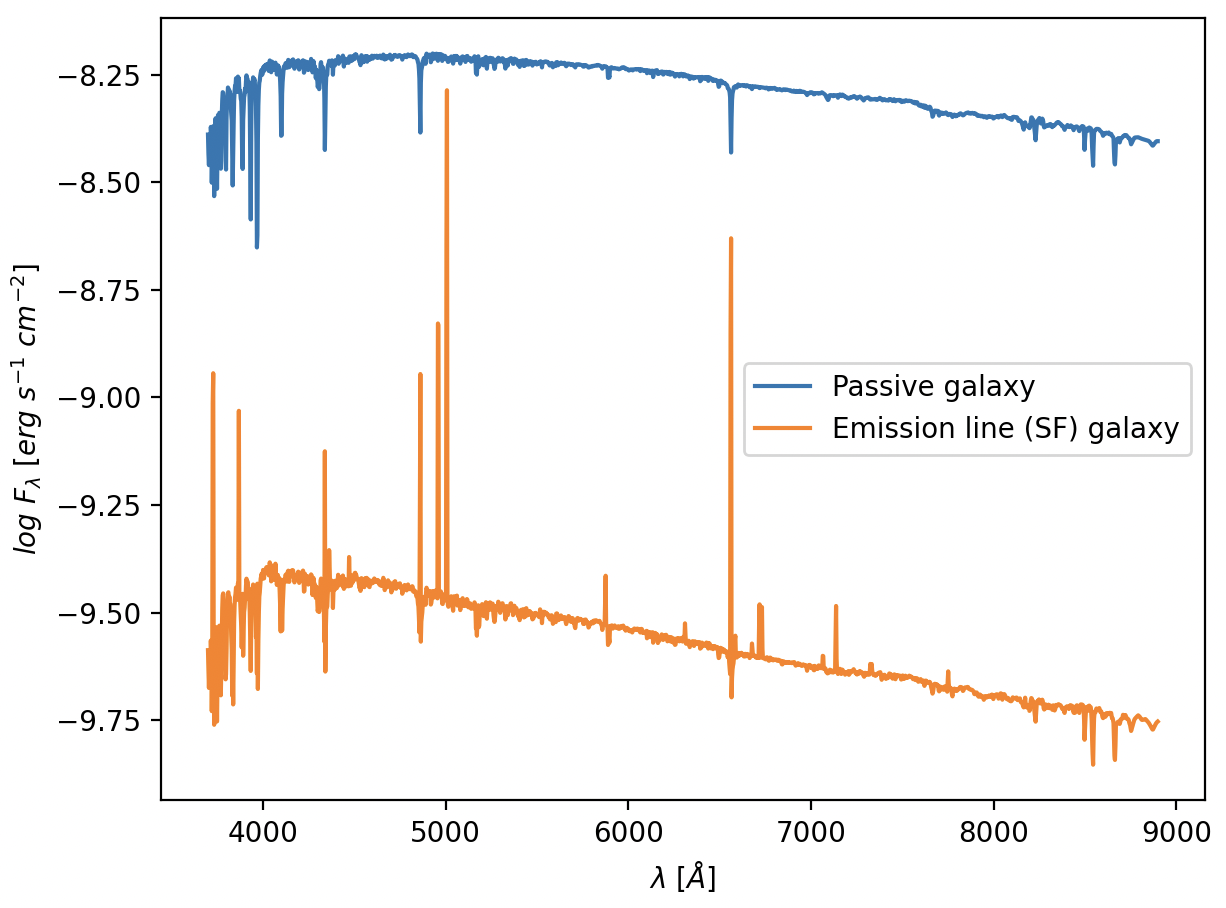}
      \caption{Synthetic spectra reproduced with the Bagpipes models. The blue spectrum represents a passive galaxy. The orange spectrum represents an SF galaxy.}
         \label{fig5}
   \end{figure}

The galaxies' redshift, $z$, was measured by cross-correlation \citep{tonrydavis} between the VIMOS spectra and synthetic ones at $z=0$ in the wavelength range from 3700$\AA$ to 8800$\AA$. In this work, we made use of the Bagpipes models\footnote{Bagpipes is a state of the art Python code for modelling galaxy spectra and fitting spectroscopic and photometric observations. https://bagpipes.readthedocs.io/en/latest/} \citep{bagpipes} to generate a standard passive and a SF synthetic spectrum, shown in Figure \ref{fig5} with blue and orange, respectively. We considered a galaxy whose spectrum presents a evolved stellar continuum with relevant absorption lines typically found in old stellar populations to represent a passive galaxy. We achieved this, generating a synthetic spectrum of a galaxy with stellar mass $M^{*} = 10^{10} M_{\odot}$ and 10 Gyr since the star formation began by using the exponential$-\tau$ model . On the other hand, an SF galaxy has a small stellar continuum with prominent emission lines: we calculated the spectrum of a galaxy with a stellar mass of $M^{*} = 10^{8} M_{\odot}$, 1Gyr after the star formation began, by using the exponential$-\tau$ model, and an ionization parameter, log($U$)=$-2.5$. 

\begin{figure}
   \centering
   \includegraphics[width=\hsize]{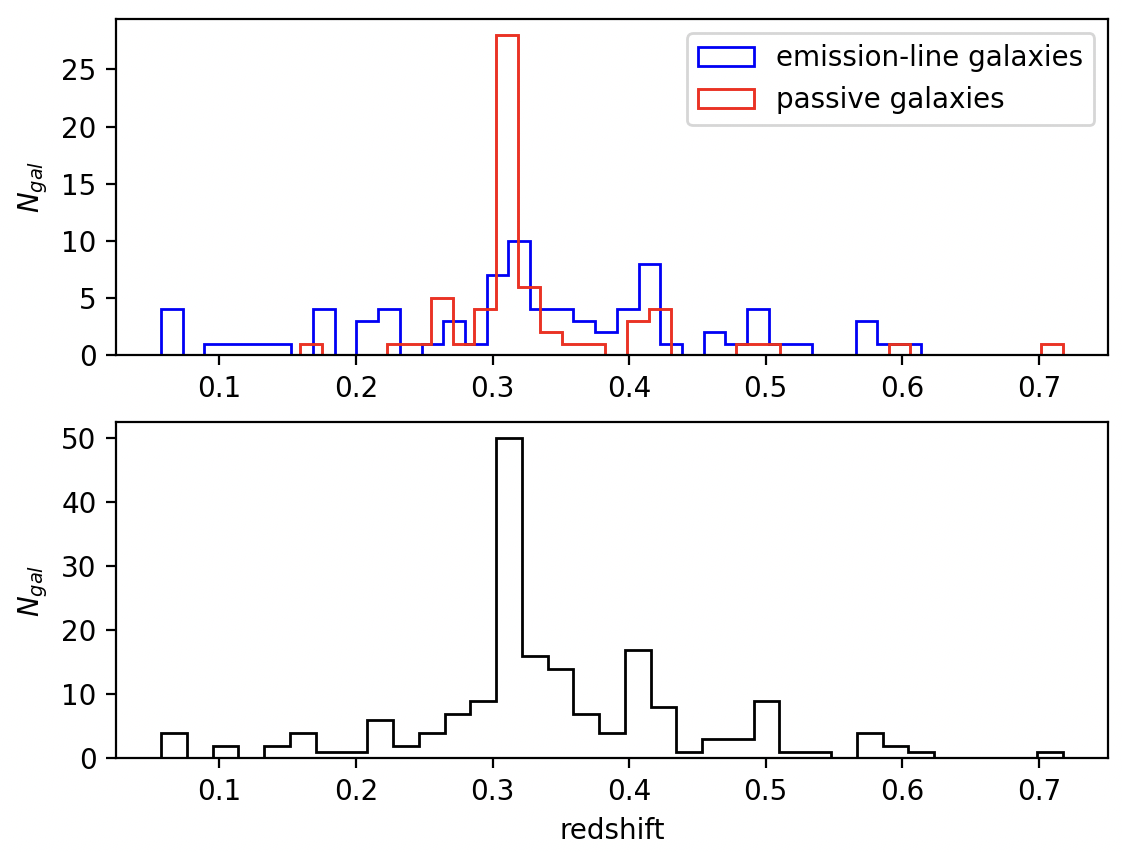}
      \caption{Redshift distribution of the galaxy sample. Upper panel: Distribution of passive galaxies (red line) and ELGs (blue line). Lower panel: Redshift distribution of the complete sample.}
         \label{fig6}
   \end{figure}

Once the synthetic spectra were ready to use, we cross-correlated them with the VIMOS spectra. We used the IRAF \footnote{IRAF is distributed by the National Optical Astronomy Observatories, which are operated by the Association of Universities for Research in Astronomy, Inc., under cooperative agreement with the National Science Foundation.} task “fxcor” to estimate the redshift of galaxies. The galaxies of the sample have $\sim0.1 < z <0.7$, with uncertainties of $0.0014$ on average. The redshift distribution of the sample is shown in Figure \ref{fig6}. Emission line galaxy and passive galaxy distributions are shown with blue and red lines, respectively. The distribution reveals that passive galaxies are found at the cluster redshift, meaning $0.28<z<0.34$ according to P15. In this region, 58 ($73\%$) of a total of 80 galaxies are passive, while 22 ($27\%$) are ELGs. Outside the cluster redshift, it is clear that there is an additional structure that is dominated by ELGs centered on $z\sim0.41$. According to P15, this background structure has a probability of $\sim 60\%$ of being a group.

\subsection{Stellar mass estimates}
\label{section:mass}

\begin{figure}
   \centering
   \includegraphics[width=\hsize]{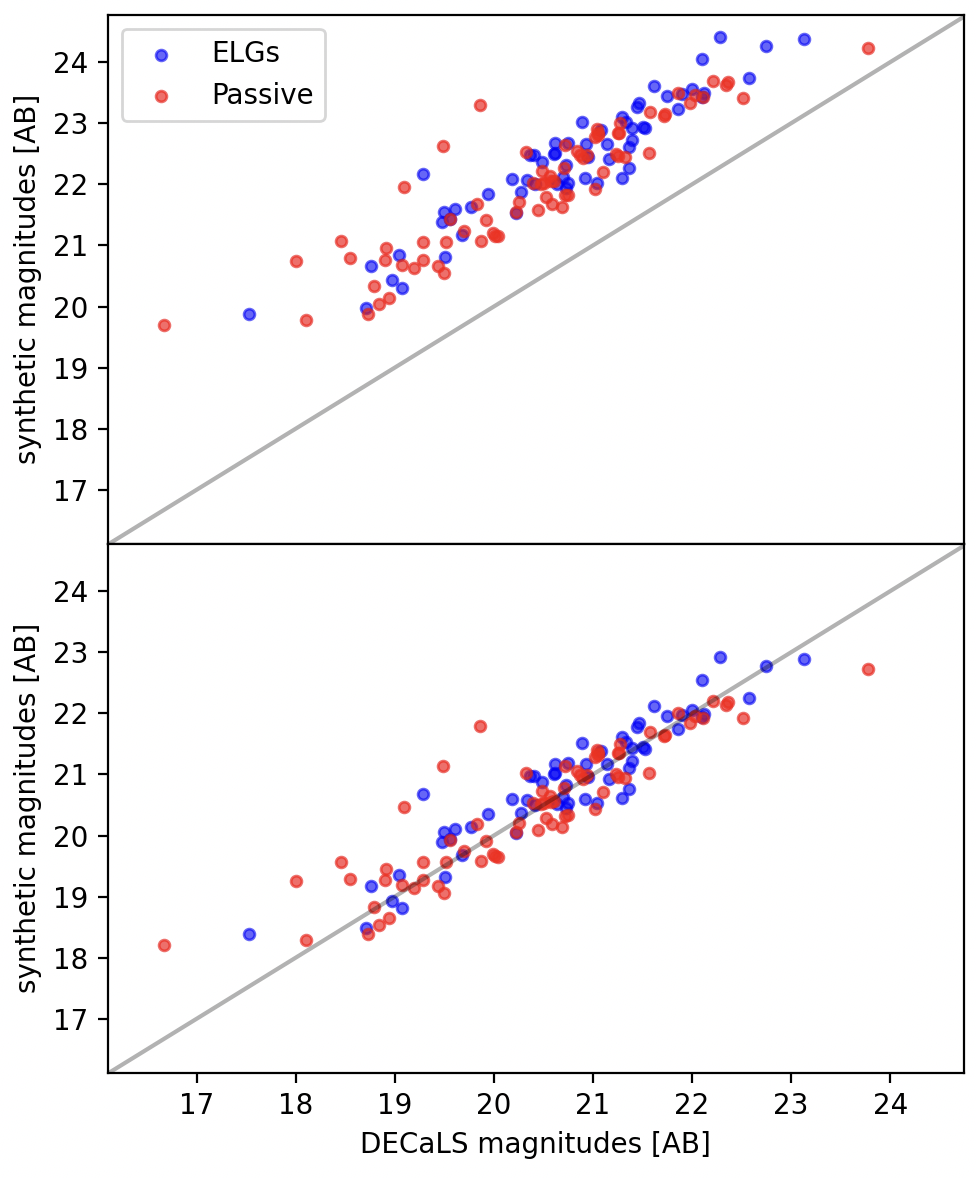}
      \caption{Comparison between DECaLS DR10 (on the x axis) and those from artificial magnitudes (on the y axis). Colors are the same as in Figure \ref{fig6}. The gray line is an identity line. Top panel: Comparison between synthetic and DECaLS magnitudes in the $i$ band. Bottom panel: Comparison between synthetic and DECaLS magnitudes in the $i$ band with the applied correction.}
         \label{fig7}
   \end{figure}

\begin{figure*}[h]
	\centering
%\floatbox[{\capbeside\thisfloatsetup{capbesideposition={right,top},capbesidewidth=6cm}}]{figure}[\FBwidth]
{\caption{Central panel: Comparison between $i$ and $Ks-$based masses. Blue and red dots represent ELGs and passive galaxies, respectively. The gray line corresponds to the 1:1 line ratio. Upper panel: Distribution of optical luminosity based stellar masses. The black line shows the mass distribution of the complete sample, whereas the blue and red areas indicate the distribution of ELGs and passive galaxies, respectively. Right panel: Distribution of near-infrared luminosity based stellar masses.}\label{fig8}}
\includegraphics[width=11cm]{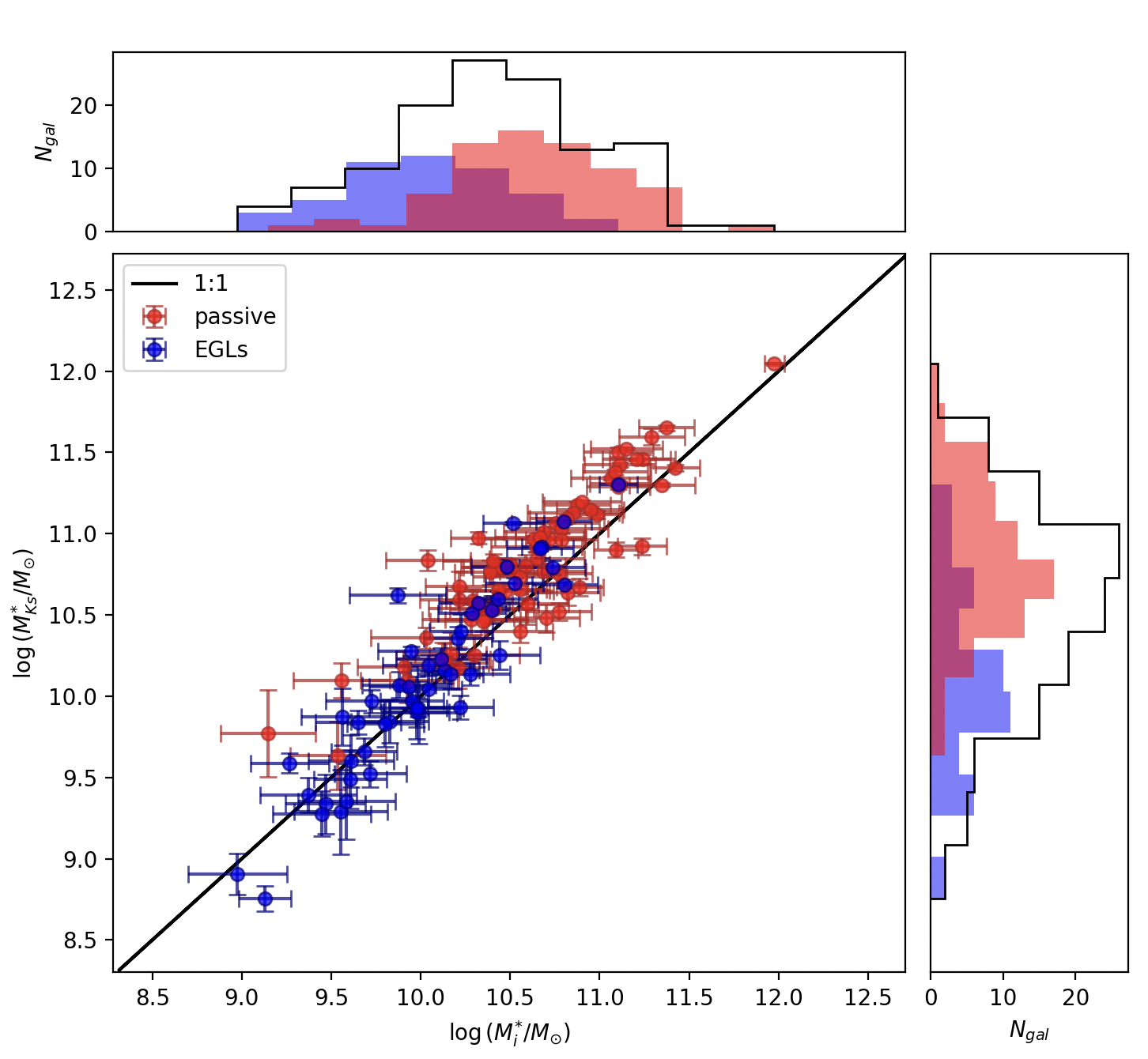} 
%{\includegraphics[width=11cm]{plots/Masses_histogram.png} }
\end{figure*}

\subsubsection{Synthetic photometry}
The masses of the galaxies were obtained by using stellar mass-to-light ratios. The data acquired from DECaLS DR10 give us photometry for $66\%$ and $76\%$ of the galaxies in the sample in the $g$ and $i$ bands, respectively. For galaxies without photometry, we obtained synthetic photometry by integrating the spectral flux of the galaxies. Optical stellar mass estimates are done via the theoretical mass-to-light ratio in the $i$ band. Therefore, synthetic photometry was estimated for this filter only because it has available photometry for more galaxies than the $g$ band does.

We made use of the PYPHOT\footnote{https://mfouesneau.github.io/pyphot/} \citep{pyphot} Python package. As a first step, we calculated the synthetic photometry of the LTT7987 standard star. We obtain magnitudes underestimated by $\Delta M_{i} = M_{standard,i}- M_{pyphot,i} = 0.16$. Therefore, this amount was applied as a correction when the artificial magnitudes were computed. In addition, the $1''$ slits used in the VIMOS observations do not entirely cover the projected area for most of the galaxies of the sample, which is $\sim 3''$. Therefore, we expect that synthetic photometry will present overestimated apparent magnitudes (underestimated luminosities). In fact, this is seen as an offset in the top panel of Figure \ref{fig7}, which presents a comparison between the artificial magnitudes and those collected from DECaLS DR10. Synthetic magnitudes are $\sim1.49$ larger in the i band, with a scatter of $\sigma= 0.46$. Hence, overestimations were corrected by subtracting this offset from the artificial magnitudes. The resulting correction is shown in the bottom panel of Figure \ref{fig7}. We adopted the scatter, $\sigma_{i}$, as a statistical error of the artificial magnitudes. This is similar to the optical magnitudes from DECaLS, which present a mean uncertainty of $0.47$ mags.

\subsubsection{Optical-based stellar masses}
To compute the absolute magnitudes, we followed the same procedure reported in S23, which converts apparent magnitudes to absolute ones, with $i-M_{i} = 5\log{D_{L}}-5$. Here, $D_{L}$ is the luminosity distance, which is a function of the redshift. A relation was computed for a few select values following \cite{cosmocalc}. Interpolating a second-degree polynomial it was found that the luminosity distances expressed in Mpc obey $D_{L} = 2245.7\ z^{2}+ 4682.4\ z-20.838$, valid only in $z\in [0.05,0.7]$ (see their Figure 11, S23).

Absolute magnitudes in the $i$ band were converted to luminosities, with $L=10^{-0.4(M_{i}-M_{i,\odot})}L_{\odot}$. Following S23, for ELGs and passive galaxies we adopted ages of 2Gyr and 10Gyr, respectively. By using the theoretical age-\textit{mass-to-light} relation from E-MILES models  (\citeauthor{emiles1} \citeyear{emiles1},\citeyear{emiles3}, \citeauthor{emiles2} \citeyear{emiles2}), the corresponding ages gives: $M/L_{i} = 0.75\ \pm \ 0.20$ (ELGs) and $2.48\ \pm \ 0.52$ (passive). These models adopt evolutionary tracks from \citet{padova00}, assuming a sub-solar metallicity of $[M/H] = -0.4$ and a Chabrier initial mass function (IMF). Finally, mass-to-light ratios together with the luminosities were converted into the  stellar masses of the galaxies. Uncertainties were propagated according to the expression $\delta f=\sqrt{\sum_{i}(\partial f/\partial x_{i})^{2}\delta x_{i}^{2}}$, where $f$ is the propagated function, $x_{i}$, and $\delta x_{i}$ are the parameters with their respective uncertainties. Optical-based mass estimations shows mean uncertainties of $\sim2\%$. Table \ref{table:2}.3 presents absolute magnitudes in the optical band used in this work, together with the respective mass estimations.  
The distribution of masses is shown in the upper panel of Figure \ref{fig8}. Masses of ELGs are distributed between  $\sim10^{8}$ and  $\sim10^{11}M_{\odot}$, whereas those of passive ones lie from $\sim10^{9}$ to $\sim10^{12}M_{\odot}$. From the upper panel of Figure \ref{fig8} it is easy to see that, in general, ELGs are less massive than passive galaxies: masses are more concentrated at $\sim 10^{9.5}$ and $\sim 10^{10.5}M_{\odot}$ for the ELGs and passive components, respectively. %It is explained because the latter harbor old stellar populations such are globular clusters, which are more massive and older stellar systems than the younger counterpart observed in active galaxies.

\subsubsection{Near-infrared-based stellar masses}
Given that we were able to obtain NIR luminosities for $66\%$ of the galaxy sample, optical-based and NIR-based stellar masses can be compared. \citet{Bell}, by using spectro-photometric models of galaxy evolution, found that there are substantial variations in the stellar mass-to-light ratios within and among galaxies as a function of the color, showing variations of a factor of three and seven in optical band passes and two in the NIR ones. These findings drive to a natural conclusion that NIR-based mass estimations are the best way to obtain the stellar masses of galaxies, because recent star formation can generate optical light that dominates the SED, leading to underestimates of the stellar mass, whereas recent star formation does not affect much NIR light.
%NIR band-passes do not cover fluxes coming from the gas-phase component, which is where the past star-formation history can be studied.

E-MILES models were used one more time to compute mass-to-light ratios as a function of the age, resulting in $M/L_{Ks} = 0.26\ \pm \ 0.09$ (ELGs) and $M/L_{Ks} = 0.91 \ \pm \ 0.13$ (passive). $K_{s}$ absolute magnitudes, together with NIR-based stellar mass estimations, are presented in Table \ref{table:2}.3. The mass distribution is also shown in the right panel of Figure \ref{fig8}. Emission line galaxies span values from $\sim10^{9}$ to $\sim10^{11}M_{\odot}$, and passives ones from  $\sim10^{10}$ and  $\sim10^{12}M_{\odot}$. Despite the lack of NIR photometry for $34\%$ of the sample, the feature that passive galaxies are more massive in general than ELGs is clearly seen. 

%\begin{figure}
 %  \centering
  % \includegraphics[width=\hsize]{plots/mass_comparison.png}
   %   \caption{Comparison between $i$ and $Ks-$based masses. Blue and red dots represents active and passive galaxies respectively. Grey line corresponds to the 1:1 line ratio.}
    %     \label{fig9}
   %\end{figure}

The central panel of Figure \ref{fig8}  shows a comparison between optical-based and NIR-based stellar masses. The comparison shows that the mass estimations agree, following a $1:1$ relation, for galaxies with $M^{*} < 10^{10.5}$, where the sample is mostly dominated by ELGs. On the other hand,  at masses  $M^{*} > 10^{10.5}$, galaxies tend to fall above the 1:1 relation by 0.23 orders of magnitude on average. This region is mostly dominated by passive galaxies. %The reason behind this small offset is the age dependency in the mass-to-light ratio: This parameter depends on both the age and the metallicity. Assuming a fix sub-solar metal content of $[M/H] < -0.4$, the offset is clearly caused by age. Representing a galaxy with an overestimated age means overestimate its stellar mass. Due to we are using age values reported in S23 for both emission line galaxies and passive galaxies, consequently, the assumption that all passive galaxies have ages of 10Gyr on average are producing ovestimations when passive galaxies are younger.
However, this offset is within the uncertainty of the mass estimates, which is $\sigma = 0.29$ mags. Therefore, it does not change our conclusions about the MZR construction. 

\subsection{Metallicity estimations}
\label{metallicities}
\subsubsection{Classification of star-forming galaxies}
   
\begin{figure}
   \centering
   \includegraphics[width=\hsize]{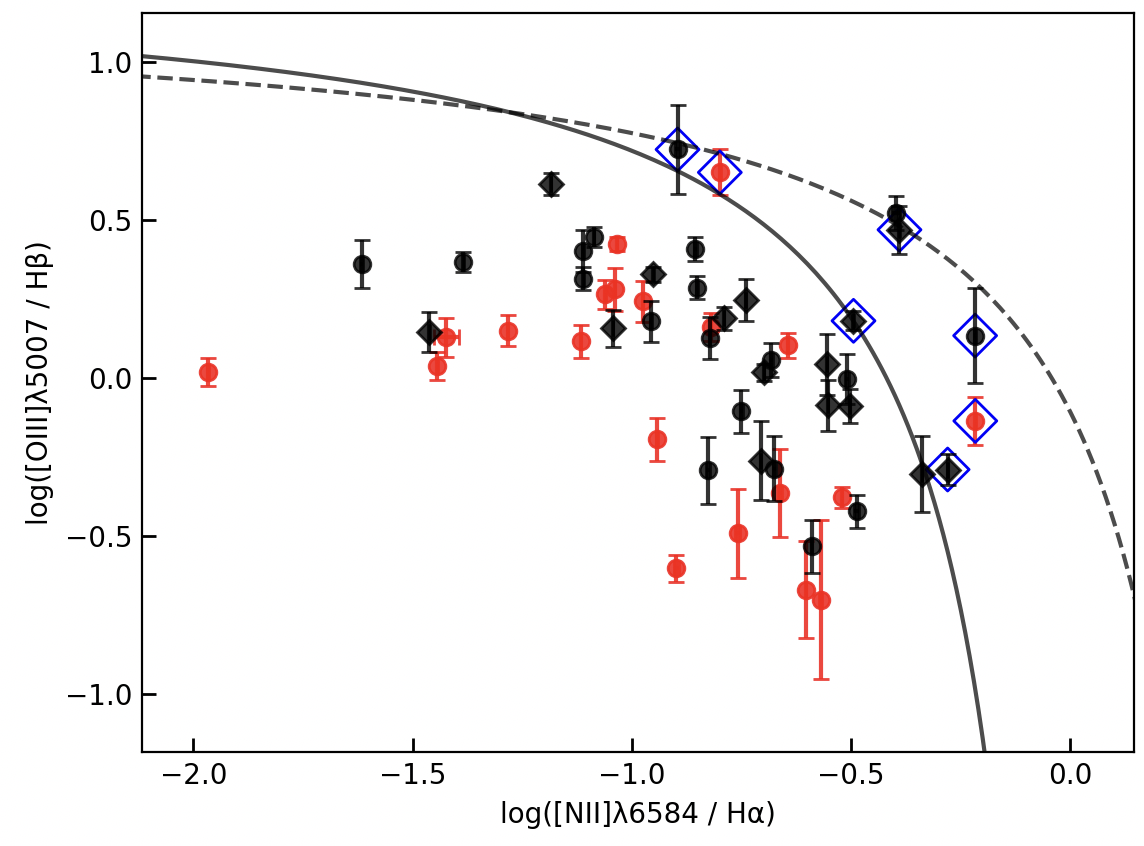}
      \caption{Baldwin-Philips-Terlevich diagram of the sample. The dashed black line is the KL at z=0 (\citealt{KLlimit}), which excludes pure AGNs. The solid black line is the \citet{kauffman03} limit, which excludes composite galaxies (ionization from star formation and AGNs). The red points are galaxies at the cluster redshift. Black diamonds correspond to foreground and background galaxies. According to both demarcation limits, all ELGs are SF galaxies. Data points with blue open diamonds are galaxies classified as composite galaxies, with two of them being composite galaxies at the cluster redshift and five of them found outside the cluster redshift.}
         \label{fig10}
   \end{figure}

In section \ref{subsection:vimos} we define ELGs by the presence of emission lines in their spectra. However, the ionization sources that trigger emission lines can be produced by active galactic nuclei (AGNs) and/or SF activity. 
To discriminate the actual nature of our galaxies, we made use of the emission line ratios, $\mathrm{[NII]\lambda6584/H\alpha}$ and $\mathrm{[OIII]\lambda5007/H\beta}$, in the Baldwin-Philips-Terlevich (BPT) diagram \citep{bpt}. In fact, the BPT diagram is a useful tool with which to identify the dominant ionization source of emission line galaxies. Figure \ref{fig10} shows the ELGs of the sample in the diagram. The red dots are cluster members and the black diamonds correspond to foreground and background galaxies. The dashed black line is the Kewley limit (KL). Galaxies found below the KL have dominant ionization from HII regions, where ongoing star formation is present. However, ELGs can have ionization from both HII regions and AGN activity, i.e. they are composite galaxies. Therefore, we used the \citet{kauffman03} demarcation limit to select pure SF galaxies and also study the possible effects of composite galaxies in the MZR. Composite galaxies are shown in Figure \ref{fig10} with open blue diamonds. Two composite galaxies are present in AC114, two belong to the foreground, and three belong to the background galaxies.

The fact that there is not a significant number of composite galaxies contaminating the galaxy sample allows us to make a direct and homogeneous comparison of the MZR for AC114, together with the foreground and background structures, without any other kind of contamination regarding the ionization sources. 

\subsubsection{Gas-phase metallicities}

The first half of the construction of the MZR was done with optical and NIR luminosities to estimate the stellar mass of the galaxies. We now continued to estimate the metallicities of the SF galaxies in the sample. 

The best technique to estimate gas-phase metallicities is through measuring intensities of emission line ratios sensitive to the electron temperature, $T_{e}$, together with the computation of line emissivities. This technique is the so-called “direct method” (\citeauthor{peimbert} \citeyear{peimbert}, \citeauthor{aller} \citeyear{aller}). However, $T_{e}-$sensitive emission line ratios involve lines such as the auroral $\mathrm{[OIII]}\lambda4363$, $\mathrm{[NII]}\lambda5755$, or $\mathrm{[SII]}\lambda\lambda 4068,4076$. These lines are $\sim 3-4$ orders of magnitude fainter than Balmer emission lines \citep{maiolinomanucci}. Therefore, auroral lines can be well measured in metal-poor or high-temperature environments. Otherwise, auroral lines get lost in the noise of the galactic spectra. 

Some alternative methods of estimating metallicities have been proposed; the so-called “strong-line methods,” calibrated on $T_{e}$-based oxygen abundances or photoionization models (e.g., \citeauthor{pyliugin} \citeyear{pyliugin}, \citeauthor{kobulnicky} \citeyear{kobulnicky}, \citeauthor{edmuns} \citeyear{edmuns}, \citeauthor{ferland} \citeyear{ferland}, \citeauthor{perezmontero} \citeyear{perezmontero}). In this work, we chose the well-known strong-line O3N2 index, defined as $\mathrm{O3N2}= \log{\left( \frac{\mathrm{[OIII]}\lambda5007/H\beta}{\mathrm{[NII]}\lambda6584/H\alpha} \right)}$, which correlates linearly with oxygen abundances in this way: $12+\log{(\mathrm{O/H})} = 8.73 - 0.32\times\textup{O3N2}$ \citep{pettini}. There exist at least ten different calibrators in the literature regarding gas-phase metallicity estimations (see \citeauthor{kewley} \citeyear{kewley} for a detailed analysis of the most common calibrators and comparisons between them). Despite the significant differences between some calibrators (up to 0.7 dex), the authors also provide consistent conversions between different calibrators within uncertainties of 0.2 dex. Those will also be used in section \ref{section:results} to make comparisons with the MZRs reported in the literature.

We chose to estimate metallicities with the O3N2 index for several reasons: (i) This index is calibrated with $T_{e}-$based oxygen abundances, which is the most reliable method despite its intrinsic biases such as temperature fluctuations, underestimations in the high metallicity range, and the way of estimating the electron temperature (\citealt{kobulnicky96}, \citealt{pilyugin16}, \citealt{cameron}). (ii) The relation between O3N2 and oxygen abundances is not bi-valued, at odds with, for instance, the R23 index; (iii) the emission lines used in the O3N2 calibrator are closer together (H$\beta - \mathrm{[OIII]}\lambda5007 $ in $\sim150\AA$ and  H$\alpha - \mathrm{[NII]}\lambda6584$ at $\sim20\AA$); hence, there are no significant noise changes in the observed spectra, the internal absorption does not affect the measurements, and the flux calibration is also less of a concern. (iv) Our constructed MZRs with field-foreground and field-background galaxies show oxygen abundances estimates consistent with those expected at their respective redshifts, despite the intrisic bias of the O3N2 index with the ionization.

Oxygen abundances of the sample of galaxies are shown in Table \ref{table:3}.4. The estimated oxygen abundances have values from $8.10\ \pm 0.06 < 12+\log{(\mathrm{O/H})}_{O3N2} < 8.77\ \pm  0.19$ for 54 ($67\%$) SF galaxies. The other SF galaxies do not have all these four emission lines to compute their respective metallicities.

%--------------------------------------------------------------------
\section{Results}
\label{section:results}
The reported statistical parameters relative to the analysis presented in this section are listed in Table \ref{table:0}
\subsection{The MZR of AC114, foreground and background galaxies}
\begin{figure*}[]
   \centering
   \includegraphics[width=\hsize]{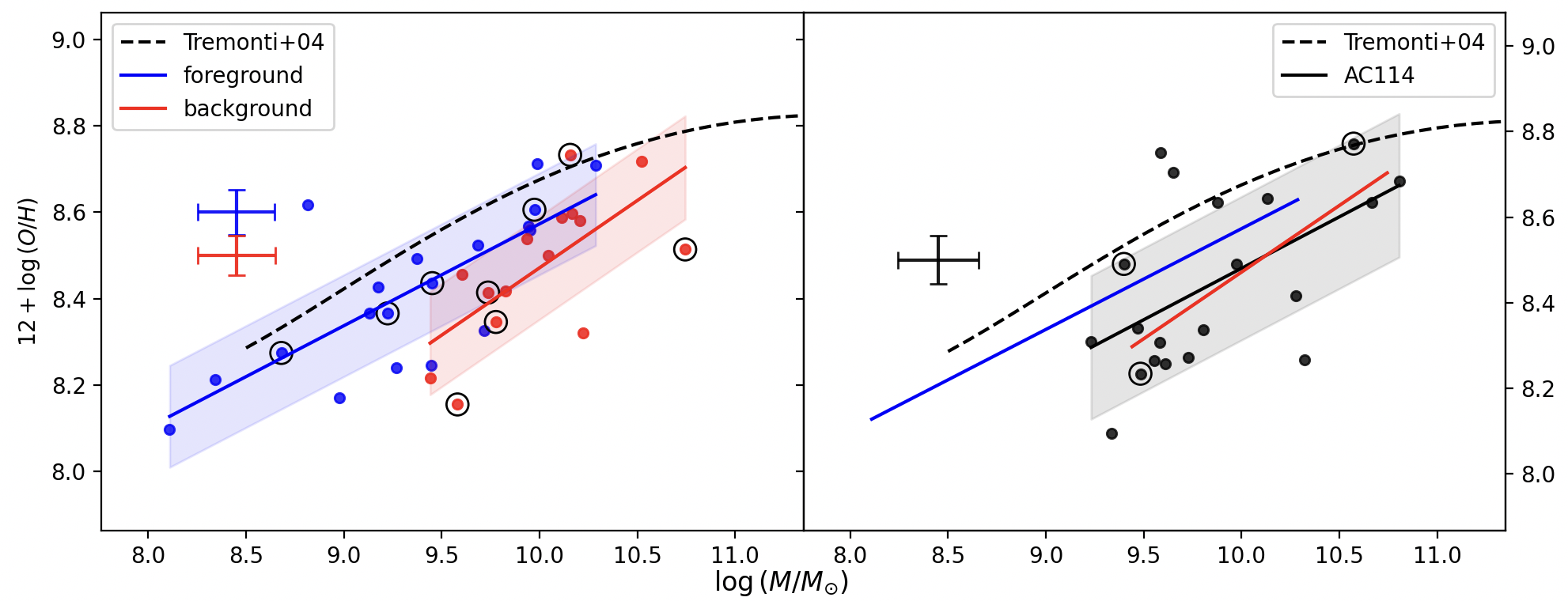}
      \caption{Mass-metallicity relations constructed with the SF galaxies of the sample compared with the local Universe MZR from \citet{tremonti}. The left panel shows the foreground MZR (fMZR) with blue dots and background MZR (bMZR) with red dots. Lines correspond to linear fits. The shaded area corresponds to the $1\sigma$ scatter for the fMZR and bMZR. Right panel: MZR with the SF members of AC114 (cMZR). The black line and shaded area are the linear fit and the respective $1\sigma$ scatter. The dashed blue and red lines are the fMZR and bMZR linear fits. The open circles (left) and open circles (right) show galaxies whose stellar mass was estimated using synthetic photometry.}
         \label{fig11}
   \end{figure*}

\begin{figure*}[h!]
   \centering
   \includegraphics[width=\hsize]{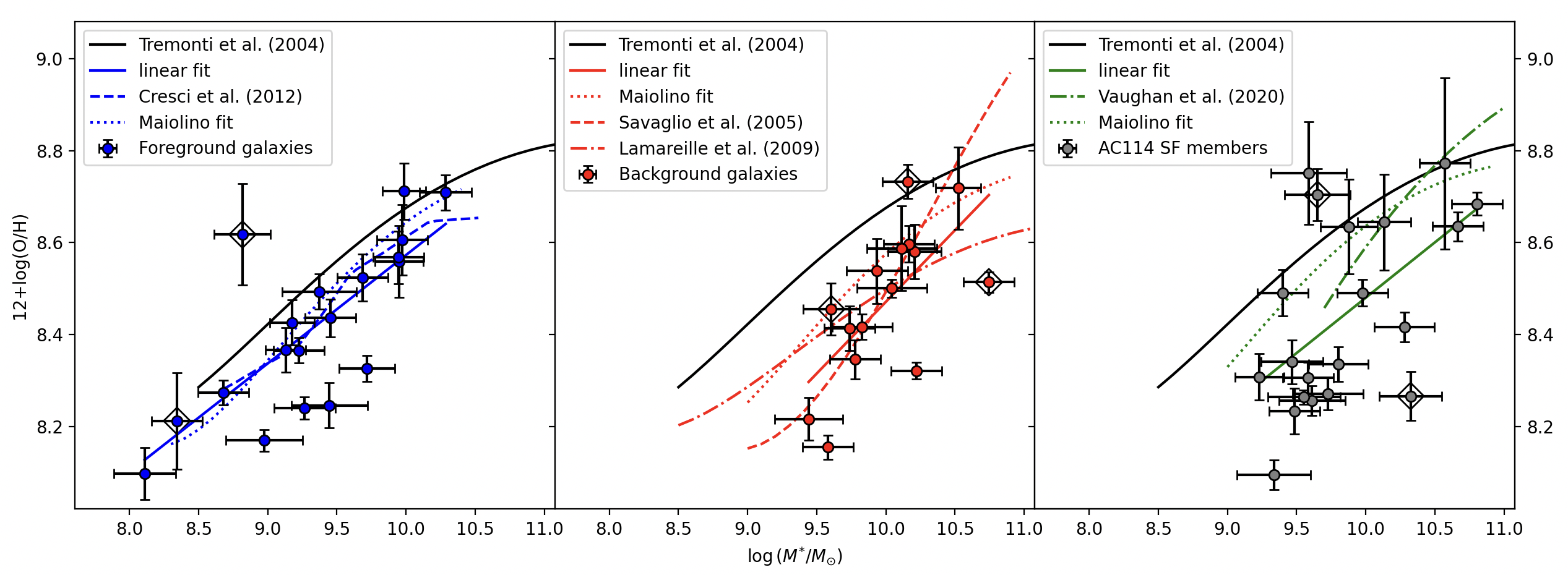}
      \caption{Comparison between the MZRs constructed in this work and the literature. Left panel: Comparison between the fMZR, \citet{cresci} MZR, and the Maiolino parameterization, both at redshift $z\sim0.25$. Middle panel: Comparison between the bMZR, \citet{savaglio}, \citet{Lamareille}, and the Maiolino parameterization at $z\sim 0.7$, respectively. Right panel: Comparison between the cMZR, \citet{vaughan}, and the Maiolino parameterization at $z\sim 0.3$. Data points with open diamonds show the galaxies classified as composite in the BPT diagram of Fig. \ref{fig10}.}
    \label{figcomparison}
   \end{figure*}

We constructed the MZRs with 54 SF galaxies for three subsamples labeled: foreground ($0.1<z<0.28$) with 19 SF galaxies, AC114 ($0.28\leq z\leq 0.34$) with 20 SF members, and background ($0.34<z<0.7$) with 15 SF galaxies. The redshift boundaries used were adopted from P15. In Figure \ref{fig11}, the MZRs are compared with the local Universe MZR from \citet{tremonti}. To do the comparison, we used the \citet{kewley} conversion to get the local Universe MZR in O3N2-based oxygen abundances. The left panel of Figure \ref{fig11} shows the MZRs of the foreground and background galaxies. Foreground and background galaxies' oxygen abundances correlate with the stellar mass, showing the existence of foreground and background MZRs (hereafter fMZR and bMZR). The non-parametric Spearman correlation test was performed for galaxies with $<1\sigma$ scatter, showing correlation coefficients of  $r=0.97$ (p-value $2.6\times10^{-8}$) and $r=0.90$ (p-value $1.5\times10^{-4}$). The $1\sigma$ scatter is $\sigma=0.118$ and $\sigma = 0.120$, showing that both the fMZR and the bMZR are tightly correlated, respectively. Additionally, they are downshifted by 0.09 and 0.20 dex in metallicities with respect to the local MZR, a clear sign of evolution with redshift of the global MZR: at a given mass, higher-redshift galaxies will present lower metal content.

The shallower slope of the fMZR ($0.236 \ \pm \ 0.003$) with respect to the bMZR ($0.312\ \pm \ 0.009$) is produced mainly by two physical phenomena: (i) Galaxies grow hierarchically across the cosmic time, and therefore wet mergers (i.e., mergers between low-mass galaxies with significant gas fractions) can enhance the gas-phase metal content, while low-mass galaxies evolve, due to the production of stars that pollute the ISM with processed material in the last stages of their evolution \citep{tissera}. (ii) Outflow processes are stronger at higher redshift, making it easier for gas particles to get expelled in shallower potential wells \citep{troncoso}. Hence, the evolution in redshift of the MZR in the low-mass range is stronger than in the high-mass range, where deeper potential wells will  produce a more efficient star formation. The different mass range of the fMZR and bMZR is a natural consequence of seeing galaxies in a range of distances with a fixed exposure time, as is the case with the DECaLS survey and AC114.

%  we shall also explain why we ignore evolution effects, which could be present for with different redshift values
We ignored evolution effects (galaxies at higher redshift are in different stages of their evolution) because we focused our analysis on studying differences in the shape and the evolution of the MZR. The evolution effects mainly affect the construction of the bMZR because it represents galaxies at redshift $0.34<z<0.7$. If we take into account the evolution effects, the scatter of the bMZR will decrease. However, we would not be able to produce a significant bMZR because there are not enough galaxies in our sample.

The right panel of Figure \ref{fig11} presents the MZR constructed at $z\sim0.3$ with SF galaxies of the AC114 galaxy cluster (hereafter cMZR). The first clear feature in the cMZR is that the slope of the linear fit ($0.241 \ \pm \ 0.008$) is slightly shallower than both the fMZR and bMZR slopes (shown with the black line), with a larger scatter of $\sigma = 0.17$ and a Spearman rank of $r = 0.65$ (p-value $=0.009$). These values suggest that physical phenomena are suppressing the chemical evolution of high-mass galaxies, increasing the observed scatter and weakening the correlation between mass and metallicity. The second important feature seen is that the high scatter of the cMZR and the effect in the high-mass range of the shallower slope with respect to the bMZR indicates that SF members of AC114 are either similar to or metal-poorer than the background galaxies; the opposite of the expected evolution of the MZR.

Classically, the MZR and its evolution in redshift reported in the literature is constructed with (i) field SF galaxies and (ii) statistically complete samples of SF galaxies at a given redshift, where the role of galaxy cluster environments is negligible. It is well known that at a given mass, gas-phase metallicities are more metal-poor at higher redshift (or at a given oxygen abundance, galaxies are less massive at higher redshift). Several physical phenomena, such as SNe and galactic winds, the infall of pristine gas, outflow processes, IMF variations, and galactic downsizing, together with the formation of galaxies through dry or wet mergers, robustly explain the shape of the MZR. However, all of these are internal physical processes. In dense environments, such as massive galaxy clusters, galaxies can suffer harassment, interaction with closer neighbours (galaxy-galaxy interaction), and/or interaction with the ICM. These external physical processes can change the gas-mass fraction, modifying the gas-phase metal content. Therefore, these effects can change the MZR and may explain the cMZR in Figure \ref{fig11}.

\subsection{Comparison of the fMZR, bMZR, and cMZR with the literature}

As a sanity check of the MZRs found above, we proceeded to compare them with those reported in the literature. The fMZR was compared with \citet{maiolino} and \citet{cresci}. The bMZR was compared with \citet{savaglio}, \citet{maiolino}, and \citet{Lamareille}. Finally, the cMZR was compared with \citet{maiolino} and \citet{vaughan}.

\citet{savaglio}, \citet{cresci}, and \citet{maiolino} use the $R_{23}$ parameter to estimate metallicities, whereas \citet{Lamareille} and \citet{vaughan} use the N2 parameter. Therefore, we followed \citet{kewley} to convert $R_{23}$ and N2 to O3N2.

In \citet{maiolino}, the authors parameterize the MZR with the expression $12+\mathrm{log(O/H)} = -0.0864(\log{M_{*}/M_{0}})^2 + K_{0}$ at redshifts 0.07, 0.7, 2.2, and 3.5, with $M_{0}$ and $K_{0}$ free variables. We interpolated both variables as a function of the redshift to get their parameterization at redshifts 0.25, 0.3, and 0.7.

The left panel of Figure \ref{figcomparison} shows the fMZR compared with field MZRs from (i) \citet{cresci} at $z\sim 0.25$, (ii) the Maiolino parameterization at $z\sim 0.25$, and (iii) the local Universe MZR of \citet{tremonti}. As we can see, there is an agreement between the fMZR and the literature. The galaxies that are not in agreement with the MZRs from the literature are, in fact, those which are beyond the $1\sigma$ scatter. Those foreground galaxies that are metal-poor with respect to the MZRs in the literature seem to follow a trend. We discarded that the feature seen is an evolutionary effect because the galaxies that fall in the MZR have redshifts up to $0.28$, covering the same range as the outlier galaxies. Three of these four outlier foreground galaxies are at the same redshift: Q4-7, Q4-8, and Q4-35, at 0.169, 0.169, and 0.170, respectively. Here, Q4-7 and Q4-8 are closer neighbors, sharing the same region of the FoV. Q4-35 is part of a crowded region in the FoV. On the other hand, Q3-11 at redshift 0.21 has a closer companion in the FoV but without available spectroscopic data. The fact that these galaxies are closer to other ones may indicate that galaxy-galaxy interactions are responsible for the metal-poor values with respect to the fMZR. This will be discussed in Section \ref{section:discussion}.

The middle panel of Figure \ref{figcomparison} shows the bMZR compared with field MZRs from (i) \citet{savaglio}, (ii) \citet{Lamareille}, and (iii) the Maiolino parameterization at $z\sim 0.7$. In this panel, our measurements are consistent with those reported in the literature in the high-mass range of the MZRs. On the other hand, the low-mass galaxies in our sample tend to follow the \citet{savaglio} MZR. Again, the background galaxies that fall below the MZRs compared are those that are beyond the $1\sigma$ scatter. From these two panels, it is clear that the fMZR and bMZR follow the field MZRs at their respective redshift ranges.

In the right panel of Figure \ref{figcomparison}, we compare the AC114-MZR (cMZR) with the Maiolino parameterization and the MZR reported in \citeauthor{vaughan} (\citeyear{vaughan}, hereafter V20), with the dotted and dash-dotted green lines, respectively. V20 studies a sample of four galaxy clusters at $0.3<z<0.6$ and compares their cluster members with respect to field galaxies in the mass-metallicity plane at the same redshift. They did not find any difference between those MZRs (see Figure 7 in V20). 

To compare the V20 data with this work, two of the four galaxy clusters were found at a similar redshift of AC114, MS2137, and MACS1931 at z=0.313 and z=0.352, respectively. We find that just seven and two galaxies have oxygen abundances estimated, belonging to MS2137 and MACS1931, respectively. The field galaxies in V20 at the redshift of AC114 ($0.28<z<0.35$) span masses from $\sim10^{10}$ to $\sim10^{11} M_{\odot}$, whereas our mass estimates in AC114 cover two orders of magnitude starting from $\sim10^{9} M_{\odot}$. Therefore, that sample is not appropriate to compare with our results.

In the right panel of Figure \ref{figcomparison}, the linear fit of the AC114 galaxy sample presents a higher downshift than the Maiolino parameterization and V20 with respect to the local Universe MZR. The cMZR is downshifted 0.19 dex with respect to the local Universe MZR, and the scatter of the cMZR is 0.17 dex.

The fact that both the fMZR and bMZR are in agreement with field MZRs at their respective redshifts, together with the cMZR showing a slightly shallower slope, a downshift similar to the bMZR, and a high scatter of 0.17 dex, indicates that there are physical phenomena behind those features, which are likely related to environmental and/or dynamical effects.

% ------------------------------------------------------------------
\subsection{Structure of AC114 and consequences for the chemical evolution of its members}
We explored the spatial distribution to gain insights into the changes seen in the cMZR with respect to the fMZR and the bMZR. We first converted the projected RA,Dec and redshift coordinates of the cluster members to physical coordinates.  \citet{cosmocalc} was used to estimate the respective arcsecond per kiloparsec scale conversion as a function of the redshift for a few select values. A second-degree polynomial was fitted, parameterizing the expression: $\frac{\mathrm{Kpc}}{''}(z) = -9.800( \pm 0.300)\ z^{2} + 16.700( \pm 0.200)\ z + 0.335( \pm 0.007)$, with $z\in[0.1,0.7]$. At the cluster distance $D_{L} = 1670 \ \pm \ 17$ Mpc, the corresponding scale is $\sim 4.6$ Kpc/$''$, and at the cluster range $0.28<z<0.34$ the scale spans values from $\sim4.3$Kpc/$''$ to $\sim4.9$Kpc$/ ''$. 

AC114  has a nonspherical shape, being elongated in the southeast-northwest direction (\citeauthor{couch} \citeyear{couch}, \citeauthor{defilippis} \citeyear{defilippis}). The dynamical study of P15 shows also that AC114 is a very elongated radial filament, spanning $12\ 000\ \mathrm{km\ s^{-1}}$ with a broad velocity dispersion of $\sim 1893_{-82}^{73} \ \mathrm{km \ s^{-1}}$. These dimensions correspond to $\sim330 \ \mathrm{Mpc}$ in the radial direction and $\sim 9 \ \mathrm{Mpc}$ in the perpendicular component, the latter being a lower limit, as the calculations are restricted to the FoV of the VIMOS observations. 

\begin{figure}
   \centering
   \includegraphics[width=\hsize]{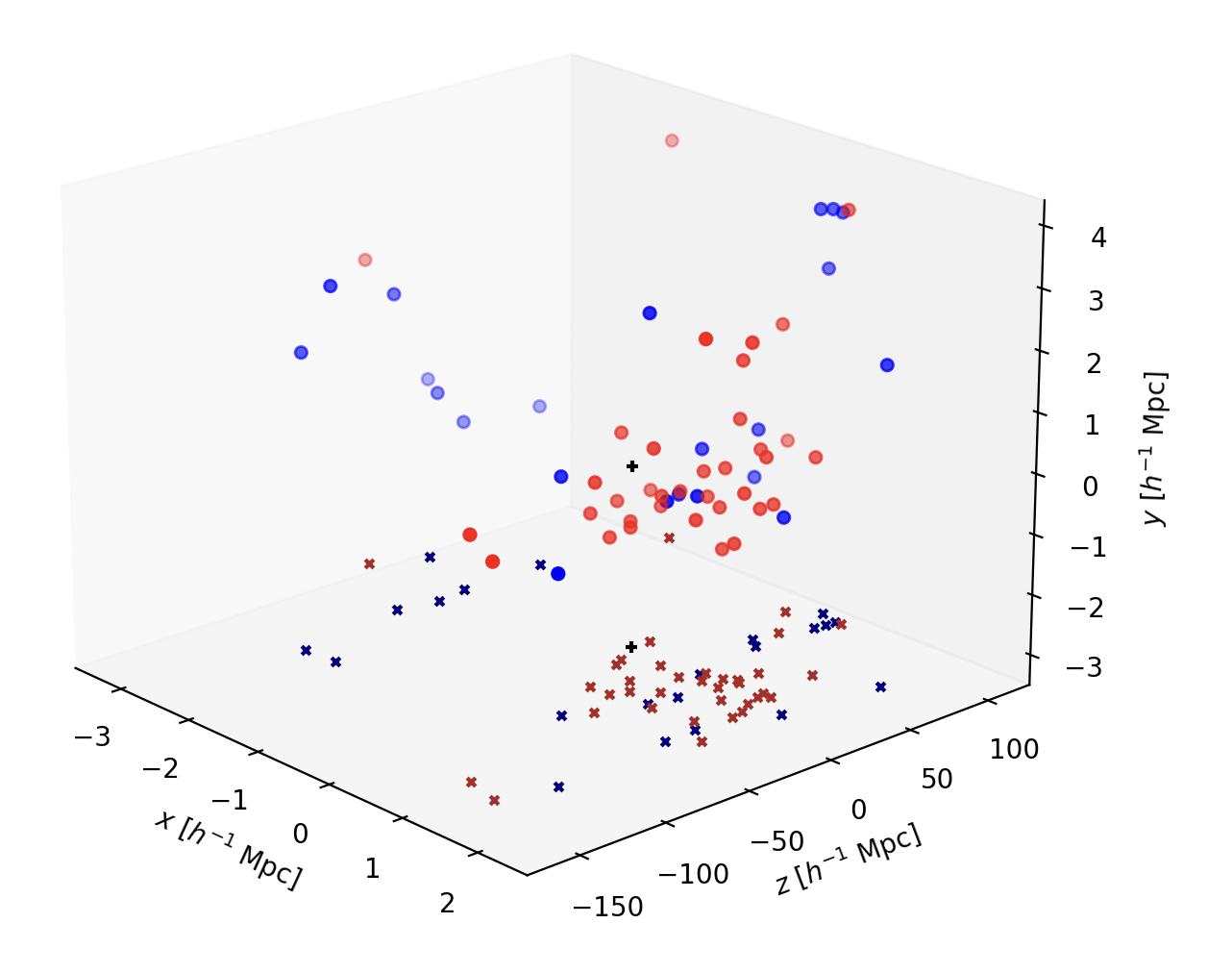}
      \caption{Three-dimensional projection of the AC114 members, expressed in Mpc with respect to the cluster center. The x, y, and z axes correspond to the RA,Dec coordinates and to the redshift, respectively. Blue and red points are SF and passive galaxies, respectively. Colored crosses are the projected distribution of the cluster sample in the xz plane. The black cross indicates the center of the galaxy cluster.}
         \label{fig12}
   \end{figure}

The three-dimensional representation of the cluster members is shown in Figure \ref{fig12} for SF and passive galaxies with blue and red dots, respectively. Figure \ref{fig12} shows that the center of the cluster (at 0,0,0 in X,Y,Z) is populated mostly by passive galaxies; SF galaxies tend to be found at the outskirts. 

\begin{figure}
   \centering
   \includegraphics[width=\hsize]{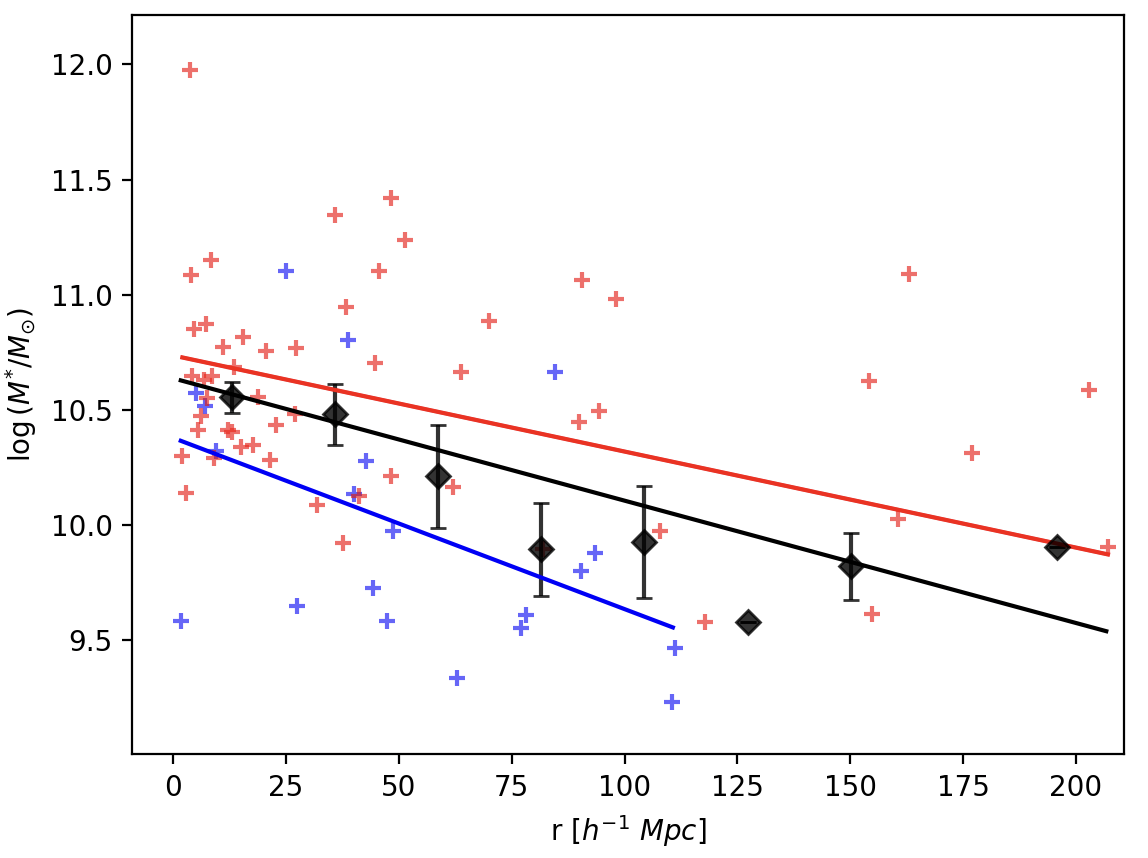}
   \includegraphics[width=\hsize]{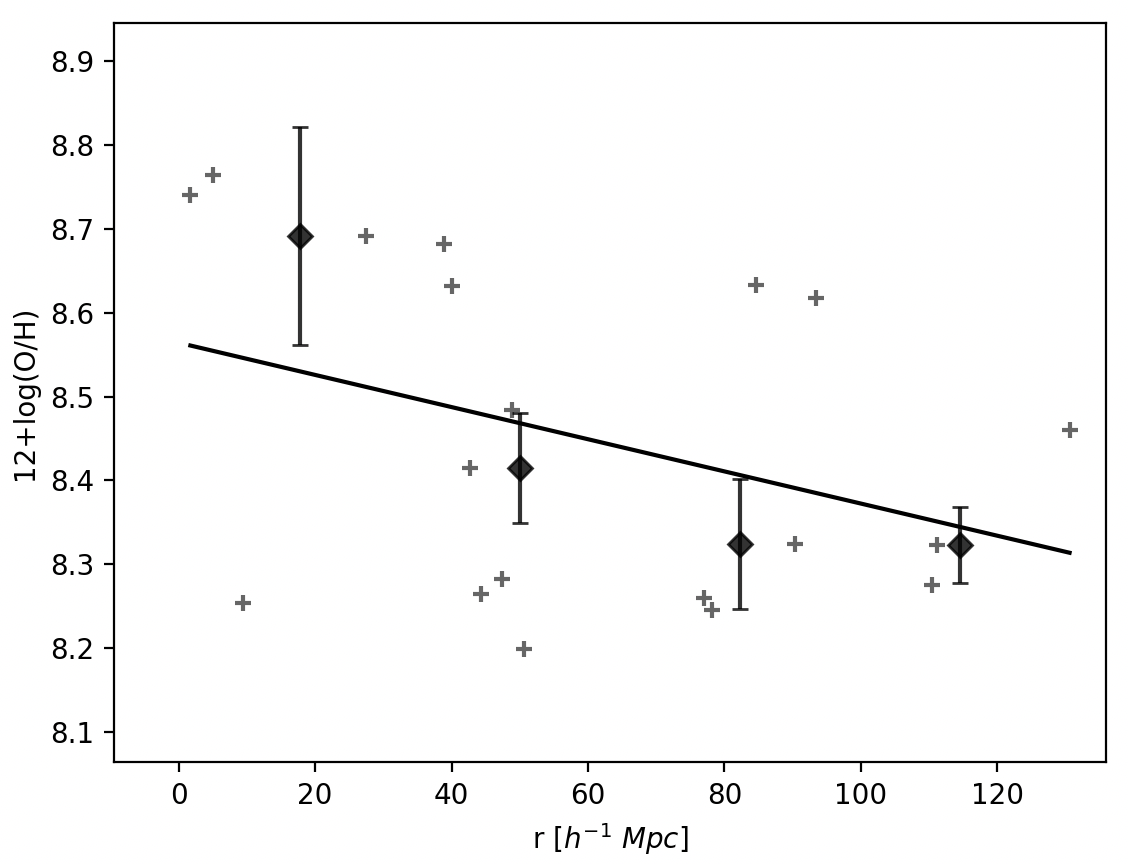}
      \caption{Radial profiles of AC114. Top panel: Stellar masses of galaxies as a function of the clustercentric radius. Blue and red crosses are SF and passive galaxies. Black diamonds represent mean mass bins of $\sim23 \ \mathrm{h^{-1}\ Mpc}$. Black, blue, and red lines are the linear fits that show a negative mass-gradient as a function of the clustercentric radius for the complete sample, SF, and passive galaxies, respectively. Bottom panel: Metallicity profile of the SF galaxies with available mean metallicity estimations. Black diamonds are mean metallicities in bins of $\sim33 \ \mathrm{h^{-1}\ Mpc}$, and the black line is the linear fit that shows a weak negative gradient.}
         \label{fig13}
   \end{figure}

The mass and metallicity profiles are presented in Figure \ref{fig13}. The top panel shows the mass profile of AC114. Just as in Figure \ref{fig13}, we observe a high concentration of galaxies toward the cluster center, which is dominated by passive galaxies. A negative radial mass gradient is detected. Black diamonds are the mean stellar mass of the galaxies in bins of $\sim23 \ $Mpc. The slope of the mass profile of $-0.005\pm0.002$ (Spearman rank is $-0.57$, p-value $= 0.005$) shows a weak but statistically significant trend to find massive galaxies in the inner regions of the cluster. This is consistent with AC114 being in an early stage of radial relaxation, as is also suggested by the large velocity dispersion estimated in P15. Furthermore, passive galaxies present a slightly flatter slope (-0.004 $\pm$ 0.003, red linear fit) compared with SF galaxies (-0.007 $\pm$ 0.003, blue linear fit), which is indicative of ongoing dynamical activity in the galaxy cluster. The bottom panel of Figure \ref{fig13} presents a weak, non-statistically significant (Spearman rank of $-0.38$ with a p-value $=0.11$) metallicity gradient of the SF galaxies, with a slope of $-0.002 \ \pm\ 0.002 $. 

\begin{figure}
   \centering
   \includegraphics[width=\hsize]{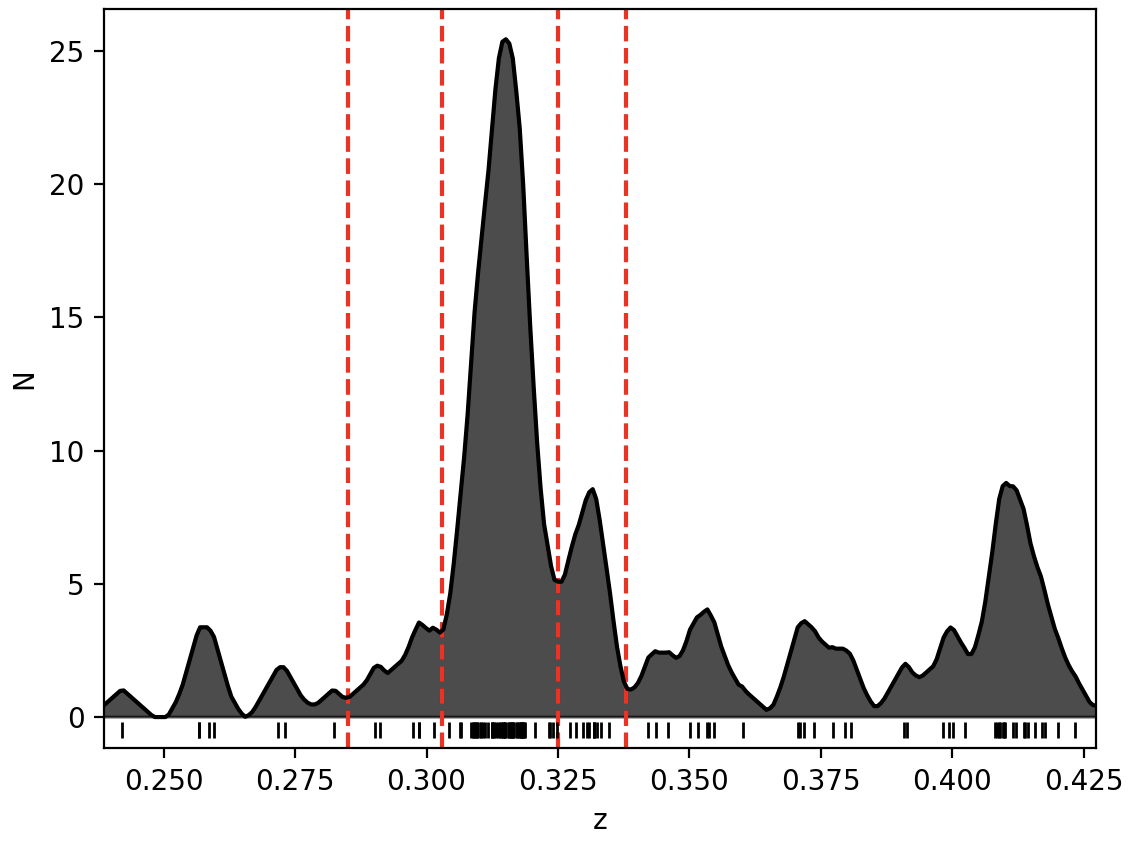}
   \includegraphics[width=\hsize]{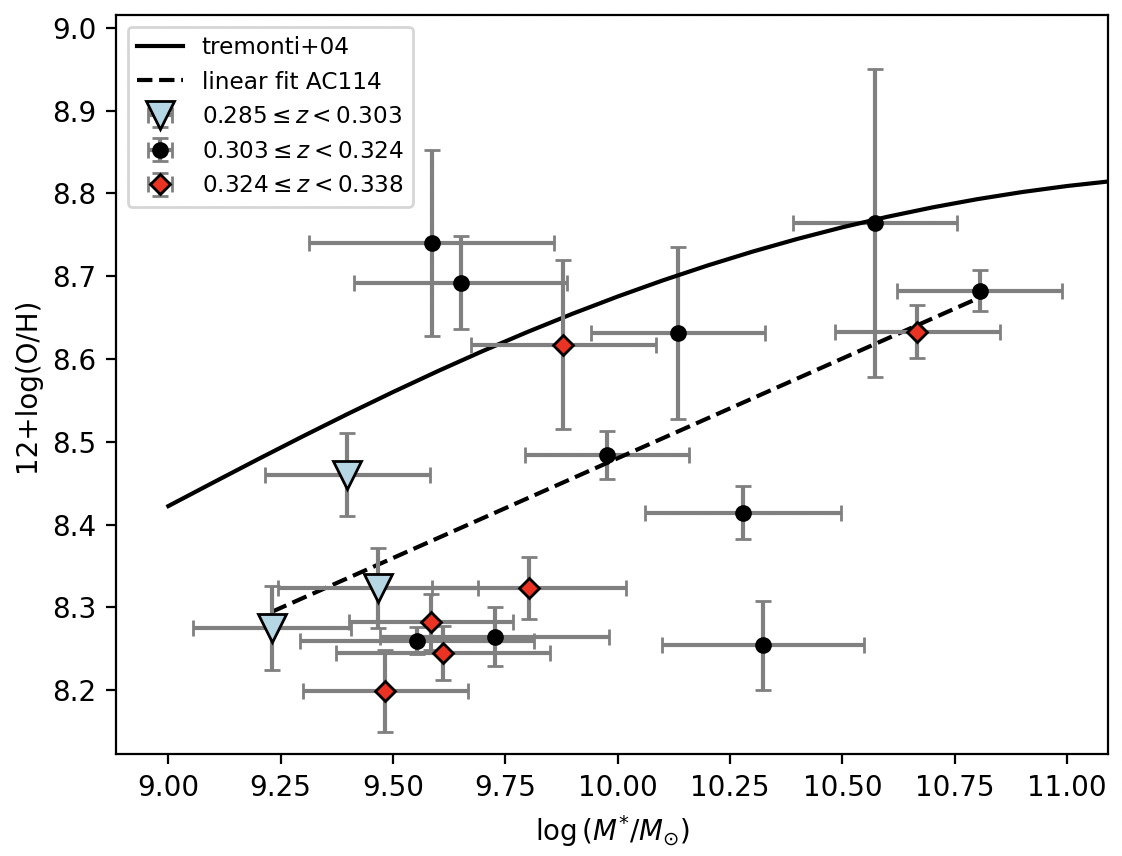}
      \caption{Top panel: Kernel density estimator in the redshift distribution of the galaxy cluster region: $0.28\leq z \leq 0.34$. Red lines show the redshift cuts of the three substructures identified: SS1 in $0.2850\leq z< 0.3029$, SS2 with $0.3029\leq z< 0.3243$, and SS3 covering $0.3243\leq z< 0.3380$. Vertical black segments are the clustercentric radial distance of the galaxies. Bottom panel: Galaxies of the three substructures in the mass-metallicity plane. Sky-blue triangles, black dots, and red dots represent SS1, SS2, and SS3, respectively. The black line is the cMZR. }
         \label{fig14}
   \end{figure}

We applied a kernel density estimator (KDE) to the redshift distribution of the galaxy cluster to search for possible substructures and construct the MZR of the sub-components. We made use of the machine-learning \textit{KernelDensity} algorithm from the Scikit-learn package\footnote{https://scikit-learn.org/stable/index.html} and used a linear kernel model. Three substructures were identified in the redshift distribution with this procedure. The red lines in the top panel of Figure \ref{fig14} show the redshift limits of the substructures, namely SS1 ($0.2850\leq z <0.3029$) being the front part of the galaxy cluster with three SF galaxies, SS2 ($0.3029\leq z <0.3243$) for the main component of AC114 with ten SF galaxies, and SS3 ($0.3243\leq z < 0.3380$) being the back part of the galaxy cluster with six SF galaxies. 

Once the substructures were identified, we went ahead to construct the respective MZRs. The bottom panel of Figure \ref{fig14} shows the SF galaxies of the substructures in the mass-metallicity plane. Sky blue, black, and red are colors for SS1, SS2, and SS3, respectively. From SS1, only three SF galaxies were identified, grouped in the same region of the mass-metallicity plane. A statistically significant positive correlation is detected for SS3 (slope of 0.26 $\pm$ 0.14, Spearman rank $r=0.860$, and p-value $=0.019$, black dots). Furthermore, most of the galaxies in the sub-sample lie below the cMZR, suggesting that outskirt galaxies of AC114 are slowly evolving due to external processes that are suppressing the chemical evolution. On the other hand, the main component of AC114 (SS2) shows more scatter (0.28 and 0.33 for SS3 and SS2, respectively) in the mass-metallicity plane (red dots). The slope of the fit of SS2 is 0.19 $\pm$ 0.13, and no statistically significant trend is seen (Spearman rank of = 0.54, p-value = 0.13). The central region of AC114 consists of ten SF galaxies and 42 passive galaxies. Therefore, SF galaxies closer to the central region of AC114 are suffering not only from interaction with the ICM, but also more likely to be suffering from interaction with their neighbors.

\begin{figure}
   \centering
   \includegraphics[width=\hsize]{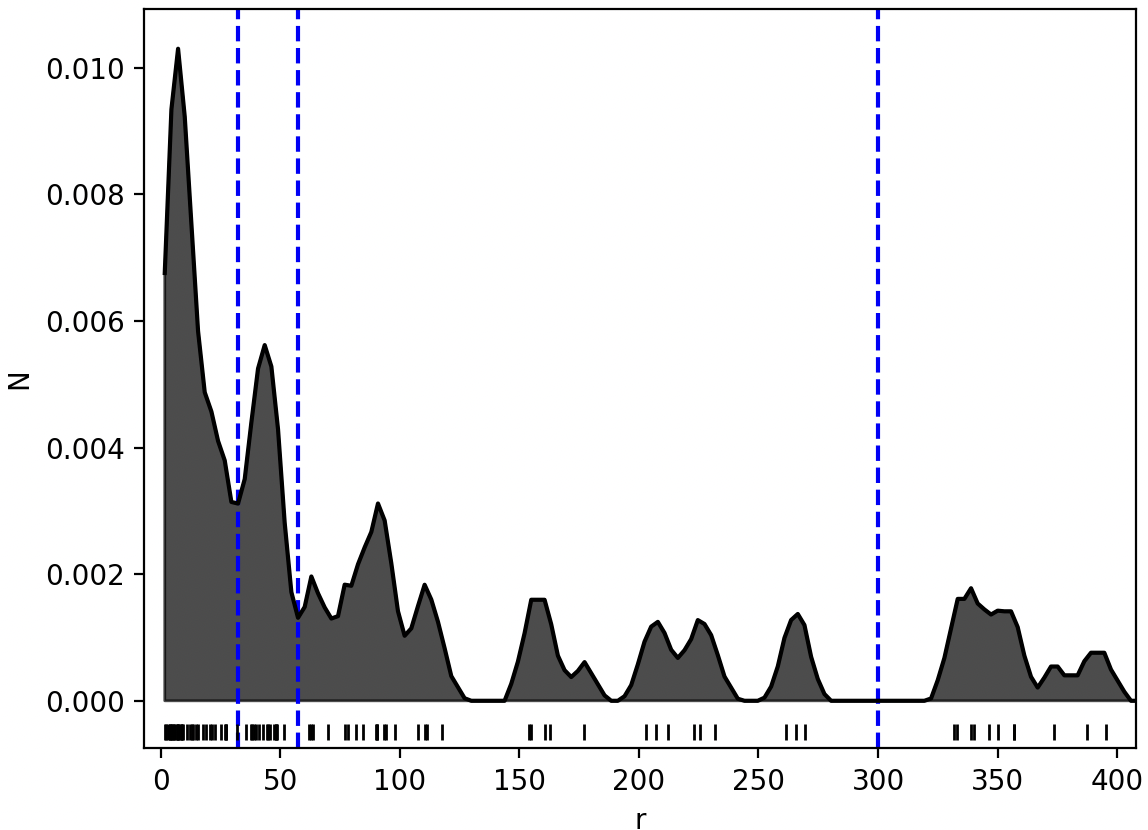}
   \includegraphics[width=\hsize]{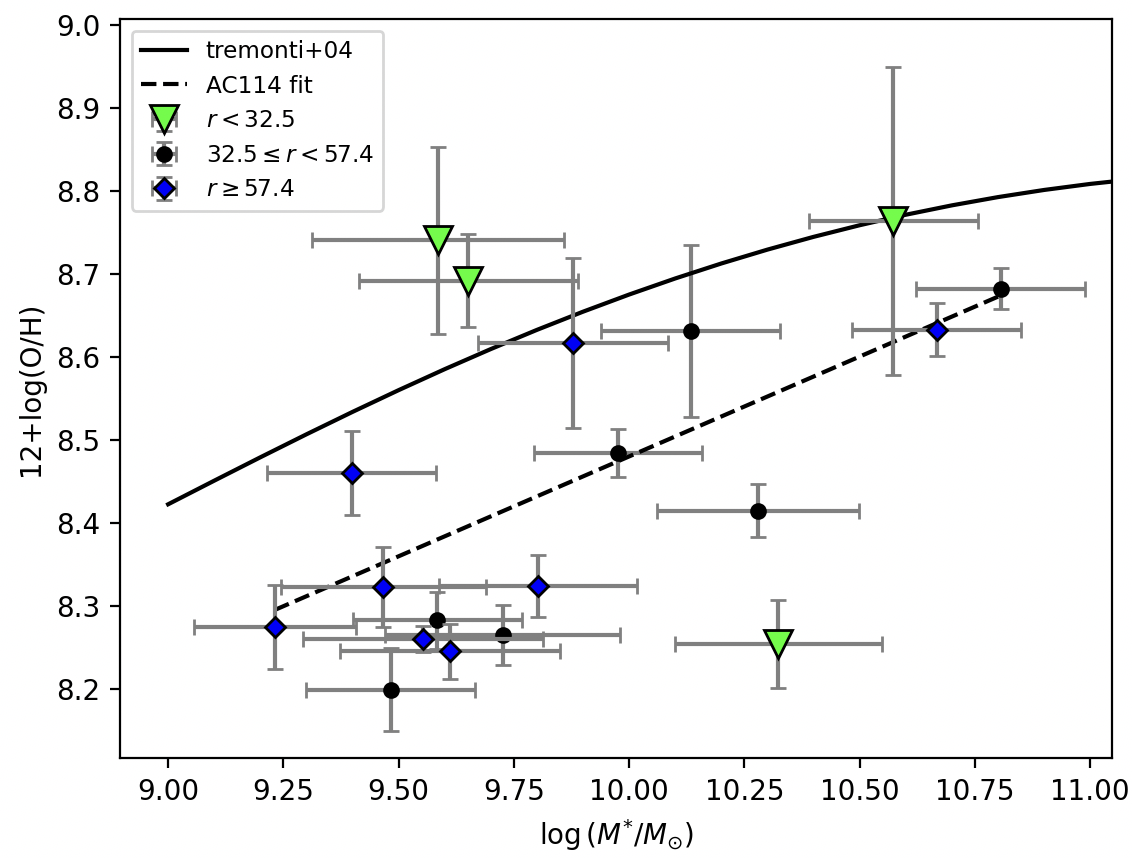}
      \caption{Same as Figure \ref{fig14}, but in this instance the KDE was applied in the radial distribution. Top panel: Kernel density estimator in the radial distribution of the galaxy cluster. Blue lines show radial cuts at $r= 32.5$, $r=57.4$, and $r=140\ $Mpc. The vertical black segments are the clustercentric radial distances of the galaxies. Bottom panel: Galaxies in the mass-metallicity plane of the three radial regions. Green triangles, black dots, and blue dots represent the central, intermediate, and outskirt regions, respectively.}
         \label{fig15}
\end{figure}

In the previous subsection we were able to estimate clustercentric radial distances, so we repeated the same KDE procedure but applying it to the clustercentric radial distribution of AC114 galaxies. We divided the galaxies according to the following radial ranges (blue lines in the top panel of Figure \ref{fig15}): the central region at $r\leq 32.5\  h^{-1}$Mpc with four SF galaxies, the intermediate region between $32.5\leq r <57.4\ $Mpc with seven SF galaxies, and the outskirts of AC114 in $r > 57.4 \ $Mpc with eight SF galaxies. 

The bottom panel of Figure \ref{fig15} shows the SF galaxies of AC114 in the mass-metallicity plane colored with green, black, and blue for the central, intermediate, and external regions, respectively. Despite the central region only having four galaxies, two of them are the metal-richer ones of the galaxy cluster. On the other hand, the intermediate and outskirt regions of AC114 show an increase in oxygen abundances with the increase in the stellar mass, with a slope of 0.35 $\pm$ 0.08, Spearman rank $r = 0.86$, and $p-$value $= 0.014$, and a slope of 0.162 $\pm$ 0.09, Spearman rank $r = 0.40$, and $p-$value $=0.319$, respectively. 

% ------------------------------------------------------------------
% ------------------------------------------------------------------
\section{Discussion}
\label{section:discussion}
P15 and S23 studied the dynamical state of AC114 and the stellar populations of the cluster members. The radial profiles in Figure \ref{fig13} show a negative mass gradient as a function of the clustercentric radius, consistent with the cluster being in an early stage of radial relaxation, as is also suggested by P15. The steeper mass gradient up to $r\sim 130 \ \mathrm{Mpc}$ of SF members suggests that these galaxies are more sensitive to the ongoing dynamical processes of AC114 because of their lower masses with respect to the passive ones. 
% This is also reflected in the weak negative mass gradient: In relaxed systems (steeper mass gradients), due to the mass dependence of the metallicity, should be noticed a steeper gradient than the observed. This disagreement may suggest that external mechanisms are playing an important role suppresing the chemical evolution.
Both the cMZR and the bMZR shown in Figure \ref{fig11} are downshifted with respect to the local Universe MZR, showing that AC114 SF members are more metal-poor than the field galaxies at the same redshift, and similar in terms of metallicity to background galaxies at a given mass.

S23, studying stellar populations of SF and passive galaxies, has shown that the stellar MZR of AC114 is steeper than for galaxies in the local Universe. However, the gas-phase cMZR in Figure \ref{fig11} has a shallower slope than galaxies in the local Universe. This could be due to galaxies quenching by (i) strangulation (\citealt{larson}, \citealt{peng}), where the gas is consumed faster when there is no fresh supply, and (ii) the ICM is boosting the metal-poor inflows. By adding our analysis to that reported in S23, stellar metallicities are lower than gas-phase metallicities because those metals were produced in the past, while the gas-phase metallicities are present-day ones. Therefore, the SF galaxies should be in an early stage of the quenching.

When cluster members were separated into substructures, important imprints of environmental effects were revealed. Figures \ref{fig14} and \ref{fig15} shows that galaxies located outside the central region set the slope of the cMZR (red and black dots in Figure \ref{fig14}, and blue and black dots in Figure \ref{fig15}). Star-forming galaxies within the main component of AC114 and the central region do not show a correlation between mass and gas-phase metallicity, with a large scatter. \\ %These properties indicate that SF galaxies are evolving slowly in the outskirts.\\
\\
The MZRs of SF galaxies in massive clusters ($>10^{14}M_{\odot}$) appear to follow those of field galaxies at their corresponding redshift. For example, \citet{vaughan}, studying four CLASH clusters, namely MACS2129, MACS1311, MACS1931, and MS2131 ($M_{200}\sim10^{14}-10^{15}\ M_{\odot}$; \citealt{Tiley}), in the redshift range $z\sim 0.2-0.6$, indicate that in these clusters the MZRs do not show any changes with respect to that constructed with field SF galaxies. Additionally, \citet{lara-lopez} and \citet{gupta}, by using SF members of the Fornax cluster ($M_{vir}\sim 7\times 10^{13}M_{\odot}$; \citealt{drinkwater}) and the RX J1532$+$30 ($z\sim 0.35$, $M_{vir} \sim 6.4 \times 10^{14}M_{\odot}$) galaxy cluster, respectively, are also consistent, with no variations with respect to the field MZR. However, these works report flatter galactic metallicity gradients, negative cluster metallicity gradients, and a negative trend in the residuals of the MZRs as a function of the clustercentric radius. Several physical processes are discussed, but the main physical phenomena that drive those trends are strangulation and RPS. %agregar cooper et al. 2008

It is well known that RPS is the physical result of the galaxy-ICM interaction that affects the outskirts of galaxy disks: in the presence of negative metallicity gradients, the stripped-out gas is metal-poorer than their nuclei, which leads to increased oxygen abundances.

Strangulation is produced when the pristine gas inflow ends and the gas reservoir is consumed faster, increasing the SFR until the quenching begins. Therefore, the H$\alpha$ disks become smaller than the stellar disks, and oxygen abundances are more metal-rich than galaxies that do not experience galaxy-ICM interactions at the same mass. 

On the other hand, galaxy clusters such as Coma, Hercules (A2151), MACSJ1115+01, and those in the SDSS DR4 survey, show enhanced metallicities with respect to the field MZR, up to $\sim 0.05$ dex in galaxies affected by RPS and/or galaxy-galaxy interactions inside the virial radius of their host dark matter halos (\citealt{ellison3}, \citealt{petropoulou2}, \citealt{petropoulou}, \citealt{gupta}). This indicates that extremely dense environments, such as galaxy clusters with $M_{vir}>10^{15}M_{\odot}$, may show variations in the cluster MZRs with respect to the field ones.

The MZR of AC114 also shows this feature for at least two cluster members when it is compared with the field MZR at $z\sim 0.3$, as is indicated in Figure \ref{figcomparison}. In this case, we used the Maiolino parameterization at redshift $0.3$ to compare with the AC114-MZR. However, the two galaxies are more metal-rich, $\sim 0.10$ dex, which is more than that reported by \citet{ellison3}. Additionally, the other galaxies (below the linear fit of the cMZR with the red line) drive the shallower slope and have lower metallicities than the field SF galaxies at the same redshift up to $\sim 0.22$ dex. The striking feature is that these galaxies are located outside the central region of AC114 (see Figure \ref{fig15}). Several reasons can explain this feature: (i) the AGN feedback, which is important at low $z$, can expel the metal-rich gas component in the central region of SF members with negative metallicity gradients (\citealt{derossi2}, \citealt{wang}); (ii) the presence of strong metal-poor inflows (\citealt{finlator}, \citealt{ceverino}), where galaxies accreting metal-poor gas decrease their average metallicities; and (iii) mergers and tidal interactions, where the interaction with closer companions can dilute the gas-phase metallicities.

In order to understand the main physical phenomena behind the metal-poor content of the SF galaxies of AC114 with respect to the field MZR at the same redshift, the AGN feedback can be discarded because our cluster sample has just two composite galaxies. One of them is over the local Universe MZR, and the other shows the largest deviation below the MZR in the cluster sample. According to \citet{kauffman03}, the location of these two composite galaxies in the mass-metallicity plane can be explained as the former being a LINER-composite galaxy and the latter a Seyfert-composite galaxy. Additionally, Seyfert galaxies have stronger [OIII]$\lambda 5007$ emission lines than SF galaxies. The O3N2 index anticorrelates linearly with the oxygen abundances, so it may explain the location of this galaxy in the mass-metallicity plane. The same applies to the LINER galaxy in our cluster sample, but in this case, LINERS have strong [NII]$\lambda 6584$ emission lines. According to observation and simulations from \citet{zahid} and \citet{wang}, despite the AGN feedback being important at low redshift, for satellite galaxies (non-cDs in galaxy clusters) this effect is negligible.

On the other hand, a strong metal-poor inflow and dynamical events can explain the observed feature. \citet{montuori}, simulating mergers and flybys between disk galaxies, showed that galaxy-galaxy interactions can dilute gas-phase metallicities and also increase the SFR, due to the accretion of metal-poor gas from galaxies with negative metallicity gradients. The estimated dilution agrees with \citet{kewley2006}, which reports that galaxy pair interaction dilutes gas-phase metallicities up to $\sim 0.2-0.3$ dex. Those values are also in line with the downshifts of the outlier foreground galaxies with respect to the fMZR, which are $\sim 0.18$ on average. Additionally, \citet{montuori} also explain that galaxy-galaxy interaction increases the gas-phase metallicities after the dilution due to the increase in the SFRs. Therefore, this mechanism increases the scatter of the MZR, as is also seen in \citet{rupke}.
 
%Galaxy-galaxy interactions can explain the scatter ($\sigma=0.13$) of the cMZR. The comparison between the cMZR with respect to the field MZR at $z\sim 0.3$ indicates that the most metal-poor galaxies at a given mass are downshifted $0.23$ dex, which is in agreement with \citet{montuori} and \citet{kewley2006}. 
This scenario is also supported by ongoing dynamical activity (P15) and by the fact that this galaxy cluster has an extended soft X-ray tail, together with the hardness ratio map, which suggests that this cluster has an ongoing merger activity with a smaller structure (\citealt{defilippis}).
%--------------------------------------------------------------------
%--------------------------------------------------------------------
\section{Summary and conclusions}
We explored the role of galaxy clusters in the MZR through SF galaxies of the AC114 galaxy cluster. We analyzed optical, intermediate-resolution VIMOS/VLT spectra for a sample of 182 galaxies in the FOV of AC114.

The search in photometric catalogues allowed us to get optical and NIR luminosities for $76\%$ and $66\%$ of the galaxies of the sample from the DECaLS DR10 and VIKING DR4 ESO surveys, respectively. Synthetic photometry was performed using the VIMOS spectra when optical photometry was missing. Optical and NIR-based stellar masses of the galaxies were estimated via theoretical stellar mass-to-light ratios computed from the E-MILES models by considering the age dependence for both sdss$-i$ and $K_{s}$ band passes. A comparison between optical and NIR-based mass estimations shows agreement between them for SF galaxies, but there is a slight deviation in massive, passive galaxies of 0.23 dex. However, the offset is within the propagated uncertainties of $\sigma = 0.29$ mags, so it does not change our results.

We classified the galaxies into ELGs and passive galaxies, based on the presence or absence of $\mathrm{[OII]\lambda3727}$, $\mathrm{[OIII]\lambda\lambda4959,5007}$, H$\beta$, and H$\alpha$ emission lines, respectively. Using the BPT diagram, together with the \citet{KLlimit} and \citet{kauffman03} demarcation limits, we found that the galaxy sample has seven composite galaxies, where two of them (a Seyfert composite and a LINER composite) are cluster members. The remaining galaxies in the sample are, in fact, pure SF galaxies.

Gas-phase metallicities were estimated from the O3N2, which are calibrated on $T_{e}-$based metallicities. 
\\
\\
Our main conclusions can be summarized as follows:
\begin{itemize}
    \item Foreground and background MZRs span masses from $10^{8.7}$ to $10^{10.3}M_{\odot}$ and $10^{9.4}$ to $10^{10.7}M_{\odot}$, respectively, with both the fMZR and the bMZR showing a scatter of $\sim$0.12 dex. The fMZR was constructed with less-massive galaxies than bMZR due to an observational bias. However, the downshifts reported show a coherent evolution in redshift of the MZR, where foreground galaxies are 0.09 dex metal-poorer, and background galaxies are 0.20 dex metal-poorer on average with respect to the local Universe galaxies.   
    
    \item The AC114-MZR presents different features with respect to field galaxies at similar redshifts. The cMZR span masses from $10^{9.2}$ to $10^{10.8}M_{\odot}$, with 0.17 dex of scatter. A downshift of 0.19 dex on average and a slightly shallower slope is seen with respect to both the fMZR and the bMZR. This shows that, in general, AC114 SF galaxies have lower metallicities at the same mass than field galaxies at the same redshift. The metallicity measurements of the SF cluster members are comparable to the background galaxies in our galaxy sample. Besides, star-forming galaxies within the main component of AC114 and the central region do not show a correlation between mass and gas-phase metallicity, with a large scatter. This gives important clues about the environmental effects that are suppressing the chemical evolution of these galaxies.

    \item The mass and metallicity profiles as a function of the clustercentric radius show that AC114 started its radial relaxation phase, in agreement with \citet{proust}. The negative mass gradient suggests that SF galaxies are the most sensitive to the dynamical state of the galaxy cluster. Additionally, the negative (although not statistically significant) metallicity gradient can be a sign that massive galaxies located near the central region of AC114 are slightly more metal-rich than those found in external regions.

    \item The substructures identified with the KDE show that for SF galaxies in the outskirts of the galaxy cluster the metallicity is dependent on the stellar mass. Moreover, these galaxies are responsible for the shallower slope of the cMZR (compared with the field MZR at $z\sim 0.3$), which means that these galaxies are also the more metal-poor ones.  

    \item When the cMZR is compared with the field MZR at $z\sim 0.3$, two galaxies are located above the field MZR by $\sim 0.10$ dex. This feature is consistent with the literature, which suggests that galaxies are suffering from RPS, strangulation, and/or galaxy-galaxy interactions.

    \item Galaxies falling below the linear fit of the cMZR are the drivers of the shallower slope. These galaxies are also found outside the central region of AC114. This means that outskirt galaxies are evolving slowly in terms of their chemical content, showing metallicities up to $0.22$ dex lower than the local Universe galaxies and on average $0.17$ dex lower than the field MZR at $z\sim 0.3$. Strong metal-poor inflows produced by dynamical events such as galaxy-galaxy interactions may explain the strikingly low metallicities of the SF members of AC114. This scenario is also supported by the dynamical state of the cluster and the ongoing merger event between the main component of AC114 and a smaller structure, observed in the X-rays. 

    \item We emphasize the role of the mass of a galaxy cluster. Environmental and dynamical effects drive the chemical evolution of members in a unique way with respect to field galaxies. However, only galaxy clusters with $M>10^{15}M_{\odot}$ appear to show observable changes in the MZR.

\end{itemize}

%--------------------------------------------------------------------
\begin{acknowledgements}
      We thank to the anonymous referee for their useful comments which improved the quality of the paper. A special thanks goes to the Dr. Piercarlo Bonifacio for the feedback presented in the early stage of this research. Additionally, we want to thanks Felipe Vivanco and Gustavo Worner, for the opportunity to discuss our results and for the valuable contributions to the development of the Python scripts used in this work.
      This publication is based on observations collected at the European Southern Observatory under ESO programme ID 083.A-0566. Also, this program has made use of data from the VIKING survey from VISTA at the ESO Paranal Observatory, programme ID 179.A-2004. Data processing has been contributed by the VISTA Data Flow System at CASU, Cambridge and WFAU, Edinburgh.
\end{acknowledgements}

\begin{appendix}
\section{Tables:}

\begin{table*}
\caption{Compilation of all statistical parameters of the fits performed in Figures 9, 12, 13, and 14. First column: Corresponding figure. Second column: Fitted structure (e.g., the cluster MZR, or the SF mass profile). Third column: Slope of the linear fit. Fourth column: 1$\sigma$ scatter of the distribution. Fifth column: Spearman correlation test. Sixth column: $p-$value of the  Spearman correlation test.}
\label{table:0}
\centering

\begin{tabular}{cccccc}
\hline\hline
Figure & Structure     & Slope              & $\sigma$ & $r$ (Spearman rank) & $p-$value              \\
\hline
9      & fMZR          & 0.236 $\pm$ 0.003  & 0.118    & 0.97                & 2.6 $\times \ 10^{-8}$ \\
9      & bMZR          & 0.312 $\pm$ 0.009  & 0.120    & 0.90                & 1.5 $\times \ 10^{-4}$ \\
9      & cMZR          & 0.241 $\pm$ 0.008  & 0.170    & 0.65                & 0.009                  \\
12     & Mass profile    & -0.005 $\pm$ 0.002 & 0.540     & -0.57               & 0.005                  \\
12     & SF mass profile     & -0.007 $\pm$ 0.003 & 0.470     & -0.50               & 0.020                  \\
12     & Pasive mass profile & -0.004 $\pm$ 0.003 & 0.520     & -0.24               & 0.090                  \\
12     & Metallicity profile   & -0.002 $\pm$ 0.002 & 0.190     & -0.38               & 0.110                  \\
13     & SS2 (cMZR)    & 0.192 $\pm$ 0.129  & 0.330     & 0.54                & 0.125                  \\
13     & SS3 (cMZR)    & 0.269 $\pm$ 0.135  & 0.280     & 0.86                & 0.019                  \\
14     & Middle region (cMZR) & 0.352 $\pm$ 0.078  & 0.270     & 0.86                & 0.014                  \\
14     & External region (cMZR)  & 0.162 $\pm$ 0.096  & 0.350     & 0.40                & 0.319                  \\ 
\hline                                   %inserts single line
\end{tabular}

\end{table*}

\begin{table*}
\caption{Compilation of photometric data collected from DECaLS DR10 and VIKING DR4. Column one indicates the IDs of the galaxies, labeled according to $\#$quadrant$-\#$slit. Galaxies that were not identified in previous works with the same spectroscopic VIMOS data are labeled with $\#$quadrant$-A$(Additional). RA and Dec coordinates are shown in columns two and three. Redshift estimates in column four. Optical apparent magnitudes are listed in columns five and six. NIR apparent magnitudes are listed in column seven. Column eight indicates the classification of passive galaxies and ELGs: passive galaxies are listed with 0 and ELGs with 1.}

%title of Tableed table:1}      % is used to refer this table in the text
\label{table:1}
\centering                          % used for centering table
\begin{tabular}{cccccccc}        % centered columns (4 columns)
\hline\hline                 % inserts double horizontal lines

ID & RA & Dec & z & $i$ & $g$ & $K_{s}$ & ELGs \\
\hline                        % inserts single horizontal line
Q1-3 & 344.889 & -34.711 & 0.057330$\pm$0.000014 & & & & 1 \\
Q1-A1 & 344.877 & -34.713 & 0.277520$\pm$0.000054 & & & & 1 \\
Q1-5 & 344.884 & -34.716 & 0.057850$\pm$0.000012 & & & & 1 \\
Q1-A2 & 344.815 & -34.721 & 0.289160$\pm$0.000022 & & & & 0 \\
Q1-8 & 344.902 & -34.724 & 0.379530$\pm$0.000015 & & & & 1 \\
Q1-9 & 344.858 & -34.726 & 0.416980$\pm$0.000011 & 22.2$\pm$0.6 & & 17.80$\pm$0.16 & 0 \\
Q1-10 & 344.827 & -34.729 & 0.219600$\pm$0.000017 & 18.7$\pm$0.5 & 20.3$\pm$0.8 & 15.78$\pm$0.03 & 1 \\
Q1-11 & 344.823 & -34.733 & 0.282340$\pm$0.000007 & 21.6$\pm$0.6 & 23.7$\pm$1.1 & 18.46$\pm$0.24 & 0 \\
Q1-12 & 344.833 & -34.736 & 0.220480$\pm$0.000016 & 19.6$\pm$0.4 & 20.4$\pm$0.8 & 17.13$\pm$0.15 & 1 \\
Q1-A3 & 344.800 & -34.740 & 0.271680$\pm$0.000007 & & & & 1 \\
Q1-15 & 344.867 & -34.741 & 0.071500$\pm$0.000013 & 21.5$\pm$0.6 & 21.9$\pm$1.0 & & 1 \\
Q1-16 & 344.870 & -34.746 & 0.408880$\pm$0.000021 & 21.7$\pm$0.5 & 23.7$\pm$1.0 & 18.16$\pm$0.24 & 0 \\
Q1-17 & 344.910 & -34.748 & 0.342210$\pm$0.000007 & & & & 0 \\
Q1-18 & 344.846 & -34.751 & 0.312460$\pm$0.000020 & 21.2$\pm$0.6 & 22.4$\pm$1.1 & 18.31$\pm$0.18 & 1 \\
Q1-19 & 344.827 & -34.754 & 0.408190$\pm$0.000009 & 21.4$\pm$0.5 & 22.5$\pm$1.0 & 18.97$\pm$0.34 & 1 \\
Q1-20 & 344.918 & -34.756 & 0.231980$\pm$0.000016 & & & & 0 \\
Q1-A4 & 344.897 & -34.757 & 0.357780$\pm$0.000010 & & & & 0 \\
Q1-22 & 344.853 & -34.760 & 0.409870$\pm$0.000019 & 19.0$\pm$0.4 & 21.5$\pm$0.8 & 15.92$\pm$0.04 & 1 \\
Q1-24 & 344.869 & -34.764 & 0.333100$\pm$0.000005 & 19.3$\pm$0.4 & 20.8$\pm$0.8 & 16.52$\pm$0.06 & 0 \\
Q1-25 & 344.896 & -34.766 & 0.414380$\pm$0.000192 & 20.5$\pm$0.5 & 22.1$\pm$1.0 & 17.73$\pm$0.17 & 1 \\
Q1-26 & 344.883 & -34.770 & 0.417650$\pm$0.000038 & 21.3$\pm$0.5 & 22.9$\pm$1.0 & 17.91$\pm$0.14 & 0 \\
Q1-27 & 344.884 & -34.773 & 0.390830$\pm$0.000017 & 22.1$\pm$0.5 & 23.5$\pm$0.9 & & 0 \\
Q1-28 & 344.859 & -34.776 & 0.343650$\pm$0.000011 & 19.1$\pm$0.3 & 19.8$\pm$0.6 & 17.15$\pm$0.12 & 0 \\
Q1-30 & 344.900 & -34.779 & 0.318050$\pm$0.000014 & 19.1$\pm$0.4 & 21.1$\pm$0.8 & 15.30$\pm$0.02 & 1 \\
Q1-31 & 344.890 & -34.780 & 0.331760$\pm$0.000021 & 20.6$\pm$0.6 & 22.6$\pm$1.0 & 17.34$\pm$0.12 & 0 \\
Q1-34 & 344.855 & -34.785 & 0.301330$\pm$0.000011 & 20.3$\pm$0.6 & 22.4$\pm$1.0 & 17.04$\pm$0.08 & 0 \\
Q1-36 & 344.892 & -34.789 & 0.498080$\pm$0.000010 & 20.7$\pm$0.5 & 22.2$\pm$0.8 & 18.75$\pm$0.20 & 0 \\
Q1-37 & 344.849 & -34.791 & 0.315480$\pm$0.000042 & 19.5$\pm$0.6 & 21.2$\pm$1.0 & 16.51$\pm$0.05 & 1 \\
Q1-A5 & 344.799 & -34.793 & 0.343630$\pm$0.000018 & 21.0$\pm$0.4 & 22.8$\pm$0.7 & 16.96$\pm$0.09 & 0 \\
Q1-38 & 344.825 & -34.794 & 0.485360$\pm$0.000018 & & & & 1 \\
Q1-40 & 344.847 & -34.799 & 0.323390$\pm$0.000019 & 19.1$\pm$0.3 & & 16.29$\pm$0.05 & 1 \\
Q1-41 & 344.820 & -34.802 & 0.576610$\pm$0.000008 & & & 17.09$\pm$0.08 & 0 \\
Q1-A6 & 344.871 & -34.808 & 0.289270$\pm$0.000014 & & & & 0 \\
Q1-44 & 344.816 & -34.812 & 0.576030$\pm$0.000009 & 21.6$\pm$0.4 & 23.1$\pm$0.7 & & 1 \\
\hline                                   %inserts single line
\end{tabular}
\end{table*}

\begin{table*}
         
% title of Table
\centering                          % used for centering table
\begin{tabular}{cccccccc}        % centered columns (4 columns)
\hline\hline                 % inserts double horizontal lines
ID & RA & Dec & z & $i$ & $g$ & $K_{s}$ & ELGs \\
\hline                        % inserts single horizontal line
Q2-1 & 344.817 & -34.561 & 0.259580$\pm$0.000022 & & & & 1 \\
Q2-2 & 344.825 & -34.564 & 0.334680$\pm$0.000010 & & & & 0 \\
Q2-3 & 344.824 & -34.566 & 0.409690$\pm$0.000012 & & & & 1 \\
Q2-4 & 344.806 & -34.570 & 0.408470$\pm$0.000019 & & & & 1 \\
Q2-5 & 344.812 & -34.573 & 0.256670$\pm$0.000559 & & & & 0 \\
Q2-6 & 344.849 & -34.576 & 0.423330$\pm$0.000025 & 18.9$\pm$0.4 & 21.0$\pm$0.7 & 15.80$\pm$0.05 & 0 \\
Q2-7 & 344.839 & -34.578 & 0.242100$\pm$0.000610 & 22.1$\pm$0.7 & 25.2$\pm$1.1 & 19.45$\pm$0.52 & 0 \\
Q2-8 & 344.803 & -34.580 & 0.445090$\pm$0.000009 & 21.6$\pm$0.6 & 22.7$\pm$1.0 & 18.88$\pm$0.29 & 0 \\
Q2-9 & 344.823 & -34.582 & 0.370950$\pm$0.000048 & 22.1$\pm$0.6 & 23.2$\pm$1.1 & & 1 \\
Q2-10 & 344.828 & -34.584 & 0.408990$\pm$0.000012 & 21.3$\pm$0.7 & 23.4$\pm$1.1 & 17.04$\pm$0.12 & 1 \\
Q2-11 & 344.820 & -34.586 & 0.298630$\pm$0.000016 & 22.1$\pm$0.4 & 22.9$\pm$0.8 & & 1 \\
Q2-12 & 344.886 & -34.590 & 0.370700$\pm$0.000016 & 21.1$\pm$0.5 & 22.9$\pm$0.9 & 17.69$\pm$0.19 & 0 \\
Q2-13 & 344.859 & -34.592 & 0.360210$\pm$0.000047 & & & & 0 \\
Q2-14 & 344.834 & -34.595 & 0.373620$\pm$0.000044 & 20.7$\pm$0.4 & 22.6$\pm$0.9 & 17.73$\pm$0.15 & 0 \\
Q2-15 & 344.815 & -34.598 & 0.410010$\pm$0.000133 & 21.7$\pm$0.6 & 23.9$\pm$1.1 & 18.37$\pm$0.20 & 0 \\
Q2-16 & 344.829 & -34.600 & 0.573910$\pm$0.000024 & 21.3$\pm$0.4 & 23.5$\pm$0.9 & 18.48$\pm$0.27 & 1 \\
Q2-A1 & 344.833 & -34.605 & 0.267590$\pm$0.000278 & & & & 0 \\
Q2-18 & 344.800 & -34.608 & 0.345920$\pm$0.000037 & 21.1$\pm$0.5 & 23.2$\pm$0.9 & 17.98$\pm$0.17 & 0 \\
Q2-19 & 344.830 & -34.611 & 0.377380$\pm$0.000030 & 19.3$\pm$0.4 & 21.5$\pm$0.8 & 16.37$\pm$0.05 & 0 \\
Q2-A2 & 344.855 & -34.612 & 0.466710$\pm$0.000057 & & & & 0 \\
Q2-21 & 344.840 & -34.615 & 0.413680$\pm$0.000105 & 20.5$\pm$0.5 & 22.5$\pm$0.8 & 17.45$\pm$0.11 & 0 \\
Q2-22 & 344.903 & -34.615 & 0.391390$\pm$0.000022 & & & & 1 \\
Q2-23 & 344.839 & -34.619 & 0.413930$\pm$0.000038 & 19.5$\pm$0.5 & 22.1$\pm$0.9 & 16.31$\pm$0.04 & 0 \\
Q2-24 & 344.903 & -34.623 & 0.380780$\pm$0.000055 & 22.1$\pm$0.6 & 23.2$\pm$1.0 & & 1 \\
Q2-A3 & 344.834 & -34.625 & 0.429260$\pm$0.000014 & & & & 0 \\
Q2-26 & 344.801 & -34.627 & 0.487890$\pm$0.000018 & 21.5$\pm$0.5 & 23.2$\pm$0.9 & 19.30$\pm$0.48 & 1 \\
Q2-27 & 344.837 & -34.629 & 0.497530$\pm$0.000033 & 20.3$\pm$0.5 & 22.1$\pm$0.9 & 17.11$\pm$0.11 & 1 \\
Q2-28 & 344.814 & -34.633 & 0.613460$\pm$0.000019 & 20.4$\pm$0.3 & 21.7$\pm$0.6 & 17.36$\pm$0.17 & 1 \\
Q2-29 & 344.823 & -34.634 & 0.523020$\pm$0.000010 & 20.6$\pm$0.2 & & 17.90$\pm$0.21 & 1 \\
Q2-31 & 344.852 & -34.638 & 0.497110$\pm$0.000022 & & & & 0 \\
Q2-A4 & 344.866 & -34.638 & 0.497370$\pm$0.000008 & & & & 1 \\
Q2-32 & 344.854 & -34.643 & 0.308930$\pm$0.000018 & & & 18.13$\pm$0.23 & 1 \\
Q2-33 & 344.819 & -34.645 & 0.579620$\pm$0.000006 & & & 17.52$\pm$0.11 & 1 \\
Q2-A5 & 344.926 & -34.648 & 0.345720$\pm$0.000017 & & & & 0 \\
Q2-35 & 344.919 & -34.651 & 0.313670$\pm$0.000056 & 19.7$\pm$0.6 & 21.9$\pm$1.1 & 16.51$\pm$0.04 & 0 \\
Q2-36 & 344.857 & -34.653 & 0.465160$\pm$0.000210 & 20.2$\pm$0.6 & 22.2$\pm$1.0 & 17.42$\pm$0.11 & 1 \\
Q2-37 & 344.808 & -34.659 & 0.139900$\pm$0.000017 & 20.4$\pm$0.4 & 21.2$\pm$0.7 & & 1 \\
Q2-38 & 344.846 & -34.663 & 0.298510$\pm$0.000033 & 21.5$\pm$0.6 & 22.7$\pm$1.0 & 19.46$\pm$0.46 & 1 \\
Q2-39 & 344.912 & -34.665 & 0.371870$\pm$0.000021 & 20.6$\pm$0.6 & 21.7$\pm$1.1 & 17.88$\pm$0.16 & 1 \\
Q2-40 & 344.901 & -34.666 & 0.420140$\pm$0.000154 & 20.9$\pm$0.6 & 23.0$\pm$1.0 & 18.49$\pm$0.31 & 0 \\
Q2-41 & 344.932 & -34.669 & 0.540080$\pm$0.000014 & 21.3$\pm$0.6 & 23.3$\pm$1.0 & 18.04$\pm$0.22 & 0 \\
\hline                                   %inserts single line
\end{tabular}
\end{table*}

\begin{table*}           
% title of Table
\centering                          % used for centering table
\begin{tabular}{cccccccc}        % centered columns (4 columns)
\hline\hline                 % inserts double horizontal lines
ID & RA & Dec & z & $i$ & $g$ & $K_{s}$ & ELGs \\
\hline                        % inserts single horizontal line
Q3-1 & 344.673 & -34.555 & 0.329830$\pm$0.000060 & 21.4$\pm$0.6 & 22.5$\pm$1.1 & 19.04$\pm$0.40 & 1 \\
Q3-2 & 344.672 & -34.557 & 0.330890$\pm$0.000003 & 18.8$\pm$0.5 & 20.1$\pm$0.8 & 15.78$\pm$0.04 & 1 \\
Q3-3 & 344.667 & -34.559 & 0.331850$\pm$0.000079 & 19.1$\pm$0.6 & 21.3$\pm$1.0 & 15.96$\pm$0.03 & 0 \\
Q3-4 & 344.667 & -34.561 & 0.212750$\pm$0.000021 & 20.9$\pm$0.7 & 22.1$\pm$1.1 & 18.49$\pm$0.26 & 1 \\
Q3-5 & 344.671 & -34.563 & 0.331820$\pm$0.000056 & 21.0$\pm$0.5 & 22.2$\pm$1.0 & 18.49$\pm$0.36 & 1 \\
Q3-6 & 344.641 & -34.564 & 0.495360$\pm$0.000019 & & & & 0 \\
Q3-A1 & 344.695 & -34.566 & 0.315910$\pm$0.000481 & & & & 0 \\
Q3-8 & 344.678 & -34.573 & 0.400050$\pm$0.000009 & & & & 1 \\
Q3-10 & 344.594 & -34.576 & 0.308500$\pm$0.000136 & 18.5$\pm$0.3 & 19.4$\pm$0.5 & 16.83$\pm$0.13 & 0 \\
Q3-11 & 344.650 & -34.579 & 0.212680$\pm$0.000023 & 20.8$\pm$0.7 & 21.5$\pm$1.1 & 18.78$\pm$0.34 & 1 \\
Q3-12 & 344.654 & -34.582 & 0.310390$\pm$0.000230 & 20.0$\pm$0.5 & 21.2$\pm$0.8 & 17.49$\pm$0.21 & 1 \\
Q3-A2 & 344.685 & -34.587 & 0.497010$\pm$0.000009 & & & & 0 \\
Q3-15 & 344.671 & -34.589 & 0.415820$\pm$0.000014 & & & & 0 \\
Q3-16 & 344.691 & -34.591 & 0.176550$\pm$0.000028 & 21.9$\pm$0.5 & 22.7$\pm$1.1 & & 1 \\
Q3-17 & 344.702 & -34.594 & 0.497500$\pm$0.000019 & 22.6$\pm$0.6 & 24.8$\pm$1.1 & 19.42$\pm$0.44 & 1 \\
Q3-18 & 344.735 & -34.596 & 0.221070$\pm$0.000019 & 19.5$\pm$0.4 & 20.5$\pm$0.8 & 17.24$\pm$0.14 & 1 \\
Q3-19 & 344.597 & -34.599 & 0.159050$\pm$0.000011 & 18.6$\pm$0.4 & 19.6$\pm$0.6 & 15.88$\pm$0.06 & 0 \\
Q3-20 & 344.635 & -34.602 & 0.354770$\pm$0.000036 & 23.1$\pm$0.5 & 24.3$\pm$1.0 & & 1 \\
Q3-21 & 344.635 & -34.604 & 0.353350$\pm$0.000020 & 19.3$\pm$0.4 & 20.8$\pm$0.7 & 16.48$\pm$0.08 & 1 \\
Q3-23 & 344.596 & -34.607 & 0.353750$\pm$0.000046 & 19.9$\pm$0.6 & 21.6$\pm$1.0 & 16.89$\pm$0.07 & 0 \\
Q3-24 & 344.621 & -34.609 & 0.271760$\pm$0.000016 & 22.4$\pm$0.7 & 24.2$\pm$1.1 & 18.57$\pm$0.27 & 0 \\
Q3-25 & 344.670 & -34.611 & 0.595430$\pm$0.000085 & 20.5$\pm$0.5 & 23.2$\pm$0.8 & 17.07$\pm$0.07 & 0 \\
Q3-26 & 344.618 & -34.613 & 0.315900$\pm$0.000024 & 20.2$\pm$0.7 & 22.1$\pm$1.1 & 17.36$\pm$0.09 & 0 \\
Q3-27 & 344.623 & -34.615 & 0.593510$\pm$0.000068 & 22.0$\pm$0.7 & 23.4$\pm$1.1 & 19.74$\pm$0.48 & 1 \\
Q3-28 & 344.601 & -34.617 & 0.258540$\pm$0.000026 & 18.1$\pm$0.5 & 20.1$\pm$0.8 & 15.05$\pm$0.02 & 0 \\
Q3-29 & 344.635 & -34.618 & 0.462490$\pm$0.000004 & 20.2$\pm$0.6 & 21.0$\pm$0.9 & 18.27$\pm$0.22 & 1 \\
Q3-30 & 344.616 & -34.621 & 0.402360$\pm$0.000025 & & & & 1 \\
Q3-31 & 344.685 & -34.623 & 0.332340$\pm$0.000228 & 20.8$\pm$0.5 & 22.0$\pm$0.9 & 17.89$\pm$0.20 & 1 \\
Q3-A3 & 344.643 & -34.626 & 0.319970$\pm$0.000009 & 23.8$\pm$0.7 & 24.7$\pm$1.1 & 19.79$\pm$0.67 & 0 \\
Q3-33 & 344.663 & -34.631 & 0.112810$\pm$0.000010 & 22.3$\pm$0.4 & 22.9$\pm$0.6 & & 1 \\
Q3-34 & 344.665 & -34.634 & 0.351610$\pm$0.000046 & & & 18.38$\pm$0.18 & 1 \\
Q3-35 & 344.640 & -34.638 & 0.060200$\pm$0.000009 & & & & 1 \\
Q3-36 & 344.631 & -34.640 & 0.202140$\pm$0.000016 & & & 17.70$\pm$0.20 & 1 \\
Q3-A4 & 344.668 & -34.642 & 0.345010$\pm$0.000078 & & & & 0 \\
Q3-38 & 344.672 & -34.644 & 0.432870$\pm$0.000022 & & & 18.63$\pm$0.28 & 1 \\
Q3-39 & 344.694 & -34.646 & 0.399430$\pm$0.000059 & & & 18.19$\pm$0.22 & 1 \\
Q3-40 & 344.642 & -34.648 & 0.318120$\pm$0.000081 & 19.5$\pm$0.5 & 21.7$\pm$0.8 & 16.26$\pm$0.04 & 0 \\
Q3-A5 & 344.610 & -34.649 & 0.479200$\pm$0.000065 & & & & 0 \\
Q3-A6 & 344.588 & -34.651 & 0.325310$\pm$0.000167 & & & & 1 \\
Q3-43 & 344.664 & -34.653 & 0.509810$\pm$0.000015 & 21.1$\pm$0.5 & 22.3$\pm$0.9 & & 1 \\
Q3-44 & 344.664 & -34.656 & 0.224270$\pm$0.000168 & 19.6$\pm$0.5 & 21.2$\pm$0.8 & 16.40$\pm$0.07 & 1 \\
Q3-45 & 344.681 & -34.659 & 0.327200$\pm$0.000073 & 21.3$\pm$0.5 & 22.9$\pm$0.9 & 18.64$\pm$0.34 & 0 \\
Q3-46 & 344.678 & -34.661 & 0.309010$\pm$0.000041 & 21.0$\pm$0.5 & 22.6$\pm$0.9 & 18.70$\pm$0.32 & 0 \\
Q3-A7 & 344.585 & -34.662 & 0.345340$\pm$0.000153 & 19.9$\pm$0.5 & 21.3$\pm$0.8 & 17.01$\pm$0.11 & 0 \\
Q3-A8 & 344.620 & -34.665 & 0.412960$\pm$0.000773 & & & & 0 \\
Q3-49 & 344.655 & -34.668 & 0.309000$\pm$0.000033 & 18.0$\pm$0.3 & 19.2$\pm$0.5 & 15.62$\pm$0.05 & 0 \\
\hline                                   %inserts single line
\end{tabular}
\end{table*}

\begin{table*}          
% title of Table
\centering                          % used for centering table
\begin{tabular}{cccccccc}        % centered columns (4 columns)
\hline\hline                 % inserts double horizontal lines
ID & RA & Dec & z & $i$ & $g$ & $K_{s}$ & ELGs \\
\hline                        % inserts single horizontal line
Q4-A1 & 344.669 & -34.704 & 0.133610$\pm$0.000008 & & & & 1 \\
Q4-2 & 344.646 & -34.706 & 0.318380$\pm$0.000216 & 20.0$\pm$0.6 & 22.0$\pm$1.0 & 16.91$\pm$0.08 & 0 \\
Q4-3 & 344.688 & -34.708 & 0.412180$\pm$0.000024 & 20.7$\pm$0.6 & 22.2$\pm$1.1 & 18.05$\pm$0.18 & 1 \\
Q4-4 & 344.648 & -34.709 & 0.310170$\pm$0.000273 & 21.3$\pm$0.7 & 23.0$\pm$1.0 & 18.57$\pm$0.25 & 0 \\
Q4-5 & 344.677 & -34.711 & 0.411570$\pm$0.000089 & 21.2$\pm$0.6 & 22.6$\pm$1.0 & 18.47$\pm$0.32 & 1 \\
Q4-A2 & 344.660 & -34.712 & 0.268420$\pm$0.000033 & & & & 0 \\
Q4-7 & 344.635 & -34.715 & 0.169450$\pm$0.000183 & 19.5$\pm$0.5 & 20.1$\pm$0.8 & 17.62$\pm$0.20 & 1 \\
Q4-8 & 344.636 & -34.717 & 0.169800$\pm$0.000028 & 21.4$\pm$0.7 & 22.2$\pm$1.1 & 19.17$\pm$0.32 & 1 \\
Q4-9 & 344.632 & -34.721 & 0.256730$\pm$0.000144 & 18.9$\pm$0.6 & 20.8$\pm$0.9 & 15.69$\pm$0.02 & 0 \\
Q4-10 & 344.605 & -34.723 & 0.311730$\pm$0.000114 & & & & 0 \\
Q4-11 & 344.635 & -34.725 & 0.291170$\pm$0.000046 & 22.4$\pm$0.7 & 24.0$\pm$1.1 & & 0 \\
Q4-13 & 344.611 & -34.728 & 0.316650$\pm$0.000101 & 21.3$\pm$0.6 & 23.2$\pm$1.0 & 18.69$\pm$0.30 & 0 \\
Q4-14 & 344.673 & -34.729 & 0.304220$\pm$0.000072 & 21.4$\pm$0.7 & 22.1$\pm$1.1 & 19.63$\pm$0.65 & 1 \\
Q4-15 & 344.608 & -34.731 & 0.306580$\pm$0.000008 & 21.9$\pm$0.7 & 22.7$\pm$1.1 & & 1 \\
Q4-16 & 344.623 & -34.732 & 0.318090$\pm$0.000212 & 20.1$\pm$0.7 & 22.1$\pm$1.2 & 16.79$\pm$0.06 & 0 \\
Q4-A3 & 344.596 & -34.733 & 0.362090$\pm$0.000022 & & & & 1 \\
Q4-18 & 344.608 & -34.735 & 0.290200$\pm$0.000282 & 21.4$\pm$0.8 & 24.2$\pm$1.2 & 18.08$\pm$0.16 & 0 \\
Q4-19 & 344.609 & -34.736 & 0.309670$\pm$0.000160 & 21.0$\pm$0.6 & 21.8$\pm$1.0 & 17.97$\pm$0.18 & 1 \\
Q4-20 & 344.664 & -34.738 & 0.306440$\pm$0.000191 & 19.9$\pm$0.6 & 22.0$\pm$1.1 & 16.67$\pm$0.05 & 0 \\
Q4-21 & 344.711 & -34.737 & 0.314790$\pm$0.000241 & & & & 0 \\
Q4-22 & 344.666 & -34.741 & 0.317210$\pm$0.000205 & 21.4$\pm$0.7 & 22.3$\pm$1.2 & 19.57$\pm$0.58 & 1 \\
Q4-23 & 344.646 & -34.742 & 0.320540$\pm$0.000446 & 21.0$\pm$0.7 & 23.0$\pm$1.1 & 18.05$\pm$0.24 & 0 \\
Q4-24 & 344.638 & -34.744 & 0.314080$\pm$0.000204 & 20.7$\pm$0.7 & 22.6$\pm$1.2 & 18.02$\pm$0.16 & 0 \\
Q4-25 & 344.626 & -34.745 & 0.297410$\pm$0.001417 & 22.5$\pm$0.8 & 24.3$\pm$1.2 & & 0 \\
Q4-26 & 344.675 & -34.747 & 0.324030$\pm$0.000127 & 19.7$\pm$0.5 & 20.4$\pm$0.9 & 17.66$\pm$0.17 & 1 \\
Q4-27 & 344.624 & -34.748 & 0.323200$\pm$0.000040 & 21.9$\pm$0.6 & 24.7$\pm$1.0 & 19.01$\pm$0.34 & 0 \\
Q4-28 & 344.612 & -34.750 & 0.314550$\pm$0.000265 & 20.8$\pm$0.7 & 22.7$\pm$1.2 & 17.83$\pm$0.17 & 0 \\
Q4-A4 & 344.600 & -34.751 & 0.358350$\pm$0.000065 & 21.6$\pm$0.5 & 22.8$\pm$0.8 & 19.53$\pm$0.45 & 1 \\
Q4-A5 & 344.625 & -34.752 & 0.427510$\pm$0.000011 & & & & 0 \\
Q4-31 & 344.652 & -34.754 & 0.317740$\pm$0.000258 & 19.5$\pm$0.6 & 21.7$\pm$1.0 & 16.39$\pm$0.04 & 0 \\
Q4-32 & 344.658 & -34.756 & 0.316410$\pm$0.000275 & 16.7$\pm$0.1 & 18.9$\pm$0.3 & 14.08$\pm$0.02 & 0 \\
Q4-33 & 344.672 & -34.758 & 0.101960$\pm$0.000016 & 19.8$\pm$0.4 & 20.6$\pm$0.7 & 18.33$\pm$0.19 & 1 \\
Q4-A6 & 344.599 & -34.759 & 0.295250$\pm$0.000024 & & & & 1 \\
Q4-35 & 344.700 & -34.762 & 0.170350$\pm$0.000007 & 20.7$\pm$0.5 & 21.4$\pm$1.1 & 17.46$\pm$0.15 & 1 \\
Q4-36 & 344.650 & -34.763 & 0.312860$\pm$0.000277 & 17.6$\pm$0.3 & 19.4$\pm$0.5 & 14.66$\pm$0.01 & 1 \\
Q4-37 & 344.624 & -34.765 & 0.273050$\pm$0.000022 & 22.8$\pm$0.5 & 24.5$\pm$0.9 & & 1 \\
Q4-38 & 344.618 & -34.767 & 0.310670$\pm$0.000239 & 19.2$\pm$0.5 & 21.2$\pm$0.9 & 16.28$\pm$0.04 & 0 \\
Q4-39 & 344.661 & -34.769 & 0.312550$\pm$0.001159 & & & & 0 \\
Q4-40 & 344.595 & -34.771 & 0.316250$\pm$0.000482 & & & & 1 \\
Q4-41 & 344.616 & -34.773 & 0.317970$\pm$0.000324 & 20.5$\pm$0.7 & 22.5$\pm$1.1 & 17.59$\pm$0.13 & 0 \\
Q4-42 & 344.624 & -34.775 & 0.497280$\pm$0.000614 & 21.1$\pm$0.6 & 22.7$\pm$1.0 & 18.42$\pm$0.34 & 0 \\
Q4-43 & 344.620 & -34.777 & 0.317200$\pm$0.000339 & 20.9$\pm$0.6 & 22.6$\pm$1.0 & 18.57$\pm$0.32 & 0 \\
Q4-44 & 344.671 & -34.779 & 0.315630$\pm$0.000309 & 18.8$\pm$0.5 & 21.0$\pm$0.9 & 15.38$\pm$0.02 & 0 \\
Q4-45 & 344.652 & -34.780 & 0.314980$\pm$0.000326 & 20.6$\pm$0.6 & 22.5$\pm$1.0 & 17.85$\pm$0.21 & 0 \\
Q4-46 & 344.712 & -34.783 & 0.328510$\pm$0.000097 & 19.5$\pm$0.3 & 20.5$\pm$0.6 & 17.61$\pm$0.13 & 0 \\
Q4-47 & 344.645 & -34.785 & 0.316340$\pm$0.000236 & 20.0$\pm$0.7 & 21.9$\pm$1.1 & 17.08$\pm$0.09 & 0 \\
Q4-48 & 344.621 & -34.786 & 0.314460$\pm$0.000288 & 19.6$\pm$0.5 & 21.5$\pm$0.8 & 16.45$\pm$0.07 & 0 \\
Q4-49 & 344.665 & -34.788 & 0.317640$\pm$0.000297 & 18.9$\pm$0.5 & 21.1$\pm$0.8 & 15.76$\pm$0.04 & 0 \\
Q4-50 & 344.700 & -34.789 & 0.330460$\pm$0.000464 & 22.0$\pm$0.7 & 23.4$\pm$1.2 & & 0 \\
Q4-51 & 344.693 & -34.792 & 0.309450$\pm$0.000221 & 18.8$\pm$0.4 & 20.9$\pm$0.8 & 15.92$\pm$0.03 & 0 \\
Q4-52 & 344.609 & -34.793 & 0.313230$\pm$0.000288 & 20.5$\pm$0.7 & 22.5$\pm$1.1 & 17.54$\pm$0.12 & 0 \\
Q4-54 & 344.641 & -34.796 & 0.350100$\pm$0.000059 & 20.4$\pm$0.5 & 21.9$\pm$0.6 & 17.44$\pm$0.17 & 0 \\
Q4-55 & 344.642 & -34.798 & 0.318480$\pm$0.000254 & 20.9$\pm$0.4 & 22.4$\pm$0.5 & 17.74$\pm$0.19 & 0 \\
Q4-56 & 344.684 & -34.799 & 0.324800$\pm$0.000005 & & & & 1 \\
Q4-57 & 344.707 & -34.800 & 0.313880$\pm$0.000143 & & & 18.17$\pm$0.18 & 0 \\
Q4-58 & 344.674 & -34.801 & 0.311070$\pm$0.000131 & & & 15.90$\pm$0.03 & 0 \\
Q4-59 & 344.690 & -34.804 & 0.316080$\pm$0.000118 & & & & 0 \\
Q4-60 & 344.685 & -34.805 & 0.312500$\pm$0.000122 & & & 17.87$\pm$0.13 & 0 \\
Q4-61 & 344.659 & -34.807 & 0.717580$\pm$0.000108 & & & 17.34$\pm$0.13 & 0 \\
Q4-62 & 344.678 & -34.809 & 0.309600$\pm$0.000172 & & & 17.94$\pm$0.21 & 0 \\
Q4-63 & 344.719 & -34.810 & 0.318810$\pm$0.000127 & & & 17.33$\pm$0.13 & 0 \\
Q4-64 & 344.631 & -34.814 & 0.314870$\pm$0.000135 & 20.6$\pm$0.7 & 22.6$\pm$0.9 & 17.10$\pm$0.11 & 0 \\
\hline                                   %inserts single line
\end{tabular}
\end{table*}

%- --------- table 2 
\begin{table*}
\label{table:2}
\caption{Optical and NIR-bases stellar mass estimations. Columns 2,3 and 4 show absolute magnitudes in the selected optical and NIR band passes, respectively. Columns 5 and 6 show optical and NIR-based stellar mass estimations in logarithmic scale. The last column shows ELGs and passive galaxies with the same definition than Table \ref{table:1}.}             
% title of Table
\centering                          % used for centering table
\begin{tabular}{ccccccc}
\hline\hline                 % inserts double horizontal lines
ID & $M_{i}$ & $M_{g}$ & $M_{Ks}$ & log(M$^{*}_{i}$/M$_{\odot}$) & log(M$^{*}_{Ks}$/M$_{\odot}$) & ELGs \\
\hline
Q1-1 & -22.3$\pm$0.5 & -21.2$\pm$0.5 & & 11.1$\pm$0.2 & & 0 \\
Q1-3 & -18.8$\pm$0.5 & -18.8$\pm$0.5 & & 9.2$\pm$0.2 & & 1 \\
Q1-A1 & -19.4$\pm$0.5 & -18.7$\pm$0.5 & & 9.5$\pm$0.2 & & 1 \\
Q1-5 & -17.5$\pm$0.5 & -17.6$\pm$0.5 & & 8.7$\pm$0.2 & & 1 \\
Q1-A2 & -19.4$\pm$0.5 & -19.0$\pm$0.5 & & 10.0$\pm$0.2 & & 0 \\
Q1-8 & -22.6$\pm$0.5 & -21.3$\pm$0.5 & & 10.7$\pm$0.2 & & 1 \\
Q1-9 & -19.6$\pm$0.6 & -18.4$\pm$0.5 & -24.01$\pm$0.16 & 10.0$\pm$0.2 & 10.84$\pm$0.06 & 0 \\
Q1-10 & -21.5$\pm$0.5 & -20.0$\pm$0.8 & -24.44$\pm$0.03 & 10.3$\pm$0.2 & 10.51$\pm$0.01 & 1 \\
Q1-11 & -19.2$\pm$0.6 & -17.2$\pm$1.1 & -22.38$\pm$0.24 & 9.9$\pm$0.3 & 10.18$\pm$0.10 & 0 \\
Q1-12 & -20.6$\pm$0.4 & -19.8$\pm$0.8 & -23.10$\pm$0.15 & 10.0$\pm$0.2 & 9.97$\pm$0.06 & 1 \\
Q1-A3 & -20.0$\pm$0.5 & -19.5$\pm$0.5 & & 9.7$\pm$0.2 & & 1 \\
Q1-15 & -16.0$\pm$0.6 & -15.6$\pm$1.0 & & 8.1$\pm$0.2 & & 1 \\
Q1-16 & -20.0$\pm$0.5 & -18.1$\pm$1.0 & -23.61$\pm$0.24 & 10.2$\pm$0.2 & 10.67$\pm$0.10 & 0 \\
Q1-17 & -21.0$\pm$0.5 & -19.3$\pm$0.5 & & 10.6$\pm$0.2 & & 0 \\
Q1-18 & -19.9$\pm$0.6 & -18.7$\pm$1.1 & -22.77$\pm$0.18 & 9.7$\pm$0.2 & 9.84$\pm$0.07 & 1 \\
Q1-19 & -20.3$\pm$0.5 & -19.2$\pm$1.0 & -22.79$\pm$0.34 & 9.8$\pm$0.2 & 9.85$\pm$0.13 & 1 \\
Q1-20 & -22.7$\pm$0.5 & -21.6$\pm$0.5 & & 11.3$\pm$0.2 & & 0 \\
Q1-A4 & -19.8$\pm$0.5 & -19.2$\pm$0.5 & & 10.1$\pm$0.2 & & 0 \\
Q1-22 & -22.8$\pm$0.4 & -20.3$\pm$0.8 & -25.85$\pm$0.04 & 10.8$\pm$0.2 & 11.07$\pm$0.01 & 1 \\
Q1-24 & -21.9$\pm$0.4 & -20.4$\pm$0.8 & -24.72$\pm$0.06 & 11.0$\pm$0.1 & 11.12$\pm$0.03 & 0 \\
Q1-25 & -21.3$\pm$0.5 & -19.7$\pm$1.0 & -24.07$\pm$0.17 & 10.2$\pm$0.2 & 10.36$\pm$0.07 & 1 \\
Q1-26 & -20.6$\pm$0.5 & -18.9$\pm$1.0 & -23.91$\pm$0.14 & 10.4$\pm$0.2 & 10.79$\pm$0.05 & 0 \\
Q1-27 & -19.6$\pm$0.5 & -18.1$\pm$0.9 & & 10.0$\pm$0.2 & & 0 \\
Q1-28 & -22.2$\pm$0.3 & -21.5$\pm$0.6 & -24.17$\pm$0.12 & 11.1$\pm$0.1 & 10.90$\pm$0.05 & 0 \\
Q1-30 & -22.1$\pm$0.4 & -20.0$\pm$0.8 & -25.83$\pm$0.02 & 10.5$\pm$0.2 & 11.06$\pm$0.01 & 1 \\
Q1-31 & -20.6$\pm$0.6 & -18.6$\pm$1.0 & -23.90$\pm$0.12 & 10.4$\pm$0.2 & 10.79$\pm$0.05 & 0 \\
Q1-34 & -20.7$\pm$0.6 & -18.6$\pm$1.0 & -23.96$\pm$0.08 & 10.5$\pm$0.2 & 10.81$\pm$0.03 & 0 \\
Q1-36 & -21.5$\pm$0.5 & -20.1$\pm$0.8 & -23.52$\pm$0.20 & 10.8$\pm$0.2 & 10.64$\pm$0.08 & 0 \\
Q1-37 & -21.6$\pm$0.6 & -19.9$\pm$1.0 & -24.60$\pm$0.05 & 10.3$\pm$0.2 & 10.57$\pm$0.02 & 1 \\
Q1-A5 & -20.3$\pm$0.4 & -18.5$\pm$0.7 & -24.36$\pm$0.09 & 10.3$\pm$0.2 & 10.97$\pm$0.04 & 0 \\
Q1-38 & -21.3$\pm$0.5 & -20.8$\pm$0.5 & & 10.2$\pm$0.2 & & 1 \\
Q1-40 & -22.8$\pm$0.5 & -19.8$\pm$0.5 & -24.89$\pm$0.05 & 10.8$\pm$0.2 & 10.68$\pm$0.02 & 1 \\
Q1-41 & -22.6$\pm$0.5 & -21.3$\pm$0.5 & -25.57$\pm$0.08 & 11.2$\pm$0.2 & 11.46$\pm$0.03 & 0 \\
Q1-A6 & -18.8$\pm$0.5 & -18.8$\pm$0.5 & & 9.7$\pm$0.2 & & 0 \\
Q1-44 & -21.1$\pm$0.4 & -19.6$\pm$0.7 & & 10.1$\pm$0.1 & & 1 \\
\hline                                   %inserts single line
\end{tabular}
\end{table*}

\begin{table*}           
% title of Table
\centering                          % used for centering table
\begin{tabular}{ccccccc}
\hline\hline                 % inserts double horizontal lines
ID & $M_{i}$ & $M_{g}$ & $M_{Ks}$ & log(M$^{*}_{i}$/M$_{\odot}$) & log(M$^{*}_{Ks}$/M$_{\odot}$) & ELGs \\
\hline
Q2-1 & -20.7$\pm$0.5 & -20.3$\pm$0.5 & & 10.0$\pm$0.2 & & 1 \\
Q2-2 & -19.4$\pm$0.5 & -19.1$\pm$0.5 & & 10.0$\pm$0.2 & & 0 \\
Q2-3 & -20.2$\pm$0.5 & -19.8$\pm$0.5 & & 9.8$\pm$0.2 & & 1 \\
Q2-4 & -19.7$\pm$0.5 & -19.5$\pm$0.5 & & 9.6$\pm$0.2 & & 1 \\
Q2-5 & -19.2$\pm$0.5 & -18.5$\pm$0.5 & & 9.9$\pm$0.2 & & 0 \\
Q2-6 & -22.9$\pm$0.4 & -20.9$\pm$0.7 & -26.05$\pm$0.05 & 11.4$\pm$0.2 & 11.65$\pm$0.02 & 0 \\
Q2-7 & -18.3$\pm$0.7 & -15.3$\pm$1.1 & -21.01$\pm$0.52 & 9.5$\pm$0.3 & 9.63$\pm$0.21 & 0 \\
Q2-8 & -20.4$\pm$0.6 & -19.3$\pm$1.0 & -23.11$\pm$0.29 & 10.4$\pm$0.2 & 10.47$\pm$0.12 & 0 \\
Q2-9 & -19.4$\pm$0.6 & -18.3$\pm$1.1 & & 9.4$\pm$0.2 & & 1 \\
Q2-10 & -20.4$\pm$0.7 & -18.4$\pm$1.1 & -24.72$\pm$0.12 & 9.9$\pm$0.3 & 10.62$\pm$0.05 & 1 \\
Q2-11 & -18.8$\pm$0.4 & -18.1$\pm$0.8 & & 9.2$\pm$0.2 & & 1 \\
Q2-12 & -20.4$\pm$0.5 & -18.6$\pm$0.9 & -23.83$\pm$0.19 & 10.4$\pm$0.2 & 10.76$\pm$0.07 & 0 \\
Q2-13 & -19.2$\pm$0.5 & -19.7$\pm$0.5 & & 9.9$\pm$0.2 & & 0 \\
Q2-14 & -20.8$\pm$0.4 & -18.9$\pm$0.9 & -23.81$\pm$0.15 & 10.5$\pm$0.2 & 10.75$\pm$0.06 & 0 \\
Q2-15 & -20.0$\pm$0.6 & -17.8$\pm$1.1 & -23.40$\pm$0.20 & 10.2$\pm$0.2 & 10.59$\pm$0.08 & 0 \\
Q2-16 & -21.3$\pm$0.4 & -19.1$\pm$0.9 & -24.17$\pm$0.27 & 10.2$\pm$0.2 & 10.40$\pm$0.11 & 1 \\
Q2-A1 & -22.3$\pm$0.5 & -20.0$\pm$0.5 & & 11.1$\pm$0.2 & & 0 \\
Q2-18 & -20.3$\pm$0.5 & -18.2$\pm$0.9 & -23.36$\pm$0.17 & 10.3$\pm$0.2 & 10.58$\pm$0.07 & 0 \\
Q2-19 & -22.2$\pm$0.4 & -20.0$\pm$0.8 & -25.19$\pm$0.05 & 11.1$\pm$0.2 & 11.31$\pm$0.02 & 0 \\
Q2-A2 & -20.5$\pm$0.5 & -20.0$\pm$0.5 & & 10.4$\pm$0.2 & & 0 \\
Q2-21 & -21.3$\pm$0.5 & -19.3$\pm$0.8 & -24.34$\pm$0.11 & 10.7$\pm$0.2 & 10.97$\pm$0.04 & 0 \\
Q2-22 & -20.1$\pm$0.5 & -20.0$\pm$0.5 & & 9.7$\pm$0.2 & & 1 \\
Q2-23 & -22.3$\pm$0.5 & -19.7$\pm$0.9 & -25.49$\pm$0.04 & 11.1$\pm$0.2 & 11.42$\pm$0.01 & 0 \\
Q2-24 & -19.4$\pm$0.6 & -18.4$\pm$1.0 & & 9.5$\pm$0.2 & & 1 \\
Q2-A3 & -20.3$\pm$0.5 & -20.2$\pm$0.5 & & 10.3$\pm$0.2 & & 0 \\
Q2-26 & -20.7$\pm$0.5 & -19.0$\pm$0.9 & -22.92$\pm$0.48 & 10.0$\pm$0.2 & 9.90$\pm$0.19 & 1 \\
Q2-27 & -22.0$\pm$0.5 & -20.1$\pm$0.9 & -25.17$\pm$0.11 & 10.5$\pm$0.2 & 10.80$\pm$0.04 & 1 \\
Q2-28 & -22.5$\pm$0.3 & -21.2$\pm$0.6 & -25.47$\pm$0.17 & 10.7$\pm$0.1 & 10.92$\pm$0.07 & 1 \\
Q2-29 & -21.8$\pm$0.2 & -20.9$\pm$0.5 & -24.50$\pm$0.21 & 10.4$\pm$0.1 & 10.53$\pm$0.08 & 1 \\
Q2-31 & -20.4$\pm$0.5 & -20.3$\pm$0.5 & & 10.4$\pm$0.2 & & 0 \\
Q2-A4 & -22.6$\pm$0.5 & -21.7$\pm$0.5 & & 10.7$\pm$0.2 & & 1 \\
Q2-32 & -20.7$\pm$0.5 & -20.5$\pm$0.5 & -22.93$\pm$0.23 & 10.0$\pm$0.2 & 9.90$\pm$0.09 & 1 \\
Q2-33 & -22.6$\pm$0.5 & -22.1$\pm$0.5 & -25.15$\pm$0.11 & 10.7$\pm$0.2 & 10.79$\pm$0.05 & 1 \\
Q2-A5 & -19.9$\pm$0.5 & -19.4$\pm$0.5 & & 10.2$\pm$0.2 & & 0 \\
Q2-35 & -21.4$\pm$0.6 & -19.2$\pm$1.1 & -24.59$\pm$0.04 & 10.8$\pm$0.2 & 11.06$\pm$0.02 & 0 \\
Q2-36 & -21.8$\pm$0.6 & -19.9$\pm$1.0 & -24.67$\pm$0.11 & 10.4$\pm$0.2 & 10.60$\pm$0.05 & 1 \\
Q2-37 & -18.7$\pm$0.4 & -17.9$\pm$0.7 & & 9.2$\pm$0.2 & & 1 \\
Q2-38 & -19.4$\pm$0.6 & -18.3$\pm$1.0 & -21.51$\pm$0.46 & 9.5$\pm$0.2 & 9.34$\pm$0.18 & 1 \\
Q2-39 & -20.9$\pm$0.6 & -19.8$\pm$1.1 & -23.64$\pm$0.16 & 10.0$\pm$0.3 & 10.19$\pm$0.06 & 1 \\
Q2-40 & -21.0$\pm$0.6 & -18.9$\pm$1.0 & -23.34$\pm$0.31 & 10.6$\pm$0.2 & 10.57$\pm$0.13 & 0 \\
Q2-41 & -21.2$\pm$0.6 & -19.1$\pm$1.0 & -24.45$\pm$0.22 & 10.7$\pm$0.2 & 11.01$\pm$0.09 & 0 \\
\hline                                   %inserts single line
\end{tabular}
\end{table*}

\begin{table*}           
% title of Table
\centering                          % used for centering table
\begin{tabular}{ccccccc}
\hline\hline                 % inserts double horizontal lines
ID & $M_{i}$ & $M_{g}$ & $M_{Ks}$ & log(M$^{*}_{i}$/M$_{\odot}$) & log(M$^{*}_{Ks}$/M$_{\odot}$) & ELGs \\
\hline
Q3-1 & -19.8$\pm$0.6 & -18.8$\pm$1.1 & -22.18$\pm$0.40 & 9.6$\pm$0.2 & 9.60$\pm$0.16 & 1 \\
Q3-2 & -22.4$\pm$0.5 & -21.1$\pm$0.8 & -25.45$\pm$0.04 & 10.7$\pm$0.2 & 10.91$\pm$0.01 & 1 \\
Q3-3 & -22.1$\pm$0.6 & -19.9$\pm$1.0 & -25.28$\pm$0.03 & 11.1$\pm$0.2 & 11.34$\pm$0.01 & 0 \\
Q3-4 & -19.2$\pm$0.7 & -18.1$\pm$1.1 & -21.66$\pm$0.26 & 9.4$\pm$0.3 & 9.39$\pm$0.11 & 1 \\
Q3-5 & -20.3$\pm$0.5 & -19.1$\pm$1.0 & -22.75$\pm$0.36 & 9.8$\pm$0.2 & 9.83$\pm$0.14 & 1 \\
Q3-6 & -19.7$\pm$0.5 & -19.8$\pm$0.5 & & 10.1$\pm$0.2 & & 0 \\
Q3-A1 & -24.5$\pm$0.5 & -24.0$\pm$0.5 & & 12.0$\pm$0.2 & & 0 \\
Q3-8 & -21.2$\pm$0.5 & -20.1$\pm$0.5 & & 10.2$\pm$0.2 & & 1 \\
Q3-10 & -22.6$\pm$0.3 & -21.7$\pm$0.5 & -24.23$\pm$0.13 & 11.2$\pm$0.1 & 10.92$\pm$0.05 & 0 \\
Q3-11 & -19.4$\pm$0.7 & -18.6$\pm$1.1 & -21.36$\pm$0.34 & 9.4$\pm$0.3 & 9.28$\pm$0.14 & 1 \\
Q3-12 & -21.1$\pm$0.5 & -19.9$\pm$0.8 & -23.58$\pm$0.21 & 10.1$\pm$0.2 & 10.16$\pm$0.08 & 1 \\
Q3-A2 & -19.6$\pm$0.5 & -19.8$\pm$0.5 & & 10.0$\pm$0.2 & & 0 \\
Q3-15 & -19.0$\pm$0.5 & -19.2$\pm$0.5 & & 9.8$\pm$0.2 & & 0 \\
Q3-16 & -17.8$\pm$0.5 & -17.0$\pm$1.1 & & 8.8$\pm$0.2 & & 1 \\
Q3-17 & -19.7$\pm$0.6 & -17.5$\pm$1.1 & -22.85$\pm$0.44 & 9.6$\pm$0.2 & 9.87$\pm$0.18 & 1 \\
Q3-18 & -20.7$\pm$0.4 & -19.8$\pm$0.8 & -22.99$\pm$0.14 & 10.0$\pm$0.2 & 9.93$\pm$0.06 & 1 \\
Q3-19 & -20.9$\pm$0.4 & -19.8$\pm$0.6 & -23.56$\pm$0.06 & 10.6$\pm$0.1 & 10.66$\pm$0.02 & 0 \\
Q3-20 & -18.2$\pm$0.5 & -17.1$\pm$1.0 & & 9.0$\pm$0.2 & & 1 \\
Q3-21 & -22.1$\pm$0.4 & -20.6$\pm$0.7 & -24.91$\pm$0.08 & 10.5$\pm$0.2 & 10.70$\pm$0.03 & 1 \\
Q3-23 & -21.4$\pm$0.6 & -19.8$\pm$1.0 & -24.50$\pm$0.07 & 10.8$\pm$0.3 & 11.03$\pm$0.03 & 0 \\
Q3-24 & -18.4$\pm$0.7 & -16.6$\pm$1.1 & -22.17$\pm$0.27 & 9.6$\pm$0.3 & 10.10$\pm$0.11 & 0 \\
Q3-25 & -22.2$\pm$0.5 & -19.5$\pm$0.8 & -25.68$\pm$0.07 & 11.1$\pm$0.2 & 11.50$\pm$0.03 & 0 \\
Q3-26 & -20.9$\pm$0.7 & -19.0$\pm$1.1 & -23.76$\pm$0.09 & 10.6$\pm$0.3 & 10.73$\pm$0.04 & 0 \\
Q3-27 & -20.7$\pm$0.7 & -19.3$\pm$1.1 & -22.99$\pm$0.48 & 10.0$\pm$0.3 & 9.93$\pm$0.19 & 1 \\
Q3-28 & -22.5$\pm$0.5 & -20.5$\pm$0.8 & -25.57$\pm$0.02 & 11.2$\pm$0.2 & 11.46$\pm$0.01 & 0 \\
Q3-29 & -21.9$\pm$0.6 & -21.1$\pm$0.9 & -23.81$\pm$0.22 & 10.4$\pm$0.2 & 10.25$\pm$0.09 & 1 \\
Q3-30 & -19.7$\pm$0.5 & -19.5$\pm$0.5 & & 9.6$\pm$0.2 & & 1 \\
Q3-31 & -20.5$\pm$0.5 & -19.2$\pm$0.9 & -23.35$\pm$0.20 & 9.9$\pm$0.2 & 10.07$\pm$0.08 & 1 \\
Q3-A3 & -17.3$\pm$0.7 & -16.5$\pm$1.1 & -21.36$\pm$0.67 & 9.1$\pm$0.3 & 9.77$\pm$0.27 & 0 \\
Q3-33 & -16.3$\pm$0.4 & -15.7$\pm$0.6 & & 8.2$\pm$0.2 & & 1 \\
Q3-34 & -21.3$\pm$0.5 & -18.4$\pm$0.6 & -23.00$\pm$0.18 & 10.2$\pm$0.2 & 9.93$\pm$0.07 & 1 \\
Q3-35 & -16.6$\pm$0.5 & -14.2$\pm$0.6 & & 8.3$\pm$0.2 & & 1 \\
Q3-36 & -20.0$\pm$0.5 & -17.1$\pm$0.6 & -22.32$\pm$0.20 & 9.7$\pm$0.2 & 9.66$\pm$0.08 & 1 \\
Q3-A4 & -19.9$\pm$0.5 & -19.2$\pm$0.5 & & 10.2$\pm$0.2 & & 0 \\
Q3-38 & -20.9$\pm$0.5 & -19.7$\pm$0.5 & -23.28$\pm$0.28 & 10.0$\pm$0.2 & 10.04$\pm$0.11 & 1 \\
Q3-39 & -21.2$\pm$0.5 & -19.5$\pm$0.5 & -23.51$\pm$0.22 & 10.2$\pm$0.2 & 10.14$\pm$0.09 & 1 \\
Q3-40 & -21.7$\pm$0.5 & -19.4$\pm$0.8 & -24.87$\pm$0.04 & 10.9$\pm$0.2 & 11.18$\pm$0.02 & 0 \\
Q3-A5 & -20.1$\pm$0.5 & -20.4$\pm$0.5 & & 10.2$\pm$0.2 & & 0 \\
Q3-A6 & -19.5$\pm$0.5 & -19.3$\pm$0.5 & & 9.5$\pm$0.2 & & 1 \\
Q3-43 & -21.2$\pm$0.5 & -20.0$\pm$0.9 & & 10.2$\pm$0.2 & & 1 \\
Q3-44 & -20.6$\pm$0.5 & -19.0$\pm$0.8 & -23.87$\pm$0.07 & 9.9$\pm$0.2 & 10.28$\pm$0.03 & 1 \\
Q3-45 & -19.9$\pm$0.5 & -18.3$\pm$0.9 & -22.56$\pm$0.34 & 10.2$\pm$0.2 & 10.26$\pm$0.14 & 0 \\
Q3-46 & -20.0$\pm$0.5 & -18.4$\pm$0.9 & -22.36$\pm$0.32 & 10.2$\pm$0.2 & 10.17$\pm$0.13 & 0 \\
Q3-A7 & -21.4$\pm$0.5 & -20.1$\pm$0.8 & -24.33$\pm$0.11 & 10.8$\pm$0.2 & 10.96$\pm$0.05 & 0 \\
Q3-A8 & -20.9$\pm$0.5 & -20.1$\pm$0.5 & & 10.6$\pm$0.2 & & 0 \\
Q3-49 & -23.0$\pm$0.3 & -21.9$\pm$0.5 & -25.44$\pm$0.05 & 11.4$\pm$0.1 & 11.41$\pm$0.02 & 0 \\

\hline
\hline                                   %inserts single line
\end{tabular}
\end{table*}

\begin{table*}          
% title of Table
\centering                          % used for centering table
\begin{tabular}{ccccccc}
\hline\hline                 % inserts double horizontal lines
ID & $M_{i}$ & $M_{g}$ & $M_{Ks}$ & log(M$^{*}_{i}$/M$_{\odot}$) & log(M$^{*}_{Ks}$/M$_{\odot}$) & ELGs \\
\hline
Q4-A1 & -17.3$\pm$0.5 & -16.9$\pm$0.5 & & 8.6$\pm$0.2 & & 1 \\
Q4-2 & -21.1$\pm$0.6 & -19.1$\pm$1.0 & -24.22$\pm$0.08 & 10.6$\pm$0.3 & 10.92$\pm$0.03 & 0 \\
Q4-3 & -21.1$\pm$0.6 & -19.6$\pm$1.1 & -23.74$\pm$0.18 & 10.1$\pm$0.3 & 10.23$\pm$0.07 & 1 \\
Q4-4 & -19.8$\pm$0.7 & -18.0$\pm$1.0 & -22.50$\pm$0.25 & 10.1$\pm$0.3 & 10.23$\pm$0.10 & 0 \\
Q4-5 & -20.6$\pm$0.6 & -19.2$\pm$1.0 & -23.31$\pm$0.32 & 9.9$\pm$0.2 & 10.06$\pm$0.13 & 1 \\
Q4-A2 & -18.7$\pm$0.5 & -18.4$\pm$0.5 & & 9.7$\pm$0.2 & & 0 \\
Q4-7 & -20.1$\pm$0.5 & -19.5$\pm$0.8 & -21.98$\pm$0.20 & 9.7$\pm$0.2 & 9.52$\pm$0.08 & 1 \\
Q4-8 & -18.2$\pm$0.7 & -17.3$\pm$1.1 & -20.44$\pm$0.32 & 9.0$\pm$0.3 & 8.90$\pm$0.13 & 1 \\
Q4-9 & -21.7$\pm$0.6 & -19.8$\pm$0.9 & -24.91$\pm$0.02 & 10.9$\pm$0.2 & 11.2$\pm$0.01 & 0 \\
Q4-10 & -19.7$\pm$0.5 & -18.6$\pm$0.5 & & 10.1$\pm$0.2 & & 0 \\
Q4-11 & -18.5$\pm$0.7 & -16.9$\pm$1.1 & & 9.6$\pm$0.3 & & 0 \\
Q4-13 & -19.8$\pm$0.6 & -18.0$\pm$1.0 & -22.43$\pm$0.30 & 10.1$\pm$0.2 & 10.20$\pm$0.12 & 0 \\
Q4-14 & -19.7$\pm$0.7 & -18.9$\pm$1.1 & -21.39$\pm$0.65 & 9.6$\pm$0.3 & 9.29$\pm$0.26 & 1 \\
Q4-15 & -19.1$\pm$0.7 & -18.4$\pm$1.1 & & 9.3$\pm$0.3 & & 1 \\
Q4-16 & -21.1$\pm$0.7 & -19.0$\pm$1.2 & -24.34$\pm$0.06 & 10.6$\pm$0.3 & 10.97$\pm$0.02 & 0 \\
Q4-A3 & -19.3$\pm$0.5 & -18.8$\pm$0.5 & & 9.4$\pm$0.2 & & 1 \\
Q4-18 & -19.5$\pm$0.8 & -16.7$\pm$1.2 & -22.82$\pm$0.16 & 10.0$\pm$0.3 & 10.36$\pm$0.07 & 0 \\
Q4-19 & -20.1$\pm$0.6 & -19.3$\pm$1.0 & -23.09$\pm$0.18 & 9.7$\pm$0.3 & 9.97$\pm$0.07 & 1 \\
Q4-20 & -21.1$\pm$0.6 & -19.0$\pm$1.1 & -24.37$\pm$0.05 & 10.7$\pm$0.3 & 10.98$\pm$0.02 & 0 \\
Q4-21 & -21.2$\pm$0.5 & -19.7$\pm$0.5 & & 10.7$\pm$0.2 & & 0 \\
Q4-22 & -19.7$\pm$0.7 & -18.8$\pm$1.2 & -21.56$\pm$0.58 & 9.6$\pm$0.3 & 9.35$\pm$0.23 & 1 \\
Q4-23 & -20.2$\pm$0.7 & -18.1$\pm$1.1 & -23.10$\pm$0.24 & 10.3$\pm$0.3 & 10.47$\pm$0.09 & 0 \\
Q4-24 & -20.3$\pm$0.7 & -18.4$\pm$1.2 & -23.08$\pm$0.16 & 10.3$\pm$0.3 & 10.46$\pm$0.06 & 0 \\
Q4-25 & -18.4$\pm$0.8 & -16.7$\pm$1.2 & & 9.6$\pm$0.3 & & 0 \\
Q4-26 & -21.5$\pm$0.5 & -20.8$\pm$0.9 & -23.51$\pm$0.17 & 10.3$\pm$0.2 & 10.14$\pm$0.07 & 1 \\
Q4-27 & -19.3$\pm$0.6 & -16.5$\pm$1.0 & -22.16$\pm$0.34 & 9.9$\pm$0.3 & 10.10$\pm$0.14 & 0 \\
Q4-28 & -20.3$\pm$0.7 & -18.4$\pm$1.2 & -23.27$\pm$0.17 & 10.3$\pm$0.3 & 10.54$\pm$0.07 & 0 \\
Q4-A4 & -19.8$\pm$0.5 & -18.6$\pm$0.8 & -21.90$\pm$0.45 & 9.6$\pm$0.2 & 9.49$\pm$0.18 & 1 \\
Q4-A5 & -19.6$\pm$0.5 & -19.5$\pm$0.5 & & 10.0$\pm$0.2 & & 0 \\
Q4-31 & -21.6$\pm$0.6 & -19.4$\pm$1.0 & -24.74$\pm$0.04 & 10.9$\pm$0.3 & 11.13$\pm$0.02 & 0 \\
Q4-32 & -24.4$\pm$0.1 & -22.2$\pm$0.3 & -27.04$\pm$0.02 & 12.0$\pm$0.1 & 12.05$\pm$0.01 & 0 \\
Q4-33 & -18.6$\pm$0.4 & -17.8$\pm$0.7 & -20.06$\pm$0.19 & 9.1$\pm$0.1 & 8.76$\pm$0.08 & 1 \\
Q4-A6 & -19.3$\pm$0.5 & -19.1$\pm$0.5 & & 9.4$\pm$0.2 & & 1 \\
Q4-35 & -18.9$\pm$0.5 & -18.2$\pm$1.1 & -22.14$\pm$0.15 & 9.3$\pm$0.2 & 9.59$\pm$0.06 & 1 \\
Q4-36 & -23.5$\pm$0.3 & -21.7$\pm$0.5 & -26.43$\pm$0.01 & 11.1$\pm$0.1 & 11.30$\pm$0.01 & 1 \\
Q4-37 & -18.0$\pm$0.5 & -16.2$\pm$0.9 & & 8.9$\pm$0.2 & & 1 \\
Q4-38 & -21.8$\pm$0.5 & -19.9$\pm$0.9 & -24.80$\pm$0.04 & 10.9$\pm$0.2 & 11.15$\pm$0.02 & 0 \\
Q4-39 & -20.7$\pm$0.5 & -18.8$\pm$0.5 & & 10.5$\pm$0.2 & & 0 \\
Q4-40 & -22.2$\pm$0.5 & -21.0$\pm$0.5 & & 10.6$\pm$0.2 & & 1 \\
Q4-41 & -20.7$\pm$0.7 & -18.7$\pm$1.1 & -23.54$\pm$0.13 & 10.5$\pm$0.3 & 10.65$\pm$0.05 & 0 \\
Q4-42 & -21.2$\pm$0.6 & -19.6$\pm$1.0 & -23.85$\pm$0.34 & 10.7$\pm$0.2 & 10.77$\pm$0.13 & 0 \\
Q4-43 & -20.2$\pm$0.6 & -18.5$\pm$1.0 & -22.55$\pm$0.32 & 10.3$\pm$0.3 & 10.25$\pm$0.13 & 0 \\
Q4-44 & -22.3$\pm$0.5 & -20.1$\pm$0.9 & -25.73$\pm$0.02 & 11.1$\pm$0.2 & 11.52$\pm$0.01 & 0 \\
Q4-45 & -20.5$\pm$0.6 & -18.6$\pm$1.0 & -23.26$\pm$0.21 & 10.4$\pm$0.2 & 10.53$\pm$0.08 & 0 \\
Q4-46 & -21.7$\pm$0.3 & -20.7$\pm$0.6 & -23.60$\pm$0.13 & 10.9$\pm$0.1 & 10.67$\pm$0.05 & 0 \\
Q4-47 & -21.1$\pm$0.7 & -19.2$\pm$1.1 & -24.04$\pm$0.09 & 10.6$\pm$0.3 & 10.85$\pm$0.04 & 0 \\
Q4-48 & -21.5$\pm$0.5 & -19.6$\pm$0.8 & -24.65$\pm$0.07 & 10.8$\pm$0.2 & 11.09$\pm$0.03 & 0 \\
Q4-49 & -22.2$\pm$0.5 & -20.0$\pm$0.8 & -25.37$\pm$0.04 & 11.1$\pm$0.2 & 11.38$\pm$0.02 & 0 \\
Q4-50 & -19.2$\pm$0.7 & -17.9$\pm$1.2 & & 9.9$\pm$0.3 & & 0 \\
Q4-51 & -22.2$\pm$0.4 & -20.2$\pm$0.8 & -25.14$\pm$0.03 & 11.1$\pm$0.2 & 11.29$\pm$0.01 & 0 \\
Q4-52 & -20.6$\pm$0.7 & -18.6$\pm$1.1 & -23.55$\pm$0.12 & 10.4$\pm$0.3 & 10.65$\pm$0.05 & 0 \\
Q4-54 & -20.9$\pm$0.5 & -19.5$\pm$0.6 & -23.93$\pm$0.17 & 10.6$\pm$0.2 & 10.80$\pm$0.07 & 0 \\
Q4-55 & -20.2$\pm$0.4 & -18.7$\pm$0.5 & -23.39$\pm$0.19 & 10.3$\pm$0.2 & 10.59$\pm$0.07 & 0 \\
Q4-56 & -19.7$\pm$0.5 & -18.8$\pm$0.5 & & 9.6$\pm$0.2 & & 1 \\
Q4-57 & -20.9$\pm$0.5 & -18.7$\pm$0.5 & -22.93$\pm$0.18 & 10.6$\pm$0.2 & 10.40$\pm$0.07 & 0 \\
Q4-58 & -22.8$\pm$0.5 & -18.7$\pm$0.5 & -25.17$\pm$0.03 & 11.3$\pm$0.2 & 11.30$\pm$0.01 & 0 \\
Q4-59 & -20.5$\pm$0.5 & -19.4$\pm$0.5 & & 10.4$\pm$0.2 & & 0 \\
Q4-60 & -21.4$\pm$0.5 & -19.4$\pm$0.5 & -23.22$\pm$0.13 & 10.8$\pm$0.2 & 10.52$\pm$0.05 & 0 \\
Q4-61 & -22.7$\pm$0.5 & -21.5$\pm$0.5 & -25.92$\pm$0.13 & 11.3$\pm$0.2 & 11.6$\pm$0.05 & 0 \\
Q4-62 & -21.2$\pm$0.5 & -19.3$\pm$0.5 & -23.12$\pm$0.21 & 10.7$\pm$0.2 & 10.48$\pm$0.08 & 0 \\
Q4-63 & -21.4$\pm$0.5 & -19.4$\pm$0.5 & -23.80$\pm$0.13 & 10.8$\pm$0.2 & 10.75$\pm$0.05 & 0 \\
Q4-64 & -20.5$\pm$0.7 & -18.5$\pm$0.9 & -24.01$\pm$0.11 & 10.4$\pm$0.3 & 10.83$\pm$0.04 & 0 \\
\hline                                   %inserts single line
\end{tabular}
\end{table*}

% ---------------------------------------------------------------------------
\begin{table*}
\label{table:3}
\caption{Oxygen abundances of the SF galaxies with available emission line measurements. Column 2 show oxygen abundances estimated by using the O3N2 strong-line method.}             
% title of Table
\centering                          % used for centering table
\begin{tabular}{cc}
\hline\hline
ID & 12+log(O/H)$_{O3N2}$ \\
\hline
Q1-3 & 8.37$\pm$0.03 \\
Q1-A1 & 8.44$\pm$0.04 \\
Q1-5 & 8.27$\pm$0.03 \\
Q1-10 & 8.71$\pm$0.04 \\
Q1-12 & 8.56$\pm$0.08 \\
Q1-15 & 8.10$\pm$0.06 \\
Q1-18 & 8.70$\pm$0.06 \\
Q1-19 & 8.42$\pm$0.03 \\
Q1-25 & 8.58$\pm$0.06 \\
Q1-37 & 8.27$\pm$0.05 \\
Q1-40 & 8.68$\pm$0.02 \\
Q2-1 & 8.61$\pm$0.08 \\
Q2-3 & 8.35$\pm$0.04 \\
Q2-9 & 8.22$\pm$0.05 \\
Q2-11 & 8.31$\pm$0.05 \\
Q2-22 & 8.41$\pm$0.05 \\
Q2-A4 & 8.51$\pm$0.02 \\
Q2-32 & 8.49$\pm$0.03 \\
Q2-37 & 8.43$\pm$0.05 \\
Q2-38 & 8.34$\pm$0.05 \\
Q2-39 & 8.50$\pm$0.02 \\
Q3-1 & 8.26$\pm$0.03 \\
Q3-2 & 8.63$\pm$0.03 \\
Q3-4 & 8.49$\pm$0.04 \\
Q3-5 & 8.34$\pm$0.04 \\
Q3-8 & 8.73$\pm$0.04 \\
Q3-11 & 8.25$\pm$0.05 \\
Q3-12 & 8.64$\pm$0.10 \\
Q3-16 & 8.62$\pm$0.11 \\
Q3-18 & 8.71$\pm$0.06 \\
Q3-21 & 8.72$\pm$0.09 \\
Q3-30 & 8.15$\pm$0.03 \\
Q3-31 & 8.63$\pm$0.10 \\
Q3-34 & 8.32$\pm$0.02 \\
Q3-35 & 8.21$\pm$0.10 \\
Q3-36 & 8.52$\pm$0.05 \\
Q3-39 & 8.60$\pm$0.04 \\
Q3-A6 & 8.23$\pm$0.05 \\
Q3-44 & 8.57$\pm$0.06 \\
Q4-3 & 8.59$\pm$0.09 \\
Q4-5 & 8.54$\pm$0.07 \\
Q4-7 & 8.33$\pm$0.03 \\
Q4-8 & 8.17$\pm$0.02 \\
Q4-14 & 8.26$\pm$0.02 \\
Q4-15 & 8.09$\pm$0.03 \\
Q4-19 & 8.27$\pm$0.04 \\
Q4-22 & 8.75$\pm$0.11 \\
Q4-26 & 8.42$\pm$0.03 \\
Q4-A4 & 8.46$\pm$0.06 \\
Q4-33 & 8.37$\pm$0.05 \\
Q4-A6 & 8.49$\pm$0.05 \\
Q4-35 & 8.24$\pm$0.02 \\
Q4-40 & 8.77$\pm$0.19 \\
Q4-56 & 8.31$\pm$0.03 \\
\hline                                   %inserts single line
\end{tabular}
\end{table*}

\end{appendix}
\end{document}